\documentclass[]{article}
\usepackage{jheppub}

\usepackage{amsmath}
\usepackage[small]{subfigure}
\usepackage{xcolor}
\usepackage{soul}

\title{NLO Oriented Event-Shape Distributions for Massive Quarks}

\preprint{\begin{flushright} IFT-UAM/CSIC-22-142\end{flushright}\vspace*{-2cm}}

\author[a,b]{Alejandro~Bris,}
\author[c]{Nestor~G.\,Gracia}
\author[c]{and~Vicent~Mateu}

\affiliation[a]{Departamento de F\'isica Te\'orica, Universidad Aut\'onoma de Madrid,\\E-28049, Madrid, Spain}
\affiliation[b]{Instituto de F\'isica Te\'orica UAM-CSIC,\\E-28049 Madrid, Spain}
\affiliation[c]{Departamento de F\'isica Fundamental e IUFFyM, Universidad de Salamanca,\\E-37008 Salamanca, Spain}

\emailAdd{alejandro.bris@uam.es}
\emailAdd{ngonzalez@usal.es}
\emailAdd{vmateu@usal.es}

\abstract{In this article we compute the cross section for the process $e^+e^- \to Q\overline Q+X$, with $Q$ a heavy quark, differential in a given event shape $e$ and the angle $\theta_T$ between
the thrust axis and the beam direction. These observables are usually referred to as oriented event shapes, and it has been shown that
the $\theta_T$ dependence can be split in two structures, dubbed the unoriented and angular terms.
Since the unoriented part is already known, we compute the differential and cumulative distributions in fixed-order for the angular part up to $\mathcal{O}(\alpha_s)$.
Our results show that, for the vector current, there is a non-zero $\mathcal{O}(\alpha_s^0)$ contribution, in contrast to the axial-vector current or for massless quarks.
This entails that for the vector current one should expect singular terms at $\mathcal{O}(\alpha_s)$ as well as infrared divergences in real- and virtual-radiation
diagrams that should cancel when added up. On the phenomenological side, and taking into account that electroweak factors enhance the vector
current, it implies that finite bottom-mass effects are an important correction since they are not damped by a power of the strong coupling and therefore cannot
be neglected in precision studies. Finally, we show that the total angular distribution for the vector current has a Sommerfeld enhancement at threshold.}

\begin{document}
\maketitle
\flushbottom

\section{Introduction}
Although the theoretical knowledge on massive event shapes still lags behind the astonishing precision achieved for massless jets, where some ingredients
necessary for next-to-next-to-next-to-next-to-log (N$^4$LL) precision have been computed in Ref.~\cite{Duhr:2022yyp,Duhr:2022cob}, recent
years have witnessed a steady progress in the subject.
On the fixed-order side, numerical results in the form of binned distributions can be obtained for unoriented cross sections up to $\mathcal{O}(\alpha_s^2)$ from
partonic Monte Carlo computer programs~\cite{Nason:1997nw,Bernreuther:1997jn,Rodrigo:1997gy,Rodrigo:1999qg}. Recently, full analytic control has been gained for the
singular structures (that is, Dirac delta or plus functions)
up to $\mathcal{O}(\alpha_s)$ while a highly efficient numerical strategy has been devised, such that machine-precision, unbinned
distributions can be obtained in fractions of a second \cite{Lepenik:2019jjk}. When it comes to resummation, factorization theorems for heavy quarks have been
established for event shapes such as two-jettiness and hemisphere masses~\cite{Fleming:2007xt,Fleming:2007qr}, which can be easily adapted to
C-jettiness~\cite{Gardi:2003iv}, a generalization of C-parameter for massive quarks~\cite{Parisi:1978eg,Donoghue:1979vi}.
In Ref~\cite{Bris:2020uyb} the computation of the NLO jet function for these observables in the P- and
E-schemes~\cite{Salam:2001bd} was carried out, and the relevant expressions for next-to-next-to-leading-log (N$^2$LL) resummation were provided. For
2-jettiness~\cite{Stewart:2009yx} and hemisphere masses, all necessary pieces to achieve next-to-next-to-next-to-log (N$^3$LL) precision are by now
known~\cite{Jain:2008gb,Gritschacher:2013pha,Pietrulewicz:2014qza,Hoang:2015vua,Hoang:2019fze}. Phenomenological studies at this order have been
carried out in Ref.~\cite{Bachu:2020nqn}, investigating the important role played by the soft-function and primary-quark mass renormalons in
robust determinations of the top quark mass at a future linear collider, and how using the MSR scheme for the quark
mass~\cite{Hoang:2008yj,Hoang:2017suc} stabilizes the peak position order by order in perturbation
theory. The {\scshape Pythia}~8.205~\cite{Sjostrand:2007gs} top quark mass parameter is calibrated at N$^2$LL in Ref.~\cite{Butenschoen:2016lpz},
showing it cannot be identified with the pole mass.

Similarly, our knowledge on cross sections in which no information on the event's orientation with respect to the beam direction is retained (that is, when only the
geometrical shape of the event is taken into account) is way more advanced than for the more differential case in which such orientation is recorded. One convenient,
infrared- and collinear-safe way of determining the event's orientation is measuring the angle formed by the $e^+e^-$ beam direction and the thrust axis, defined as
the unit vector $\hat n$ that maximizes the sum appearing in the thrust event-shape's definition~\cite{Farhi:1977sg}:
\begin{equation}\label{eq:axisDef}
\tau = 1 - \max_{\hat n}\frac{\sum_i |\vec{p}_i\!\cdot \hat{n}|}{\sum_i |\vec{p}_i|}\,,
\end{equation}
where the index $i$ runs over all particles in the final state. This angle will be denoted by $\theta_T$ in what follows.
An alternative possibility emerges in this case:~measuring only the orientation but not the shape of
the event itself. This gives rise to the so-called total oriented cross section $R_{\rm or}$, an interesting observable which is more sensitive to $\alpha_s$ than
the total cross section but suffers from milder hadronization corrections than differential event-shape distributions. Hence it emerges as a viable candidate for a
competitive determination of the strong coupling.

Early studies of orientation in hadron production go back to Ref.~\cite{Lampe:1992au}, in which $\mathcal{O}(\alpha_s)$ analytical and $\mathcal{O}(\alpha_s^2)$
numerical results were provided for $R_{\rm or}$. In the more recent work of Ref.~\cite{Mateu:2013gya} it was shown that the differential cross section in
$\cos(\theta_T)$ can be decomposed into structures with orbital angular momentum $0$ and $1$. It is however more convenient to consider linear combinations of
those such that one of them is the unoriented cross section and the other one vanishes upon integration over all angles:
\begin{align}\label{eq:decomposition}
\frac{1}{\sigma_0} \frac{{\rm d} \sigma}{{\rm d} \!\cos (\theta_T) {\rm d} e} &=\frac{3}{8}[1+\cos^2(\theta_T)]
\frac{1}{\sigma_0} \frac{{\rm d} \sigma}{{\rm d} e}
+[1 - 3 \cos^2(\theta_T)]\frac{1}{\sigma_0} \frac{{\rm d} \sigma_{\rm ang}}{{\rm d} e}\,,\\
\frac{1}{\sigma_0} \frac{{\rm d} \sigma}{{\rm d} e} &=\int_{-1}^1 {\rm d}\! \cos (\theta_T) \frac{1}{\sigma_0} \frac{{\rm d} \sigma}{{\rm d} \!\cos (\theta_T) {\rm d} e}\,, \nonumber\\
\frac{1}{\sigma_0} \frac{{\rm d} \sigma_{\rm ang}}{{\rm d} e} &=\frac{3}{8}\! \int_{-1}^1 {\rm d} \!\cos (\theta_T)
[2-5 \cos^2(\theta_T)] \frac{1}{\sigma_0} \frac{{\rm d} \sigma}{{\rm d} \!\cos (\theta_T) {\rm d} e} \,,\nonumber\\
R_{\rm ang} &= \int {\rm d}e \frac{1}{\sigma_0} \frac{{\rm d} \sigma_{\rm ang}}{{\rm d} e}\,,\nonumber
\end{align}
where $\sigma_0$ is the Born cross-section, that shall be defined later in this section, and $R_{\rm ang}$ is the total angular cross-section, which does not
depend on any particular event shape $e$.
This result holds for massive or massless particles, and is valid for hadronic or partonic cross sections. In Ref.~\cite{Mateu:2013gya} it was shown that for massless quarks
the angular term is zero at lowest order and at $\mathcal{O}(\alpha_s)$ analytic results were given for a number of event-shapes. Moreover, event-shape distributions do not
have singular terms. Event2~\cite{Catani:1996vz} was used to obtain binned distributions at $\mathcal{O}(\alpha_s^2)$, what enabled a numerical determination
of $R_{\rm ang}$ at this order by an extrapolation of the cumulative distribution for a set of event shapes. Remarkably, the obtained results were not compatible with the
numbers quoted in Ref.~\cite{Lampe:1992au} and to date the discrepancy stands.

Finally, in Ref.~\cite{Hagiwara:2010cd} a factorization theorem for the thrust angular distribution was derived in Soft-Collinear Effective
Theory (SCET)~\cite{Bauer:2000ew, Bauer:2000yr, Bauer:2001ct, Bauer:2001yt, Bauer:2002nz}, involving the known soft and jet functions but also
additional hard and jet functions. Since, as already mentioned, the angular distribution is not singular, this
new jet function is sub-leading in the SCET power counting
and does not involve distributions. The new ingredients were computed at next-to-leading order (NLO), allowing next-to-leading-log (NLL) resummed precision.

Measurements of event shape distributions differential in $\theta_T$ are available from the DELPHI
collaboration since long, see e.g.\ Ref.~\cite{DELPHI:2000uri}, where also a determination of the
strong coupling is presented. Fixed-order theoretical expressions at $\mathcal{O}(\alpha_s^2)$ were
used, accounting for hadronization effects trough parton shower Monte Carlos.
A direct measurement of the angular cross section was performed by the OPAL
collaboration, see Ref.~\cite{OPAL:1998tla}. To the best of our knowledge, no full fledged analysis beyond
$\mathcal{O}(\alpha_s^2)$, including resummation and with a consistent treatment of non-perturbative
power corrections exists. There are, however, ongoing efforts to
determine $\alpha_s$ from measurements of the total angular cross-section.

In this work we take a first look at oriented event shapes initiated by massive jets. Exploring the fixed-order structure of the distribution at NLO is a necessary step before
adapting the factorization theorem derived in Ref.~\cite{Hagiwara:2010cd}. We find that, in contrast to the massless situation, the vector current generates a (singular)
contribution to the oriented cross section already at $\mathcal{O}(\alpha_s^0)$. Therefore one expects (even more) singular structures at higher perturbative
orders. In fact, we find the exact same structure as in Ref.~\cite{Lepenik:2019jjk}:
\begin{align}\label{eq:general-diff}
\frac{1}{\sigma^C_0} \frac{{\rm d} \sigma^C_{\rm ang}}{{\rm d} e} =\,&R^{0,C}_{\rm ang}(\hat m)\,\delta [e - e_{\rm min}(\hat m)] +
C_F \frac{\alpha_s(\mu)}{\pi} A^{{\rm ang},C}_{e}({\hat m})\delta [e - e_{\rm min}(\hat m)] \\
& + C_F \frac{\alpha_s(\mu)}{\pi} B^{{\rm ang},C}_{\rm plus}({\hat m})\biggl[\frac{1}{e - e_{\rm min}(\hat m)} \biggr]_+
+ C_F \frac{\alpha_s(\mu)}{\pi} F^{\rm ang}_{C,e} (e, \hat{m}) + \mathcal{O}(\alpha_s^2)\,,\nonumber
\end{align}
with $F^{\rm ang}_{C,e}$ containing only non-singular terms and $\hat m = m/Q$ standing for the quark's
reduced mass. Here $\sigma^C_0$ with $C=V,A$ for the vector and axial-vector currents, respectively, is the Born
cross section, which we define as the lowest-order cross section for producing massless quarks mediated by a photon and a $Z$-boson, hence accounting for
electroweak factors and the fact that quarks are produced in $N_c=3$ colors:
\begin{align}\label{eq:EW}
\sigma^V_0 = \,& \frac{N_c}{3} \frac{4 \pi \alpha_{\rm em}^2}{s} \!
\left[ Q^2_q + \frac{v^2_f (v_e^2 + a_e^2)}{(1 - \hat{m}_Z^2)^2 +\!
\bigl( \frac{\Gamma_Z}{m_Z} \bigr)^{\!2}} + \frac{2 Q_q v_e v_f (1 -
\hat{m}_Z^2)}{(1 - \hat{m}^2_Z)^2 + \bigl( \frac{\Gamma_Z}{m_Z} \bigr)^{\!2}}
\right]\!, \\
\sigma^A_0 = \,& \frac{N_c}{3} \frac{4 \pi \alpha_{\rm em}^2}{s} \!
\left[ \frac{a_f^2 (v_e^2 + a_e^2)}{(1 - \hat{m}_Z^2)^2 + \bigl( \frac{\Gamma_Z}{m_Z} \bigr)^{\!2}} \right]\!. \nonumber
\end{align}
Here $s=(p_1+p_2)^2\equiv Q^2$ is the center-of-mass energy squared with $p_{1,2}$ the $4$-momenta of the initial-state leptons as shown in Fig.~\ref{fig:Tree},
$\hat{m}_Z = m_Z/Q$ the reduced mass of the $Z$-boson and $\Gamma_{\!Z}$ its width, $\alpha_{\rm em}$ the fine structure constant, $Q_q$ the electric
charge of the quark (not to be confused with the center-of-mass energy $Q$), and $v_f$ and $a_f$ the vector and axial-vector charges for the fermion $f$
\begin{equation}
v_f = \dfrac{T^f_3 - 2 Q_f \sin^2 (\theta_W)}{\sin (2\theta_W)},
\qquad a_f = \dfrac{T^f_3}{\sin (2\theta_W)},
\end{equation}
with $T^f_3$ the third component of weak isospin and $\theta_W$ Weinberg's angle. In our case $f=q,e$ for quarks and electrons, respectively.
For the axial-vector current one has $R^{0,A}_{\rm ang}=A^{{\rm ang},A}_{e}=B^{{\rm ang},A}_{\rm plus}=0$ in Eq.~\eqref{eq:general-diff}. Therefore,
for simplicity we adopt the convention $R^{0}_{\rm ang}\equiv R^{0,V}_{\rm ang}$, $A^{\rm ang}_{e}\equiv A^{{\rm ang},V}_{e}$ and
$B^{\rm ang}_{\rm plus} =B^{{\rm ang},V}_{\rm plus}$, and do not refer to the axial-current coefficients anymore.

The Feynman diagram with a virtual gluon shown in Fig.~\ref{fig:virtual} will also have a non-vanishing contribution. Since the massive quark form factor
contains infrared (IR) divergences, they must also be present in the real-radiation contribution of Fig.~\ref{fig:real}, such that the sum of both contributions,
after integrating their respective phase spaces, must remain finite. Therefore, we carry out the computation in $d=4-2\varepsilon$
dimensions to regularize these divergences. For the axial-vector
current the entire computation can be performed in $4$ dimensions, but as a cross check we also kept $d=4-2\varepsilon$. Our strategy
will be to project out the angular pieces using the third line of Eq.~\eqref{eq:decomposition} at very early stages of the calculation. As a cross check, we have also
computed directly the complete $\theta_T$-differential distribution, verifying that dimensional regularization does not introduce additional angular structures and
finding the same result as the direct computation presented in the bulk of this manuscript.

This paper is organized as follows: in Sec.~\ref{sec:general} we sketch the general structure of the computation, giving explicit expressions at each
order in the strong coupling; the $d$-dimensional phase space for two and three particles, differential in the polar angle of the particles' momenta, is derived in Sec.~\ref{sec:phaseSpace}, along with a discussion on the projection into the thrust axis and some angular master integrals;
in Sec.~\ref{sec:LO} the result at $\mathcal{O}(\alpha_s^0)$ is computed, together with the Born cross-section for massless quarks in $d$ dimensions
(that is, our normalization); the virtual radiation contribution is computed in Sec.~\ref{sec:virtual}, while the real radiation, which is the most involved computation,
is contained in Sec.~\ref{sec:real}. The final form of the differential cross section is derived in Sec.~\ref{sec:differential},
while in Sec.~\ref{sec:analytic} we present analytic results for the $2$-jettiness and heavy-jet-mass differential
distributions, and closed integral forms for their cumulative counterparts. A number of consistency checks on our computations and an extended numerical analysis
is to be found in Sec.~\ref{sec:numerical}, while Sec.~\ref{sec:conclusions} contains our conclusions.

\section{General Structure of the Computation}\label{sec:general}
The observables under study are inclusive in the number of particles produced, and therefore can be written as the incoherent sum of exclusive cross sections in which
$n\ge2$ partons are produced. The amplitude for each one of these $n$-particle cross sections is the coherent sum of diagrams having the same external legs
but different internal propagators. Since individual $n$-parton contributions are IR divergent, to ensure an IR-finite result one has to consistently truncate the
coherent and incoherent sums such that only terms up to a given power of $\alpha_s$ in the incoherent cross section are retained.

Since we consider electroweak interactions at Born level only, Feynman diagrams at any order in perturbation theory and with an arbitrary number of partons
in the final state will have the factors involving the initial-state leptons and $\gamma/Z$-boson propagator in common. We can therefore factorize those ahead of time.
The amplitude with $n$ partons can be written as (we omit the dependence on the particles momenta) 
\begin{equation}\label{eq:MC}
\mathcal{M}^{C}_{n\lambda}= \frac{g_{\rm em}^2}{s} L_{p\mu} H_{n\lambda}^{C,\mu}\,,\qquad H_{n\lambda}^{C,\mu} = \!\biggl[\frac{\alpha_s(\mu)}{\pi}\biggr]^{\!\frac{n-2}{2}}
\sum_{i=0}^\infty\biggl[\frac{\alpha_s(\mu)}{\pi}\biggr]^i h_{ni\lambda}^{C,\mu}\,,
\end{equation}
with $g_{\rm em}$ and $g_s$ the electromagnetic and strong couplings, respectively, and $L_p^\mu$ the leptonic part of the diagrams. The subscripts $p$ and $\lambda$
stand for the polarization of the initial- and final-state particles, respectively. The hadronic vector $H_{n\lambda}^\mu$
accounts for all the quantum corrections $h_{ni\lambda}^{C,\mu}$ with $i$ loops, for a fixed number $n$ of external partons. For convenience we have explicitly factored
out all powers of the strong coupling. Removal of ultraviolet (UV) divergences can be carried out at the level of the amplitudes, therefore
the strong coupling, quark masses (which are not explicitly shown) and $h_{ni\lambda}^{C,\mu}$ appearing in Eq.~\eqref{eq:MC} are already renormalized
(and as such, $\mu$ dependent).
The matrix element squared for the cross section with $n$ particles, averaged (summed) over the initial (final)
polarizations
can be written as
\begin{align}
M_n &\equiv \frac{1}{4} \sum_\lambda |\mathcal{M}^C_{n\lambda}|^2 =
L_{\mu\nu}\frac{g_{\rm em}^4}{s^2} \biggl[\frac{\alpha_s(\mu)}{\pi}\biggr]^{\!n-2} H_n^{C,\mu\nu},
&H_n^{C,\mu\nu} &= \sum_{i=0}^\infty \biggl[\frac{\alpha_s(\mu)}{\pi}\biggr]^{\!i} H_{ni}^{C,\mu\nu},\\
H^{C,\mu\nu}_{ni} &= \sum_{j=0}^{i} \sum_\lambda h_{nj\lambda}^{C,\mu} \bigl[h_{n,i-j,\lambda}^{C,\nu}\bigr]^\dagger,\qquad
&L_{\mu\nu} & =\frac{1}{4}\sum_p L_p^\mu L_p^{\nu\dagger} = p_1^{\mu} p_2^{\nu} + p_2^{\mu} p_1^{\nu} - \frac{s}{2} g^{\mu \nu}.\nonumber
\end{align}
We can now deal with the incoherent sum over channels with different number of particles in the final state.
At this point we add the flux factor and, in order to measure an event shape denoted generically by $e$,
insert a Dirac delta function:\footnote{In practice, to compute $R_n^{{\rm ang},C}$ one can simply
use the integral formula for $f_{e,n}^C$ but dropping $\delta[e - e(Q_n)]$.}
\begin{align}
\frac{1}{\sigma_0^V}\frac{{\rm d} \sigma^C_{\rm ang}}{{\rm d} e} & = \sum_{n=0}^\infty
\biggl[\frac{\alpha_s(\mu)}{\pi}\biggr]^n f_{e,n}^{{\rm ang},C}\biggl(\hat m,e,\frac{\mu}{Q}\biggr)\,, \quad\quad
R_{\rm ang}^C = \sum_{n=0}^\infty \biggl[\frac{\alpha_s(\mu)}{\pi}\biggr]^{\!n} R_n^{{\rm ang},C}\!\biggl(\hat m,\frac{\mu}{Q}\biggr),\\
f_{e,n}^{{\rm ang},C}\biggl(\hat m,e,\frac{\mu}{Q}\biggr) & = \frac{9\pi L_{\mu\nu}}{4N_cs^2} \sum_{i=0}^{n} \int\! {\rm d}Q_{n-i+2} H_{n+2-i,i}^{C,\mu\nu}
\delta[e - e(Q_n)][2-5\cos^2 \theta_T(Q_n)]\,,\nonumber\\
R_n^{{\rm ang},C}\!\biggl(\hat m,\frac{\mu}{Q}\biggr) & = \int {\rm d}e f_{e,n}^{{\rm ang},C}\biggl(e,\hat m,\frac{\mu}{Q}\biggr)\,,\nonumber
\end{align}
where ${\rm d}Q_n$ stands for the Lorentz-invariant $n$-particle phase space. Here $e(Q_n)$ and $\theta_T(Q_n)$ return respectively the value of the
event shape $e$ and angle $\theta_T$ at the phase space point $Q_n$. In the previous equation, as anticipated, we have projected out the
angular piece. For $n=2$ one has that $e(Q_2)=e_{\rm min}$, the lowest possible value of the event shape and $\theta_T(Q_2)=\theta$ is the angle
formed by the massive quark and the beam. In the next sections we compute the first two perturbative orders:
\begin{equation}
H^C_0 \equiv L_{\mu\nu}\!\!\int\! {\rm d}Q_2 H_{20}^{C,\mu\nu},\qquad
H^C_1 \equiv L_{\mu\nu}\Biggl[\int\! {\rm d}Q_2 H_{21}^{C,\mu\nu}+\int \!{\rm d}Q_3 H_{30}^{C,\mu\nu}\Biggr],
\end{equation}
with $H_{21}^{C,\mu\nu}=2{\rm Re}\{\sum_\lambda h_{20\lambda}^{C,\mu} \bigl[h_{21\lambda}^{C,\nu}\bigr]^\dagger\}$ and
$H_{30}^{C,\mu\nu} = \sum_\lambda h_{30\lambda}^{C,\mu} \bigl[h_{30\lambda}^{C,\nu}\bigr]^\dagger$. Finally, $h_{30\lambda}^{C,\mu}$ has
contributions from two different Feynman diagrams.

\section{\boldmath Phase Space in $d=4-2\varepsilon$ Dimensions}\label{sec:phaseSpace}
Given that the real- and virtual-radiation contributions at NLO are afflicted by IR divergences that cancel when adding up the two, one needs to
regularize these singularities in individual terms. Since QCD is a non-abelian gauge theory, regulators such as a gluon mass are not advisable
as they explicitly break gauge symmetry. Moreover, a gluon mass would translate into an additional scale and complicate computations
unnecessarily. On the other hand, dimensional regularization is gauge invariant and does not introduce additional energy scales, but causes some
spurious terms that cancel when adding up all contributions.

For the tree-level and virtual-radiation contributions we will need the $2$-body phase space differential in the polar angle
for $d=4-2\varepsilon$ dimensions. We therefore consider the $z$ axis pointing in the beam direction and the two final-state particles with equal
non-zero mass $m$. Including the flux factor one gets:
\begin{equation}\label{eq:2body}
\frac{1}{2s} \frac{{\rm d} Q_2}{{\rm d}\! \cos (\theta)} = \frac{\beta^{1 - 2 \varepsilon} \sin^{- 2 \varepsilon} (\theta)}{2^{5 - 4
\varepsilon} s^{1 + \varepsilon} \Gamma (1 - \varepsilon) \pi^{1 -
\varepsilon}}\,,
\end{equation}
with $\beta\equiv\sqrt{1-4\hat m^2}$ the quark velocity in the center of mass frame. Here $\theta$ stands for the angle defined by the quark's $3$-momentum and the beam direction,
which is identified with $\theta_1$ ---\,which defines the $z$ axis\,---, the first polar angle of the $(3-2\varepsilon)$-dimensional spherical
coordinates.
Upon integration over $\theta$, which ranges from $0$ to $\pi$, one recovers the well-known result for the totally integrated phase
space, as given for instance in Eq.~(3.5) of Ref.~\cite{Lepenik:2019jjk}. As expected, if $\varepsilon=0$
the flux-normalized $2$-particle phase space has dimensions of an area in natural units.

For the real-radiation contribution at NLO we need the $3$-particle phase space. We consider now two particles with equal mass $m$ (quarks, labeled $1$ and $2$)
and a massless particle (gluon, labeled $3$). We introduce the dimensionless variables $x_i\equiv 2E_i/Q$
with $E_i$ the energy of the $i$-th particle measured
in the center-of-mass frame and $i=1,2, 3$. Conservation of energy implies $x_1+x_2+x_3=2$.
We again define our $z$ axis in the beam direction, such that momentum conservation in the $z$ direction implies
\begin{equation}\label{eq:P3cons}
\beta_1 \cos (\theta_1) + \beta_2 \cos(\theta_2) + x_3 \cos (\theta_3) = 0\,,
\end{equation}
where we have defined $\beta_i\equiv2|\vec{p}_i|/Q=\sqrt{x_i^2-4\hat m_i^2}$
---\,not to be confused with the particle's velocity\,--- with $\hat m_1 = \hat m_2=\hat m$ and $\hat m_3=0$.
One has that $x_i\geq 2\hat m$ within the phase space.
\begin{figure}[t]\centering
\includegraphics[width=0.4\textwidth]{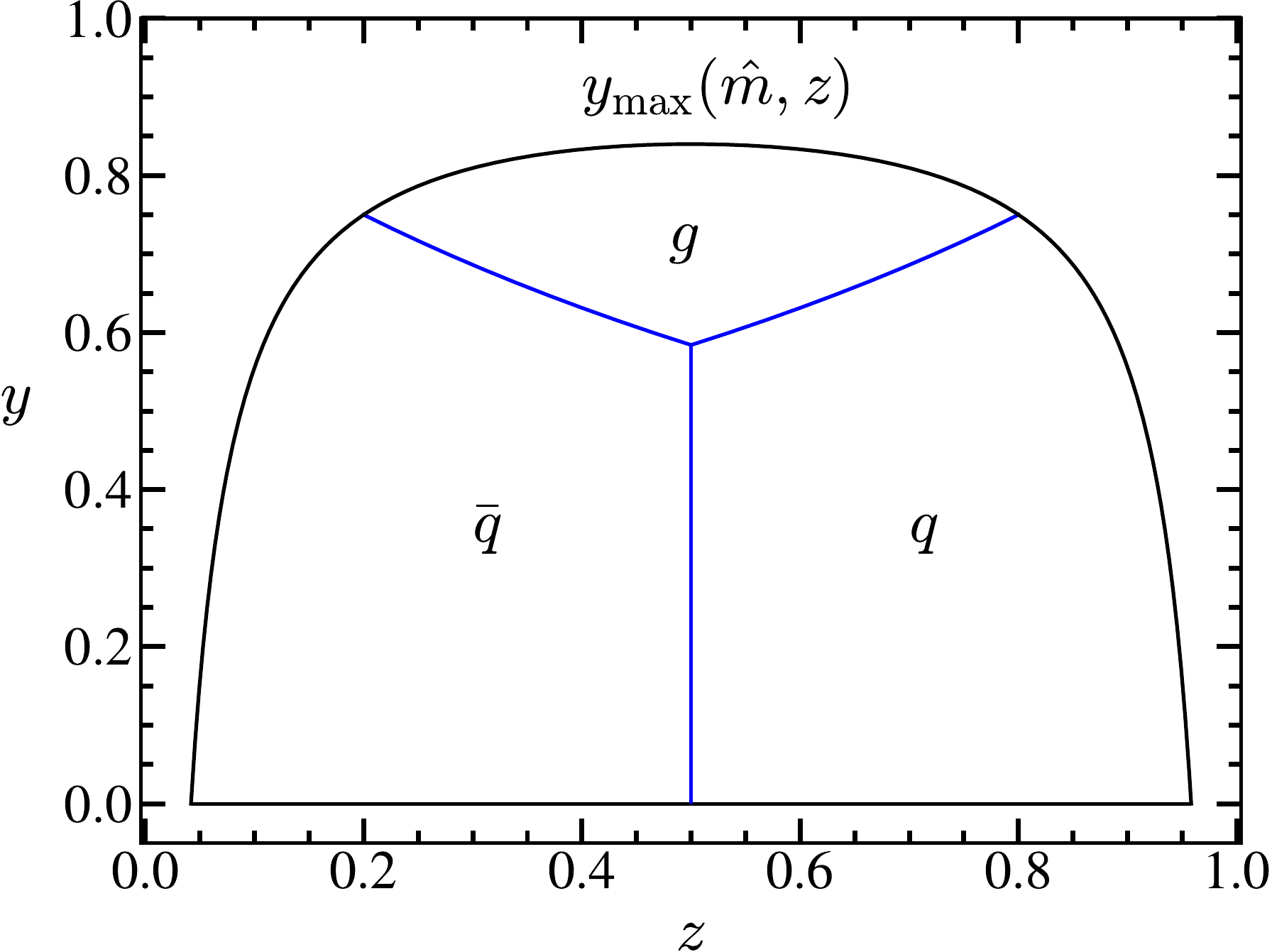}
\caption{Dalitz region in $(z,y)$ coordinates for two massive quarks and a gluon. In addition to the phase-space boundaries, in black, we show
in blue the borders between the regions in which the thrust axis points into the direction of the quark, anti-quark or gluon $3$-momentum.
To generate the plot we use $m/Q=0.2$.
\label{fig:Phase-space}}
\end{figure}

For simplicity we define the $x$ axis such that $\vec{p}_1$ has no $y$ component and a positive
projection on the $x$ axis (that is, through the Gram-Schmidt process):
\begin{equation}
\hat u_x = \frac{\vec{p}_1 - (\vec{p}_1\!\cdot\hat u_z)\hat u_z}{\sqrt{|\vec{p}_1|^2 - (\vec{p}_1\!\cdot\hat u_z)^2}}\,.
\end{equation}
Here $\hat u_i$ with $i=x,y,z$ are three unitary vectors pointing in the direction of the respective coordinate axes.
To define the $y$ axis we use once again the Gram-Schmidt procedure:
\begin{equation}
\hat u_y = \frac{\vec{p}_2 - (\vec{p}_2\!\cdot\hat u_z)\hat u_z
- (\vec{p}_2\!\cdot\hat u_x)\hat u_x}
{\sqrt{|\vec{p}_2|^2 - (\vec{p}_2\!\cdot\hat u_z)^2- (\vec{p}_2\!\cdot\hat u_x)^2}}\,,
\end{equation}
such that, by construction, $\hat u_y\!\cdot\vec{p}_2>0$, which is exactly what we need to define spherical
coordinates in a coherent way in our $(3-2\varepsilon)$-dimensional euclidean vector space. This choice greatly
simplifies the computations but, however, implies
$\hat u_x \times \hat u_y = {\rm sign}[\hat u_z\! \cdot (\vec{p}_1\times\vec{p_2})]\hat u_z$,
so that the axes orientation is not always standard. Since there are no outer products in our matrix elements this
fact is irrelevant. Moreover, the inner product $\vec{p}_1\!\cdot \vec{p}_2$ can take positive and negative values.
In any case, to avoid this issue, whenever $\hat u_z \!\cdot (\vec{p}_1\times\vec{p_2})<0$ one can use $\vec{p}_2$
first to define the $x$ axis followed by $\vec{p}_1$ that fixes the $y$ axis.

To compute oriented
event shapes we need the phase space differential in the quark and anti-quark energies, as well as in the angles
$\theta_i$ and $\theta_j$ defined by the $3$-momenta of particles $i$ and $j$, and the beam. The indices $i$ and $j$ can be
chosen freely and do not necessarily need to coincide with the quark energies (that is, we are not forced to choose
$i=1$ and $j=2$, but of course $i\neq j$). The angles $\tilde \theta_{ij}$ formed by the $3$-momenta of any two
different particles in the final state do not depend on $\theta_i$ or $\theta_j$ (ergo, do not depend on the
orientation), and can be expressed in terms of masses and energies as follows:
\begin{align}\label{eq:sinCos12}
\sin^2 (\tilde{\theta}_{ij}) =\,& \frac{4 [(1 - x_1) (1 - x_2) (x_1+x_2 - 1) -
\hat{m}^2 x_3^2]}{\beta^2_i\beta^2_j} \equiv \frac{4 \xi(\hat m, x_1,x_2)}{\beta^2_i\beta^2_j}\,,\\
\cos (\tilde{\theta}_{ij})=\,& \frac{x_i x_j-2(x_i+x_j-1)+4\hat m_i \hat m_j}{\beta_i\beta_j}\,.\nonumber
\end{align}
The first line result shows that $\beta_i\beta_j \sin (\tilde{\theta}_{ij}) = \sqrt{4\xi(\hat m, x_1,x_2)}$ is independent of
$i$ and $j$ as long as $i\neq j$. Since $0\leq \theta_{ij}\leq\pi$ one has that $\sin(\tilde{\theta}_{ij})\geq 0$
anywhere in the phase space such that
the square root can be computed unambiguously.
Including the flux factor one finds the following result for the $3$-particle phase space in $d=4-2\varepsilon$ dimensions:
\begin{align}\label{eq:Q3}
\frac{{\rm d} Q_3}{2 s} =\,& \frac{4^{\varepsilon} s^{- 2 \varepsilon}}{2
(4 \pi)^{4 - 2 \varepsilon} \Gamma (1 - 2 \varepsilon)} \! \int \!
{\rm d} x_1 {\rm d} x_2 {\rm d} \!\cos (\theta_i) {\rm d}\! \cos (\theta_j)
\frac{\beta_i^{-2\varepsilon}\beta_j^{-2\varepsilon}\theta(h_{i j})}{h_{i j}^{1 / 2 + \varepsilon}}\,,\\
h_{ij} = \,& \sin^2 (\tilde{\theta}_{ij}) - \cos^2 (\theta_i) - \cos^2 (\theta_j) + 2 \cos
(\tilde{\theta}_{ij}) \cos (\theta_i) \cos (\theta_j)\nonumber\\
\equiv\,& [\cos (\theta_i) - \cos (\theta_{ij}^-)] [\cos (\theta_{ij}^+) - \cos(\theta_i)]\,,\nonumber\\
\cos (\theta_{ij}^{\pm}) = \,&\cos (\tilde{\theta}_{ij}) \cos (\theta_j) \pm \sin (\tilde{\theta}_{ij})\sin (\theta_j)
= \cos (\tilde{\theta}_{ij} \mp \theta_j)\,.\nonumber
\end{align}
As expected, the flux-normalized $3$-particle phase space is dimensionless for $\varepsilon\to0$.
Here $\theta_i$ is identified with the first polar angle $\theta_i^1$ in the $(3-2\varepsilon)$-dimensional spherical
coordinates that specify the direction of the $i$-th particle's $3$-momentum. For simplicity we carry out our
discussion for the choice $i=1$, $j=2$, but the result is valid for any other pair of values, as shall be proven later.
Our axes choice is such that, as far as particle $1$ is concerned, there is no angular dependence except for
$\theta_1^1\equiv \theta_1$, therefore we can integrate $\theta_1^{n>1}$ getting simply a solid angle.
There is, however, dependence on $\theta_2^1\equiv \theta_2$, and $\theta_2^2$, the two polar angles that specify the direction of
$\vec{p}_2$. We stress that since $\hat u_y \!\cdot \vec{p}_2>0$ one has $0\leq\theta_2^2\leq\pi$, such that $\theta_2^2$
is necessarily a polar angle, not azimuthal. We therefore can integrate $\theta_2^{n>2}$ getting again a solid angle.
We note that in $d-1$ dimensions there is a single azimuthal angle $\phi\equiv\theta_{d-1}$ that is always integrated
over in our computations. The dependence on
$\theta_2^2$ comes solely from the scalar product
\begin{equation}\label{eq:cos12}
\vec{p}_1\!\cdot \vec{p}_2 = |\vec{p}_1||\vec{p}_2|\cos(\theta_{12}) = \frac{Q^2}{4}\beta_1\beta_2\,[\,\sin (\theta_1) \sin (\theta_2) \cos (\varphi_2) + \cos (\theta_1) \cos(\theta_2)\,]\,,
\end{equation}
that appears in the Dirac delta function enforcing energy conservation.
We integrate $\theta_2^2$ against this delta function to obtain the result in Eq.~\eqref{eq:Q3}.\footnote{Enforcing $4$-momentum
conservation in Eq.~\eqref{eq:cos12} one obtains the result in the second line of Eq.~\eqref{eq:sinCos12}.}

Before we go on, we pause and show that
$\int\! {\rm d}\!\cos(\theta_i){\rm d}\!\cos(\theta_j)\beta_i^{-2\varepsilon}\beta_j^{-2\varepsilon}/h_{ij}^{1/2+\varepsilon}$
does not depend on the values of $i$ and $j$ as long as $i\neq j$. To that end we need to use the following relations:
\begin{equation}\label{eq:cosId}
1 + \frac{\beta^2_k}{\beta^2_\ell} - \frac{2 \beta_k}{\beta_\ell} \cos(\tilde\theta_{12}) = \frac{x_3^2}{\beta_\ell^2} \,,\qquad
\frac{\beta_k}{\beta_\ell} + \cos (\tilde\theta_{12}) = - \frac{x_3}{\beta_\ell} \cos (\tilde\theta_{13})\,,
\end{equation}
with $k,\ell \leq 2$
and $k\neq \ell$. Using Eq.~\eqref{eq:P3cons} to express $\cos(\theta_2)$ [$\cos(\theta_1)$] as a linear
combination of
$\cos(\theta_1)$ [$\cos(\theta_2)$] and $\cos(\theta_3)$,
with the help of Eq.~\eqref{eq:cosId} it is trivial to show \mbox{$h_{12} = (x_3/\beta_2)^2 h_{13}=(x_3/\beta_1)^2 h_{23}$} and
${\rm d}\!\cos(\theta_1){\rm d}\!\cos(\theta_2)=(x_3/\beta_2){\rm d}\!\cos(\theta_1){\rm d}\!\cos(\theta_3)=(x_3/\beta_1){\rm d}\!\cos(\theta_2){\rm d}\!\cos(\theta_3)$.
The first result implies that $\theta(h_{ij})$ does not depend on $i$ or $j$, and together with the second it is immediate to check that
\begin{equation}
\int\! {\rm d}\!\cos(\theta_1){\rm d}\!\cos(\theta_2)\frac{\beta_1^{-2\varepsilon}\beta_2^{-2\varepsilon}}{h_{12}^{1/2+\varepsilon}}
=\!\int \!{\rm d}\!\cos(\theta_1){\rm d}\!\cos(\theta_3)\frac{\beta_1^{-2\varepsilon}\beta_3^{-2\varepsilon}}{h_{13}^{1/2+\varepsilon}}
=\!\int\! {\rm d}\!\cos(\theta_2){\rm d}\!\cos(\theta_3)\frac{\beta_2^{-2\varepsilon}\beta_3^{-2\varepsilon}}{h_{23}^{1/2+\varepsilon}}\,.
\end{equation}

The Heaviside function $\theta(h_{ij})$ makes that, for a fixed value of $\theta_j$, the integration limits for $\theta_i$ coincide with
$\theta_{ij}^{\pm}$ (note that $\theta_{ij}^{\pm}\neq \theta_{ji}^{\pm}$ even though $h_{ij}=h_{ji}$ and $\tilde \theta_{ij}=\tilde \theta_{ji}$). Let us
provide some master integrals that will become necessary for projecting out the angular structure when dealing with real radiation [\,for simplicity
the step function $\theta(h_{12})$ is over understood\,]:\footnote{To obtain these results we use the fact that
$\int\! {\rm d}\!\cos(\theta_i){\rm d}\!\cos(\theta_j)\beta_i^{-2\varepsilon}\beta_j^{-2\varepsilon}/h_{ij}^{1/2+\varepsilon}$
does not depend on the values of $i$ and $j$ as long as $i\neq j$ and the following integrals:
\begin{align}\label{eq:cosInt}
\frac{1}{\Gamma (1 - 2 \varepsilon)}\! \int\! \frac{{\rm d}\! \cos(\theta_i) \cos^n(\theta_i) }
{\beta_i^{2\varepsilon}\beta_j^{2\varepsilon}h_{ij}^{1 / 2 + \varepsilon}}\!=\,&
\frac{\pi\xi^{-\varepsilon}(\hat m, x_1,x_2)\! \sin^{- 2\varepsilon} (\theta_j)}{\Gamma (1 - \varepsilon)^2}\times\!
\left\{ \begin{array}{ll}
\! 1 & n=0 \\
\! \cos (\tilde{\theta}_{ij}) \cos (\theta_j) & n=1 \\
\! \frac{\cos^2(\theta_2) [2(1- \varepsilon) -(3-2 \varepsilon )\! \sin^2(\tilde \theta_{ij})]+\sin^2(\tilde \theta_{ij})}{2(1-\varepsilon)}& n=2
\end{array} \right.\! \!, \nonumber\\
\frac{4^\varepsilon}{\Gamma (1-\varepsilon )^2}\!\int_{-1}^1\! {\rm d}x (1-x^2)^{-\varepsilon} x^{2k} =\,&
\frac{2^{1-2 k} (2 k)!}{k! \Gamma (2-2\varepsilon ) \left(\frac{3}{2}-\varepsilon \right)_k}\,.
\end{align}}
\begin{align}\label{eq:master}
\frac{4^{\varepsilon}}{\Gamma (1 - 2 \varepsilon)} \int \frac{{\rm d} \!\cos(\theta_1) {\rm d}\! \cos (\theta_2) \cos^{2k}(\theta_j) }
{\beta_1^{2\varepsilon}\beta_2^{2\varepsilon}h_{12}^{1 / 2 + \varepsilon}}
&\,= \frac{2 \pi (2 k)! \xi^{-\varepsilon}(\hat m, x_1,x_2)}{4^k k!\bigl(\frac{3}{2}-\varepsilon\bigr)_{\!k}\Gamma(2 - 2\varepsilon)} \,,\\
\frac{4^{\varepsilon}}{\Gamma (1 - 2 \varepsilon)} \int \frac{{\rm d}\! \cos(\theta_1) {\rm d}\! \cos (\theta_2) \cos(\theta_i)\cos^{2k+1}(\theta_j) }
{\beta_1^{2\varepsilon}\beta_2^{2\varepsilon}h_{12}^{1 / 2 + \varepsilon}}
&\,= \frac{2\pi[2(k+1)]! \cos(\tilde \theta_{ij}) \xi^{-\varepsilon}(\hat m, x_1,x_2)}{4^{k+1} (k+1)!\bigl(\frac{3}{2}-\varepsilon\bigr)_{\!k+1}\Gamma(2-2\varepsilon)}\,,\nonumber\\
\frac{4^{\varepsilon}}{\Gamma (1 - 2 \varepsilon)} \int \frac{{\rm d} \!\cos(\theta_1) {\rm d}\! \cos (\theta_2) \cos^2(\theta_i)\cos^{2k}(\theta_j) }
{\beta_1^{2\varepsilon}\beta_2^{2\varepsilon}h_{12}^{1 / 2 + \varepsilon}}
&\,= \frac{\pi (2 k)! [1+2 k \cos^2(\tilde{\theta}_{ij})] \xi^{-\varepsilon}(\hat m, x_1,x_2)}
{4^k k! (\frac{3}{2}-\varepsilon)_{k+1}\Gamma(2-2\varepsilon)}\,,\nonumber
\end{align}
with $k$ a non-negative integer number, $i,j=1,2,3$ and $(a)_n=\Gamma(a+n)/\Gamma(a)$ the Pochhammer symbol.
Of course one
has $\cos(\tilde \theta_{ii})=1$, and in that sense the first line is contained in the second and third if one sets $i=j$. Likewise, for $i=j$
the second and third lines become equal, as can be easily checked. Setting $k=0$ in the third line is identical to setting $k=1$ in the
first. Finally, if the power of $\cos(\theta_j)$ on the upper or lower (middle) lines is set to an odd (even) number, the integral vanishes.
Using the first line of Eq.~\eqref{eq:master} with $k=0$ one can integrate $\theta_i$ and $\theta_j$ in Eq.~\eqref{eq:Q3} to recover the
known result for the angular-integrated $3$-particle phase space in $d$ dimensions, as given in Eq.~(3.6) of
Ref.~\cite{Lepenik:2019jjk}.

While the Dalitz region looks somewhat awkward when expressed in terms of the $x_i$ variables, it takes a much simpler and more symmetric form if the
following change of variables is implemented: \mbox{$x_1 = 1 - (1 - z) y$}, $x_2 = 1 - z y$, making the soft limit $y \rightarrow 0$ apparent as
$y=x_3$ is proportional to the gluon energy. The Dalitz region is now specified by the conditions $z_- \leq z \leq z_+$ and \mbox{$0\leq y \leq y_{\rm max} (\hat m, z)$},
with $z_\pm$ and the symmetric function $y_{\rm max} (\hat m, z)= y_{\rm max} (\hat m, 1-z)$ defined as:
\begin{equation}
y_{\rm max} (\hat m, z) = 1 - \frac{\hat{m}^2}{z (1 - z)}\,,\qquad
z_{\pm} \equiv \frac{1 \pm \beta}{2} \,.
\end{equation}

Since we aim to obtain a distribution differential in $\theta_T$ and the thrust axis coincides with the
direction of the particle with largest $3$-momentum magnitude, it is clear that $\theta_T=\theta_i$ if $\beta_i=\max\{\beta_1,\beta_2,\beta_3\}$.
Hence we can design a function that will project out the correct value of $\theta_T$ depending on the phase-space point:
\begin{align}\label{eq:delta3}
\delta^{(3)}_T = \,& \theta (2z-1) \theta [y_{\tau} (\hat{m}, 1-z) - y ]
\delta [\cos (\theta_T) - \cos (\theta_1)] \\
& + \theta (1-2z) \theta [y_{\tau} (\hat{m}, z) - y ] \delta [\cos (\theta_T) - \cos (\theta_2)] \nonumber\\
& + \theta [y - y_{\tau} (\hat{m}, z) ] \theta [y - y_{\tau} (\hat{m}, 1-z)] \delta [\cos (\theta_T) - \cos
(\theta_3)] \equiv \delta^{(1)}_T+\delta^{(2)}_T+\delta^{(3)}_T , \nonumber
\end{align}
where we assume that $\delta^{(3)}_T$ acts only inside the Dalitz region. We have defined the function
\begin{equation}
y_{\tau} (\hat{m}, z) = \frac{\sqrt{1 - 4 \hat{m}^2 (1 - z^2)} - z}{1 - z^2}\,,
\end{equation}
which, for $\hat m \leq z \leq 1/2$ sets the limit between the regions in which the thrust axis points into the anti-quark or gluon momenta. Likewise,
$y_{\tau} (\hat{m}, 1-z)$ for $1/2 \leq z \leq 1-\hat m$ is the limit between the regions in which it points into the momenta of the
quark and gluon. For completeness, the boundary between the regions in which it points in the same direction as the quark
or anti-quark momenta is parametrized by $z=1/2$ and
$0\leq y \leq 4(\sqrt{1-3\hat m^2}-1/2)/3\equiv y_{\rm middle}(\hat m)$. The Dalitz region, along with these borders,
is depicted in Fig.~\ref{fig:Phase-space}.\footnote{The equivalent Fig.~5 in Ref.~\cite{Lepenik:2019jjk} has the labels $q$ and $\bar q$ swapped.}

Since in this article we exclusively deal with the angular distribution, we can project out this term using Eq.~\eqref{eq:delta3} and the third
line of Eq.~\eqref{eq:decomposition} to obtain the following integration kernel for $3$-particle contributions:
\begin{align}\label{eq:angOut}
& \!\!K(\theta_i,y,z)=\frac{3}{8}\! \int_{- 1}^1\!\! {\rm d} \!\cos (\theta_T) [2 - 5 \cos^2 (\theta_T)]
\delta^{(3)}_T \! = \frac{3}{8} \biggl\{\! \theta (2z-1) \theta [y_{\tau\!} (\hat{m}, 1-z) - y ] [2 - 5 \cos^2\! (\theta_1)] \\
&\qquad\quad~ +\!\theta (1-2z) \theta [y_{\tau\!} (\hat{m}, z) - y ] [2 - 5 \cos^2\! (\theta_2)]
+\! \theta [y - y_{\tau\!} (\hat{m}, z) ] \theta [y - y_{\tau\!} (\hat{m}, 1-z)] [2 - 5 \cos^2 \!(\theta_3)]\!
\biggr\} . \nonumber
\end{align}

\section{Lowest Order Result}\label{sec:LO}
\begin{figure*}[t!]\centering
\subfigure[]
{\includegraphics[width=0.4\textwidth]{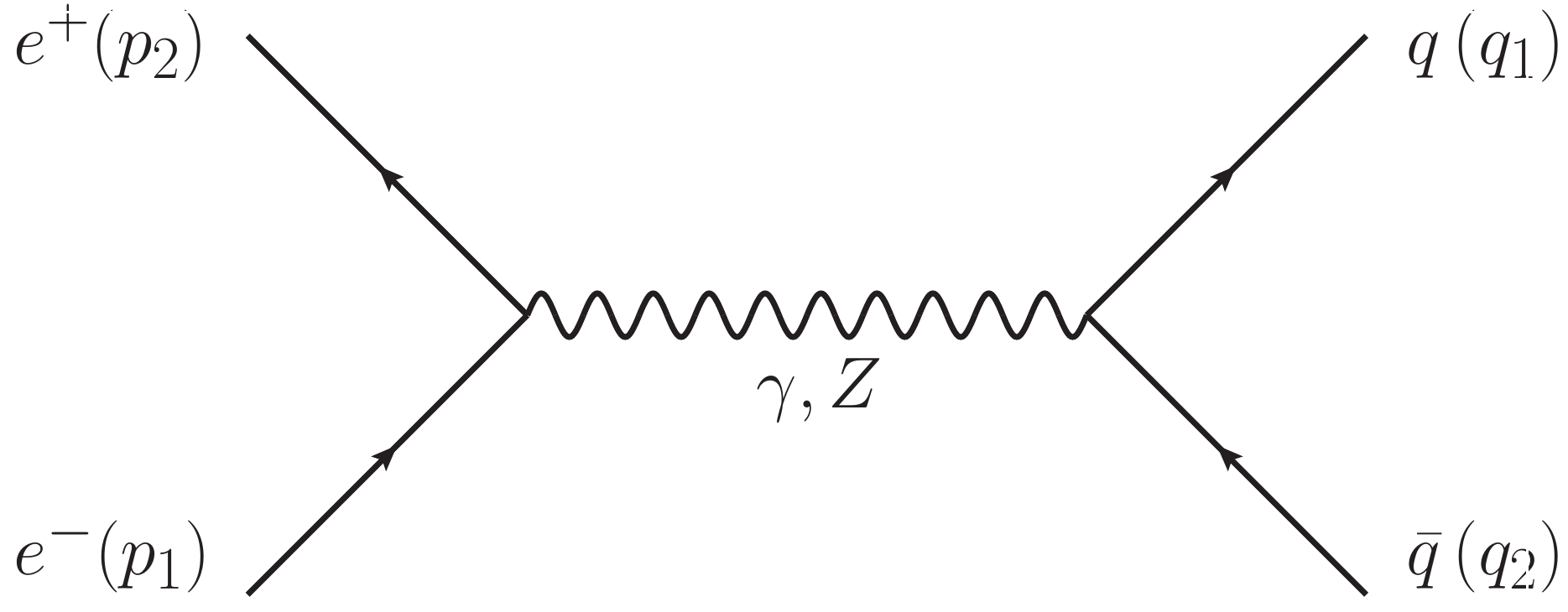}
\label{fig:Tree}}~~~~~~~~
\subfigure[]{\includegraphics[width=0.21\textwidth]{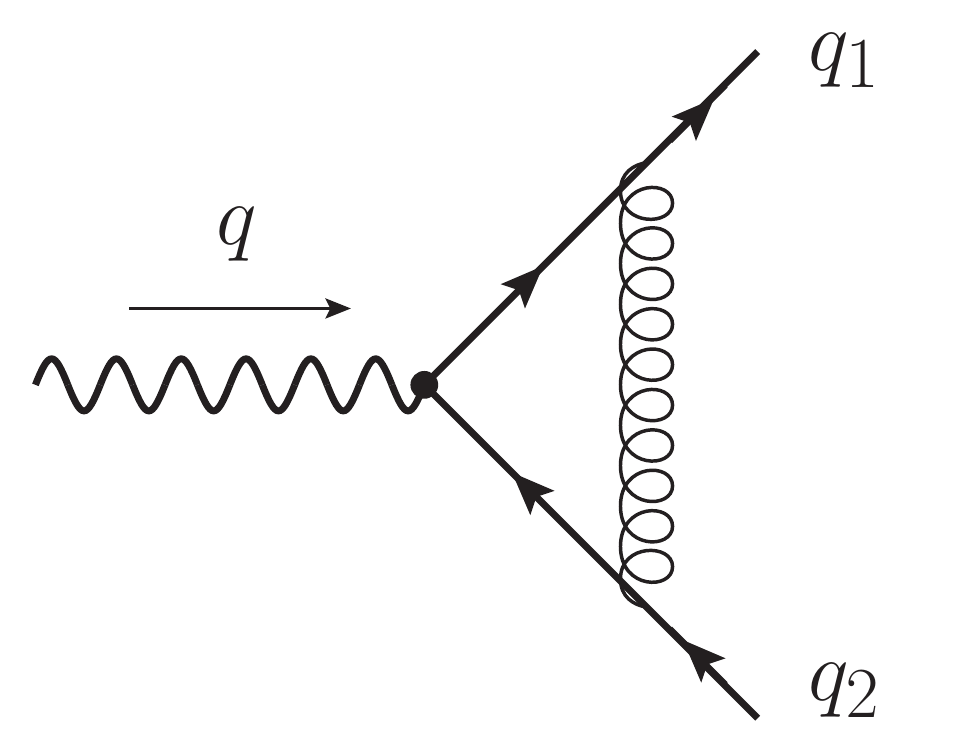}
\label{fig:virtual}}
\caption{Panel (a): Lowest-order Feynman diagram contributing to oriented event shapes for massive quarks. The distribution is proportional
to a Dirac delta function. Panel (b): Feynman diagram representing the vector and axial-vector form factors for massive quarks.}
\label{fig:TreeVirt}
\end{figure*}
Although the results for the massive cross section at lowest order are known since long, we sketch the computation as it sets the basis for the more
complex NLO case. Furthermore, the results presented in this section with the quark mass set to zero constitute the normalization of the cross section at any
order. To make each step of the computation free from spurious logarithms with dimensionful arguments that would otherwise appear when expanding the
results in $\varepsilon$ ---\,as an artifact of having $d$-dimensional phase space integrals\,---, we normalize the distributions with the $d$-dimensional
Born cross-section. A standard computation yields the following result for the hadronic tensor
\begin{equation}
H_{2,0}^{C,\mu\nu} = 4 \Bigl\{ q_2^{\mu} q_1^{\nu} + q_1^{\mu} q_2^{\nu} - \frac{1}{2} \bigl[s + 2(1\mp 1) m^2\bigr]
g^{\mu \nu} \Bigr\},
\end{equation}
where in $\mp$ the upper (lower) part corresponds to the vector (axial-vector) current. Taking the massless limit, including the flux factor and integrating
over the phase space one obtains the $d$-dimensional point-like cross section (which is nothing else that the Born cross-section for massless quarks
if only a virtual photon is exchanged):
\begin{equation}
\sigma_B = N_c Q_q^2 \frac{(4 \pi)^{1 + \varepsilon} (1 - \varepsilon) \Gamma (2 -
\varepsilon) \alpha_{\rm em}^2}{(3 - 2 \varepsilon)
\Gamma (2 - 2 \varepsilon) s^{1 + \varepsilon}}.
\end{equation}
If quark masses are not neglected and the polar angle is left unintegrated one obtains the following result for the
vector and axial-vector currents:
\begin{figure*}[t!]
\subfigure[]
{\includegraphics[width=0.45\textwidth]{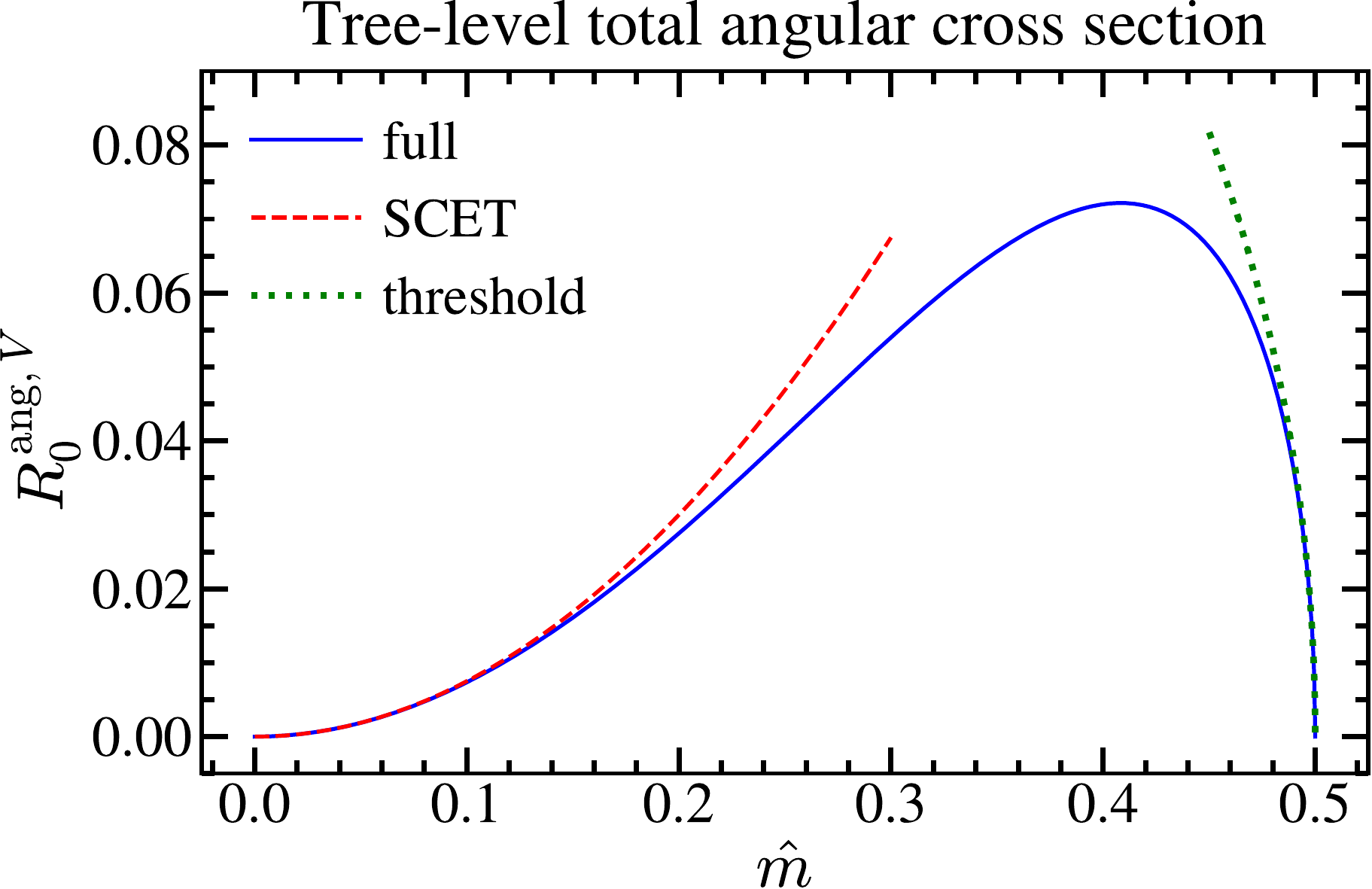}
\label{fig:R-tree}}~~~~~~
\subfigure[]{\includegraphics[width=0.45\textwidth]{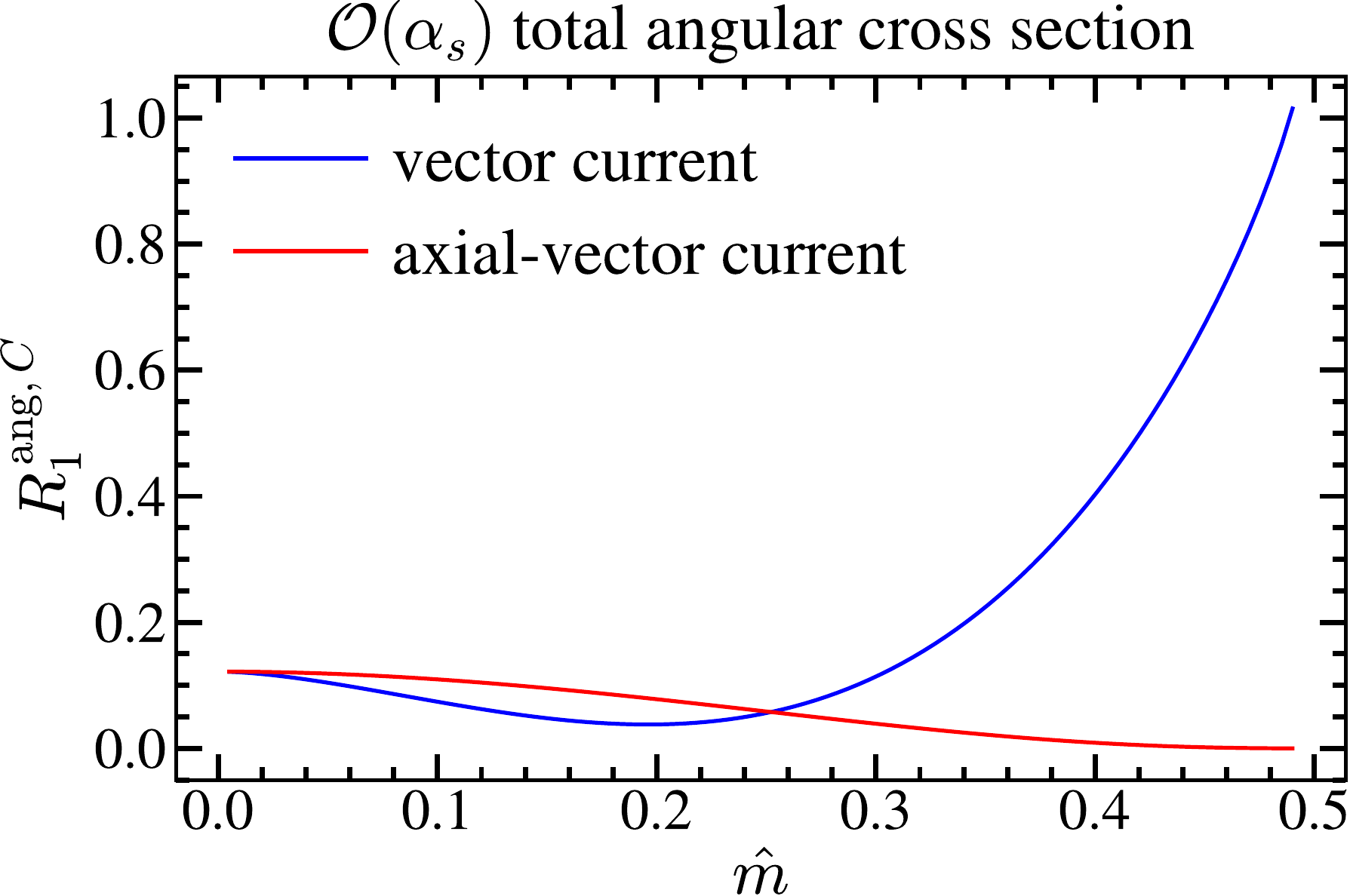}
\label{fig:R-loop}}
\caption{Total angular cross-section at $\mathcal{O}(\alpha_s^0)$ [\,panel (a)\,]
for the vector current $R_0^{{\rm ang},V}(\hat m)$ in solid blue, together with its SCET (dashed red)
and threshold (dotted green) approximations, and at $\mathcal{O}(\alpha_s)$ [\,panel (b)\,] for vector (blue)
and axial-vector (red) currents. The $R_0^{{\rm ang},V}(\hat m)$ cross section vanishes in the massless limit
$\hat m = 0$ and at threshold $\hat m = 1/2$, as opposed to $R_1^{{\rm ang},C}(0)$. While
$R_1^{{\rm ang},A}(1/2)=0$, one has that $R_1^{{\rm ang},C}(0)$ and $R_1^{{\rm ang},V}(1/2)$
are both non-zero.}
\label{fig:R-tree-loop}
\end{figure*}
\begin{align}
\frac{1}{\sigma_0^V} \, \frac{{\rm d} \sigma_{\rm tree}^V}{{\rm d}\! \cos(\theta)} = &
\frac{ (3 - 2 \varepsilon) \Gamma (2 - 2 \varepsilon) }{2^{3 - 2 \varepsilon} \Gamma^2
(2 - \varepsilon)} \beta^{1 - 2 \varepsilon} \sin^{- 2 \varepsilon} (\theta) [2 (1 - \varepsilon) - \beta^2 \sin^2 (\theta)]\,, \\
\frac{1}{\sigma_0^A} \, \frac{{\rm d} \sigma_{\rm tree}^A}{{\rm d}\! \cos (\theta)} = &
\frac{(3 - 2 \varepsilon) \Gamma (2 - 2 \varepsilon)}{2^{3 - 2 \varepsilon}
\Gamma^2 (2 - \varepsilon)} \beta^{3 - 2 \varepsilon} \sin^{- 2 \varepsilon}
(\theta) [1 - 2 \varepsilon + \cos^2 (\theta)] \,. \nonumber
\end{align}
Both results shown above coincide for $\beta=1$. We can project out the total angular cross-section using the last
line of Eq.~\eqref{eq:decomposition} and the second line of Eq.~\eqref{eq:cosInt}, obtaining
\begin{align}
R_0^{{\rm ang},V} \,& = \frac{3 \beta^{1 - 2 \varepsilon}}{16 (1 - \varepsilon)} \frac{[5 + 8
\varepsilon^2 - 22 \varepsilon - \beta^2 (5 - 4 \varepsilon)]}{5 - 2
\varepsilon} \xrightarrow[\varepsilon = 0]{} \frac{3 \hat m^2\beta}{4}\,,\\
R_0^{{\rm ang},A} \,&= - \frac{3 \beta^{3 - 2 \varepsilon}}{8} \frac{\varepsilon (9 - 4
\varepsilon)}{(1 - \varepsilon) (5 - 2 \varepsilon)} \xrightarrow[\varepsilon = 0]{} 0\,,\nonumber
\end{align}
where, the tree-level Born-normalized differential distribution is simply
$f^{{\rm ang},C}_{e,0} = R_0^{{\rm ang},C} \delta[e-e_{\rm min}(\hat m)]$. A graphical representation of
$R_0^{{\rm ang},V}$ is shown in Fig.~\ref{fig:R-tree}, together with its massless and threshold expansions.
For the axial-vector current we obtain a vanishing result, but the vector result only becomes zero in the
$\hat m\to 0$ limit.
As anticipated, this will significantly complicate the NLO computation.

\section{Virtual Contribution}\label{sec:virtual}
As long as IR singularities are handled in dimensional regularization, the computation of the virtual contribution is very similar to the lowest-order
term outlined in Sec.~\ref{sec:LO}. We take advantage of the well known results for the so-called vector and axial-vector form factors for massive
quarks shown in Fig~\ref{fig:virtual}, which, after accounting for the wave function renormalization $Z^{\rm OS}_q$, should be UV finite due to
current conservation, making the present $1/\varepsilon$ pole of IR origin. The general form of the wave-function-corrected form factors up to
one loop is as follows\,\footnote{Mass renormalization is carried out in the OS scheme such that, unless otherwise stated, all quark masses appearing in the various expressions
are understood in the pole scheme.}
\begin{align}
V^{\mu} & = \biggl[ 1 + C_F \frac{\alpha_s}{\pi} A (\hat{m}) \biggr]
\gamma^{\mu} + C_F \frac{\alpha_s}{\pi} \frac{B (\hat{m})}{2 m} (q_1 - q_2)^{\mu} \,, \\
A^{\mu} & = \biggl[ 1 + C_F \frac{\alpha_s}{\pi} C (\hat{m}) \biggr]
\gamma^{\mu} \gamma_5 + C_F \frac{\alpha_s}{\pi} \frac{D (\hat{m})}{2 m}\, \gamma_5\, q^{\mu}\,, \nonumber
\end{align}
with $q = q_1 + q_2$ the photon or $Z$-boson momentum, and $q_i$ with $i=1,2$ the quark and anti-quark momenta, respectively. Vector current conservation implies
$q_\mu V^\mu = 0$ and also ensures that the term proportional to $D (\hat{m})$ from the axial-vector current will vanish when contracted with
the leptonic tensor. For our purposes we only need the real part of the $A$, $B$ and $C$ coefficients that can be written as~\cite{Jersak:1981sp, Harris:2001sx}
\begin{align}
{\rm Re} [A (\hat{m})] &= \biggl( \frac{1+\beta^2}{2\beta}L_\beta -\frac{1}{2} \biggr)\!\biggl[\frac{1}{\varepsilon} - 2\log\biggl( \frac{m}{\mu} \biggr)\biggr]\!+
A_{\rm reg}(\hat{m})\,,\\
A_{\rm reg}(\hat{m}) & =\frac{3}{2} \beta L_\beta - 1 + \, \frac{1+\beta^2}{4\beta} \biggl[ \pi^2 - 2
L_\beta^2 - 2\, {\rm Li}_2\biggl( \frac{2 \beta}{1 + \beta} \biggr)\! \biggr]\nonumber,\\
{\rm Re} [C (\hat{m})] & = {\rm Re} [A (\hat{m})] + \frac{4\hat{m}^2 }{\beta}L_\beta \,,\nonumber
\end{align}
where we have defined $L_\beta \equiv \log[ (1 + \beta)/(2 \hat{m})]$.
For the vector and axial-vector current, after taking the $\varepsilon\to 0$ limit we find (we factor out $3C_F/4=1$ for $N_c=3$)
\begin{align}\label{eq:virtual}
\frac{3C_{\!F}}{4}R_{21}^{{\rm ang},C} \,& \equiv \frac{9\pi L_{\mu\nu}}{4N_cs^2}
\int {\rm d}Q_{2} H_{21}^{C,\mu\nu} [2-5 \cos^2 (\theta)]\,,\\
R_{21}^{{\rm ang},V} \,& = \frac{\beta}{2}\biggl\{\! \biggl[ 1 - \frac{2 }{\beta}(1 - 2 \hat{m}^2) L_{\beta}
\biggr]\! \biggl[ 2 \hat{m}^2 \log\! \biggl(
\frac{m}{\mu} \biggr) - \frac{\hat{m}^2}{\varepsilon} + \frac{3}{10}(3 - 2 \hat{m}^2) + 2 \hat{m}^2 \log
(\beta) \biggr] \nonumber\\
& + 2 \hat{m}^2 {\rm Re} [A_{\rm reg} (\hat{m})] - \hat{m}^2 \beta
L_{\beta} \biggr\} ,\nonumber\\
R_{21}^{{\rm ang},A} \,& = \frac{9 \beta^2}{5} \biggl[ \frac{\beta}{2} -
(1 - 2 \hat{m}^2) L_{\beta} \biggr],\nonumber
\end{align}
with $\theta$ defined after Eq.~\eqref{eq:2body}. Since there are two particles in the final state, the contribution of the virtual radiation to
the differential cross section is once again $f^{{\rm ang},C}_{e,21} = R_{21}^{{\rm ang},C} \delta[e-e_{\rm min}(\hat m)]$.
Surprisingly, we find a non-zero result for the axial-vector current. Likewise, the vector-current result does not vanish in the massless limit.
These are artifacts of dimensional regularization, and once the real-radiation contribution is added, there will be no term proportional to
$\delta(e-e_{\rm min})$ for the axial-vector current, and the coefficient of such delta will vanish as $\hat m\to 0$.

\section{Real Radiation and Total Angular Cross-Section}\label{sec:real}
\begin{figure}[t]\centering
\includegraphics[width=0.65\textwidth]{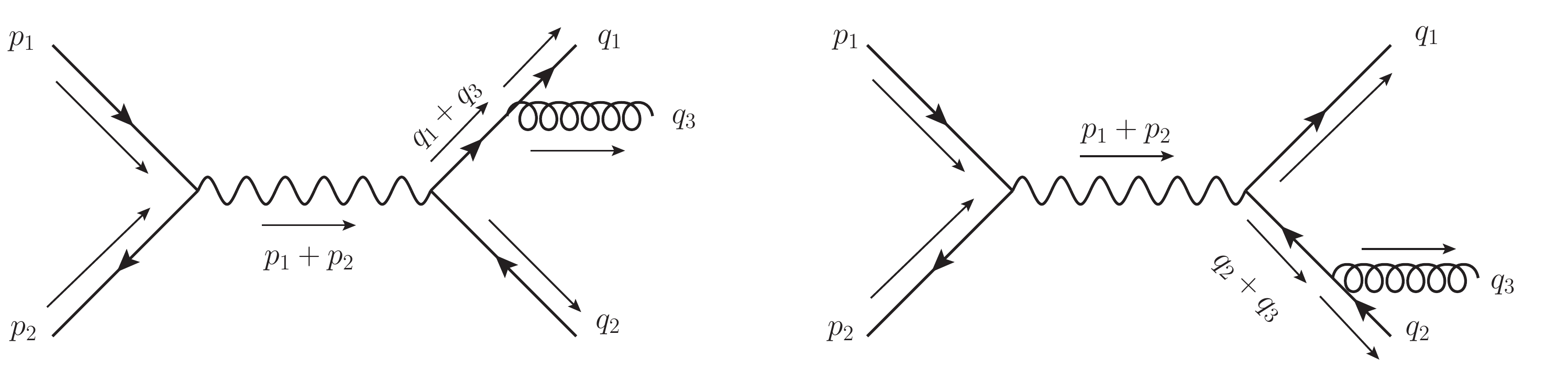}
\caption{Feynman diagrams for the real-radiation contribution at NLO.\label{fig:real}}
\end{figure}
The last terms that contribute at $\mathcal{O}(\alpha_s)$ come from the two diagrams shown in Fig.~\ref{fig:real} in which a real gluon is emitted.
The complete contribution consists on the modulus squared of each diagram plus the interference of the two. We compute the relevant traces
using \texttt{TRACER}~\cite{Jamin:1991dp}, and organize the result in the following angular structures:
\begin{align}\label{eq:cosDecom}
L_{\mu\nu}H_{30}^{C,\mu\nu} = \,& \frac{8 \pi^2 C_F}{s}\biggl(\frac{\mu^2e^{\gamma_E}}{4\pi}\biggr)^{\!\!\varepsilon}
[A^C_0 + A^C_1 \beta_1^2 \cos^2 (\theta_1) + A^C_2 \beta_2^2 \cos^2 (\theta_2)
+ A^C_{12} \beta_1 \beta_2 \cos (\theta_1) \cos (\theta_1)]\,, \nonumber\\
A^V_0 (x_1, x_2) = \,& \frac{(1 - 2 \varepsilon)(x_1^2 + x^2_2 - x_3^2 \varepsilon)}{(1 - x_1) (1 - x_2)}
- \frac{8 \hat{m}^4 x_3^2}{(1 - x_1)^2 (1 - x_2)^2} - \frac{2 \hat{m}^2}{(1 - x_1)^2 (1 - x_2)^2} [8 - 2 \varepsilon x_3^2\nonumber\\
& + 5 (x^2_1 + x^2_2) + 14 x_1 x_2 - 4 (x_1 + x_2) (3 + x_1 x_2)]\,,\nonumber\\
A^A_0 (x_1, x_2) = \,& \frac{(1 - 2 \varepsilon)(x_1^2 + x^2_2 - x_3^2 \varepsilon)}{(1 - x_1) (1 - x_2)}
- \frac{8 \hat{m}^4 x_3^2}{(1 - x_1)^2 (1 - x_2)^2}- \frac{2 \hat{m}^2}{(1-x_1)^2(1-x_2)^2} \{ -
8 (1 - \varepsilon) \nonumber\\
& +\! 2 (1 - \varepsilon) [x_1^3 (1 - x_2) + x_2^3 (1 - x_1)] + x_1 [20 - 13 x_1 - 8 (3 - 2 x_1) \varepsilon] + x_2 [20 \nonumber\\
& -\! 13 x_2 - 8 (3 - 2 x_2) \varepsilon]
+ 2 x_1 x_2 [- 19 + 24 \varepsilon + (9 - 11 \varepsilon)
(x_1 + x_2) - 2 (1 - \varepsilon) x_1 x_2] \}\,,\nonumber\\
A^V_1 (x_1, x_2) = \,& A^V_2 (x_2, x_1)= \frac{1 - \varepsilon}{(1 - x_1) (1 - x_2)} - \frac{2 \hat{m}^2}{(1 - x_1)^2}\,,\nonumber\\
A^A_1 (x_1, x_2) = \,& A^A_2 (x_2, x_1) = \frac{1 - \varepsilon + 4 \hat{m}^2 \varepsilon}{(1 - x_1)(1 - x_2)}
+ \frac{2 \hat{m}^2 (3 - 2 x_1 - x_2)}{(1 - x_1)^2 (1 - x_2)}\,, \nonumber\\
A^V_{12} (x_1, x_2) = \,& \frac{2 (2 \hat{m}^2 -\varepsilon)}{(1 - x_1) (1 - x_2)}\,,\qquad
A^A_{12} (x_1, x_2) = - \frac{2 [2 \hat{m}^2 + \varepsilon (1 - 4 \hat{m}^2)]}{(1 - x_1) (1 - x_2)}\,.
\end{align}
Of course, both currents yield the same result if $\hat m=0$, the functions $A^C_0$ and $A^C_{12}$ are symmetric under the exchange of its two arguments,
and, as expected, $A^C_{12}$ vanishes for $d=4$ if the quark mass is set to zero. With this result we find for the $4$-times differential distribution
at $\mathcal{O}(\alpha_s)$ the following expression (for conciseness, in what follows we omit the arguments of the $A_i^C$ functions)
\begin{align}
\frac{1}{\sigma^C_0} \frac{{\rm d}^4 \sigma^C_{\alpha_s}}{{\rm d} x_1 {\rm d} x_2 {\rm d}\! \cos
(\theta_i) {\rm d}\! \cos (\theta_j)} =\, & \frac{4^{\varepsilon} \alpha_s
C_F}{16 \pi^2} \frac{(3 - 2 \varepsilon) (1 - 2 \varepsilon)}{(1 -
\varepsilon) \Gamma (2 - \varepsilon)} \biggl( \frac{\mu^2e^{\gamma_E}}{s}
\biggr)^{\!\!\varepsilon} \frac{\beta_i^{-2\varepsilon} \beta_j^{-2\varepsilon}}{h_{i j}^{1 / 2 + \varepsilon}} \\
& \times\![A^C_0 + A^C_1 \beta_1^2 \cos^2 (\theta_1) + A^C_2 \beta_2^2 \cos^2 (\theta_2)
+ A^C_{12} \beta_1 \beta_2 \cos (\theta_1) \cos (\theta_1)]\,.
\nonumber
\end{align}
As a cross check, we can integrate the polar angles to obtain the unoriented cross section,
differential in the dimensionless variables $y$ and $z$ already
defined:\footnote{Note that in these variables one has $\xi(\hat m, x_1,x_2) = y^2 [ (1 - y) (1 - z) z \, - \hat{m}^2 ]$
and ${\rm d} x_1{\rm d}x_2=y\, {\rm d}y {\rm d}z$.}
\begin{align}
\frac{1}{\sigma_0^C} \frac{{\rm d}^2 \sigma^C_{\alpha_s}}{{\rm d} y {\rm d} z} =\, &
\frac{\alpha_s C_F}{8 \pi} \frac{y^{1 - 2 \varepsilon}}{(1 - \varepsilon)
\Gamma (2 - \varepsilon)} \biggl( \frac{\mu^2e^{\gamma_E}}{s}
\biggr)^{\!\!\varepsilon} [ (1 - y) (1 - z) z \, - \hat{m}^2 ]^{-
\varepsilon} \\
& \times \{ (3 - 2 \varepsilon) A_0^C + \beta^2_1 A_1^C + \beta^2_2 A_2^C + A_{12}^C [y +
y^2 (1 - z) z + 4 \hat{m}^2 - 1] \} \,. \nonumber
\end{align}
If the $A_i^C$ coefficients given in Eq.~\eqref{eq:cosDecom} are substituted in the previous
expression, full agreement with Ref.~\cite{Lepenik:2019jjk} is found. On the other hand, projecting out the angular
distribution differential in $y$ and $z$ through the integration kernel in Eq.~\eqref{eq:angOut} yields
\begin{align}\label{eq:realProj}
\frac{1}{\sigma_0^C} \frac{{\rm d}^2 \sigma^{\alpha_s,C}_{\rm ang}}{{\rm d} z {\rm d} y}
= \,&\frac{1}{\sigma_0}\! \int\! {\rm d}\!\cos(\theta_1){\rm d}\!\cos(\theta_2) K(\theta_i,y,z)
\frac{{\rm d}^4 \sigma^C_{\alpha_s}}{{\rm d} z {\rm d} y {\rm d} \!\cos(\theta_1) {\rm d}\! \cos (\theta_2)}\\
= \,& \frac{3 \alpha_s C_F}{8 \pi} \frac{y^{1 - 2 \varepsilon}}{(1 -
\varepsilon)^2 \Gamma (1 - \varepsilon)} \biggl( \frac{\mu^2e^{\gamma_E}}{s}
\biggr)^{\!\!\varepsilon}[ (1 - y) (1 - z) z \, - \hat{m}^2 ]^{-
\varepsilon} \nonumber \\
&\times \biggl\{ A^C_q(\hat m, y, z) \theta \biggl(\!z-\frac{1}{2}\biggr) \theta [y_{\tau} (\hat{m}, 1-z) - y ]
+ A^C_{\bar q}(\hat m, y, z) \theta \biggl(\frac{1}{2}-z\!\biggr) \theta [y_{\tau} (\hat{m}, z) - y ]\nonumber\\
&+ A^C_g(\hat m, y, z) \theta [y - y_{\tau} (\hat{m}, z) ] \theta [y - y_{\tau} (\hat{m}, 1-z)] \biggr\},
\nonumber
\end{align}
with $A^C_g(\hat m, y, z)=A^C_g(\hat m, y, 1-z)$ and $A^C_q(\hat m, y, z)=A^C_{\bar q}(\hat m, y, 1-z)$. A tedious but straightforward computation
yields
\begin{align}
8A_q^C(\hat m,y,z) =\, & (1 - 4 \varepsilon) A_0^C - \frac{5 + 4 \varepsilon}{5 - 2 \varepsilon}
\Bigl\{ \bigl([1 - y (1 - z)]^2 - 4 \hat{m}^2\bigr) A_1^C + \bigl[(1 - y z)^2 - 4 \hat{m}^2\bigr] A_2^C \\
& + A_{12}^C \bigl[y + y^2 (1 - z) z + 4 \hat{m}^2 - 1\bigr] \Bigr\} + \frac{40
A_2^C}{5 - 2 \varepsilon} \frac{y^2 [(1 - y) (1 - z) z - \hat{m}^2]}{[1 - y (1 -
z)]^2 - 4 \hat{m}^2} \,,\nonumber\\
8A_g^C(\hat m,y,z) =\, & (1 - 4 \varepsilon) A_0^C - \frac{5 + 4 \varepsilon}{5 - 2 \varepsilon}
\Bigl\{ \bigl([1 - y (1 - z)]^2 - 4 \hat{m}^2\bigr) A_1^C + \bigl[(1 - y z)^2 - 4 \hat{m}^2\bigr] A_2^C \Bigr\} \nonumber\\
& + \frac{40 [(1 - y) z (1 - z) - \hat{m}^2]}{5 - 2 \varepsilon} (A_1^C + A_2^C + 2 A_{12}^C) \nonumber\\
& + \frac{A_{12}^C}{(5 - 2 \varepsilon)} \{ (5 + 4 \varepsilon) [1 - y - y^2
z (1 - z)] - 120 z (1 - z) (1 - y) + 4 (25 - 4 \varepsilon)\hat{m}^2 \} . \nonumber
\end{align}
Plugging the expressions for $A_i^C$ in Eq.~\eqref{eq:cosDecom} and setting both $\hat m=\varepsilon=0$
one recovers the results displayed in Eq.~(1.3) of Ref.~\cite{Mateu:2013gya}.

\subsection{Axial-vector current}\label{sec:axial}
Since in this case there are no IR singularities, neither in the virtual-radiation term (as long as one sets
$d=4$ in the phase-space right away) nor in the real-radiation one, for conciseness we show results with $\varepsilon=0$ only:
\begin{align}
A_q^A(\hat m,y,z) = \,& \frac{(1 - y) z^2 - \hat{m}^2 z \{ 2 - y^2 + z [2
+ y (y - 2)] \} \!+ 2 \hat{m}^4}{z^2 \{ [1 - y (1 - z)]^2 - 4 \hat{m}^2 \}},\\
A_g^A(\hat m,y,z) = \,&\frac{2 (1 - y) (1 - z)^2 z^2 - \hat{m}^2 (1 - z)
z (4 - y^2 - 2 y) + 2 \hat{m}^4}{y^2 (1 - z)^2 z^2} . \nonumber
\end{align}
As anticipated, $A_q^A(\hat m,y,z)$ is finite as $y\to 0$, therefore no soft singularity is present. On the other hand, $y\,A_g^A(\hat m,y,z)$
diverges if $y=0$, but the Heaviside functions that multiply this term in Eq.~\eqref{eq:realProj} impose $y>y_{\rm middle}(\hat m)$ which is
a positive number in the physical range \mbox{$0\leq\hat m< 1/2$}, and therefore screens the soft singularity. This entails that for any event shape, the
angular axial-vector distribution will have no singular structures
at $\mathcal{O}(\alpha_s)$. Only a
non-singular distribution will remain, that can be computed analytically or numerically depending on the event shape. We will explore this further
in subsequent sections.

Since, as we just discussed, for the axial-vector current only the real radiation contributes, we can already provide a closed form for the total
angular cross section, simply integrating $A_q^A$, $A_{\bar q}^A$ and $A_g^A$ in their respective patches within the phase space.
Since there is a mirror symmetry with respect to the $z=1/2$ vertical axis, it is enough to integrate between $z=z_-$ and $z=1/2$ and
double the result. Finally, the region in which the thrust axis points into the anti-quark direction has two distinct upper boundaries:
$y_{\rm max}(\hat m, z)$ for $z_-< z < \hat m$ and $y_\tau(\hat m, z)$ for $\hat m < z < 1/2$, and we split the corresponding integral accordingly:
\begin{align}\label{eq:RangAx}
R_1^{{\rm ang},A} (\hat{m}) = \,& \frac{3 C_F}{4} \biggl\{\int_{z_-}^{\hat{m}} {\rm d} z \tilde{A}_{\bar q}^A [\hat{m}, y_{\max} (\hat{m},z), z]
+\! \int_{\hat{m}}^{\frac{1}{2}} {\rm d} z \tilde{A}_{\bar q}^A [\hat{m}, y_{\tau} (\hat{m},z), z]\\
&\qquad~ + \!\int_{\hat{m}}^{1 / 2} {\rm d} z \tilde{A}_g^A [\hat m, y_{\max} (\hat{m},z), y_{\tau} (\hat{m},z), z] \biggr\}, \nonumber\\
\tilde{A}_{\bar q}^A (\hat{m},y, z) =\, & \frac{1}{4 \hat{m} (1 - z)^2 z^3}
\biggl\{ [1 - z - 2 \hat{m}^2 (2 - z)] \biggl[ (1 - 2 \hat{m}) (1 - z - \hat m)^2
\log \biggl( 1 - \frac{y z}{1 - 2 \hat{m}} \biggr) \nonumber\\
& \qquad\qquad\qquad~ -(1 + 2 \hat{m}) (1 - z + \hat{m})^2
\log \biggl(\! 1 - \frac{y z}{1 + 2 \hat{m}} \biggr)\! \biggr]\! \nonumber \\
&\qquad\qquad\qquad~ - 2 \hat{m} y (1 - z) z \{ 2 (1 - z) - \hat{m}^2 [8 - (4 - y) z] \} \!\biggr\},\nonumber\\
\tilde{A}_g^A (\hat{m}, y_1, y_2, z) =\,& \frac{1}{2 (1 - z)^2
z^2} \biggl\{ 4 [(1 - z) z - \hat{m}^2]^2 \log \biggl(
\frac{y_{1}}{y_{2}} \biggr) \nonumber\\
& \qquad\qquad\quad~+ (1 - z) z (y_1 - y_2) [\hat{m}^2
(y_1 + y_2 + 4) - 4 (1 - z) z] \biggr\},\nonumber
\end{align}
where the functions $\tilde{A}_q^A$ and $\tilde{A}_g^A$ are defined as
\begin{equation}\label{eq:AtildeDef}
\tilde{A}_{\bar q}^A(\hat m, y, z) = \!\int_0^y{\rm d}h\,h A_{\bar q}^A(\hat m,h,z)\,,\qquad
\tilde{A}_g^C(\hat m, y_1,y_2, z) = \!\int^{y_1}_{y_2}{\rm d}h\,h A_g^C(\hat m,h,z)\,.
\end{equation}
Even though the definition of $\tilde{A}_g^C$ can be used both for vector and axial-vector currents, due to soft singularities we need to
define $\tilde{A}_{\bar q}^A$ and $\tilde{A}_{\bar q}^V$ separately.
While the integrals in $y$ have been carried out analytically, we have not found simple expressions for the $z$ integrations.\footnote{We
found extremely lengthy analytical expressions in terms of polylogarithms and have not been
able to simplify them to an amenable size. Therefore it is unpractical to code these and we instead opt for a numerical
implementation.} Instead, we
carry out these (along with similar ones for the vector current or cumulative cross sections, to be discussed in Sec.~\ref{sec:differential}) numerically,
and to that end we have implemented our results in \texttt{Mathematica}~\cite{mathematica} and
\texttt{Python}~\cite{10.5555/1593511}, and found agreement within $15$ decimal places.
For special functions and
quadrature in \texttt{Python} we use the \texttt{NumPy}~\cite{harris2020array} and \texttt{SciPy}~\cite{2020SciPy-NMeth} modules,
while in \texttt{Mathematica}
we simply employ native functions. All plots in this article have been produced with the \texttt{Python} module \texttt{Matplotlib}~\cite{Hunter:2007}.
A graphical representation of $R_1^{{\rm ang},A}$ is shown in Fig.~\ref{fig:R-loop}, where it can be realized
that the $\mathcal{O}(\alpha_s)$ total angular cross-section vanishes for $\hat m=1/2$. This is easy to understand: $\tilde{A}_{q,g}^A$ are
finite for $\hat m = 1/2$ but $z_- \to 1/2$, such that all lower and higher integration limits coincide at threshold.
Results for the differential distribution shall be provided in Sec.~\ref{sec:differential}.

\subsection{Vector current}\label{sec:vector}
Due to the non-vanishing tree-level result, the vector-current matrix element diverges in the soft limit and the linear dependence on
$\varepsilon$ must be retained. However, we only need to keep track of this parameter in the terms of $A_{q,\bar q}^V$ which do
diverge when $y\to 0$. Accordingly we define
\begin{align}
V^{\rm div} (\hat{m}, z, \varepsilon) = & \lim_{y \rightarrow 0} y^2
A_q^V(\hat{m},y,z) = \lim_{y \rightarrow 0} y^2 A_{\bar q}^V(\hat{m},y,z) = V^{\rm div} (\hat{m},z, 0) +
V^{\varepsilon} (\hat{m}, z) \varepsilon +\mathcal{O} (\varepsilon^2)\,, \\
V^{\rm div} (\hat{m},z, 0) = & - \!2 \hat{m}^2 M_V^1 (\hat{m},z)\,,\qquad V^{\varepsilon} (\hat{m},z) =
\frac{9 + 14 \hat{m}^2}{5} M_V^1 (\hat{m},z)\,, \nonumber\\
M_V^1 (\hat{m},z) = & -\! \frac{(1 - z) z - \hat{m}^2}{(1 - z)^2 z^2}\,,
\nonumber
\end{align}
where the function $M_V^1 (\hat{m},z)$ is present also in the computation of the unoriented cross section, see Eq.~(3.18)
of Ref.~\cite{Lepenik:2019jjk}. We note
$V^{\rm div} (\hat{m}, z, \varepsilon) = V^{\rm div} (\hat{m}, 1 - z, \varepsilon)$ and also that
$V^{\rm div} (\hat{m}, z, 0)$ vanishes in the
massless limit. The fact that $V^{\varepsilon} (0, z)\neq 0$ is an artifact of dimensional regularization that leaves
no trace once the virtual-radiation contribution is added. Keeping only the necessary (linear) dependence on $\varepsilon$ to
carry out the computation we end up with the following expressions:
\begin{align}
A_q^V(\hat{m},y,z) =\, & \frac{V^{\rm div} (\hat{m}, z, 0) +
V^{\varepsilon} (\hat{m},z) \varepsilon}{y^2} + V^{\rm fin} (y, z,
\hat{m})\,, \\
A_g^V(\hat{m},y,z) = \,& \frac{2 [(1 - y) z (1 - z) - \hat{m}^2]}{y^2 (1 -
z) z}, \nonumber\\
V^{\rm fin} (\hat{m},y, z) =\, & \frac{(1 - y) y (1 - z) z^2 - \hat{m}^2
z [2 - y (1 - 2 z^2)] + 2 \hat{m}^4 [y (1 - z) + 4 z]}{y (1 - z) z^2 \{ [1 -
y (1 - z)]^2 - 4 \hat{m}^2 \}} \,. \nonumber
\end{align}
It is simple to see that $y$ integrals with zero lower integration limit (such as the total cross section or any
cumulative distribution) will produce a $1/\varepsilon$ pole.

As for the axial-vector current, we postpone the computation of the differential distribution to the next section
and show now results for the total angular cross-section, discussing how the cancellation takes place. 
One has to integrate $A_{\bar q}^V$ between the lowest part of the phase space, \mbox{$y=0$}, and the upper
boundary of the region in which the thrust axis points in the same direction as the anti-quark's \mbox{3-momentum}. We define $A^R_{\rm sing}(\hat m)$
as the contribution to the total angular cross-section coming from the terms inversely proportional to $y^2$ in
$A_q^V$. As we discussed in the previous
section, the $z$ integration can be restricted to $z<1/2$ such that one can compactly write the $y$ upper integration
limit as $y_{\rm top}(\hat m, z)\equiv \min [y_{\tau}(\hat m, z), y_{\max}(\hat m, z)]$. With this definition one
can compute $A^R_{\rm sing}$ analytically as follows (we do not include the prefactor $3C_F/4$ that equals unity in
QCD):
\begin{align}\label{eq:ARdelta}
A^R_{\rm sing}(\hat m) =\,&\frac{1}{(1 - \varepsilon)\Gamma (2 - \varepsilon)} \biggl( \frac{\mu^2e^{\gamma_E}}{s}
\biggr)^{\!\!\varepsilon}\!\int_{z_-}^{\frac{1}{2}}\!{\rm d}z
V^{\rm div} (\hat{m}, z, \varepsilon)[(1 - z) z \, - \hat{m}^2]^{-\varepsilon}\!\!
\int_0^{y_{\rm top}(\hat m, z)}\!\!\frac{{\rm d}y}{y^{ 1 + 2 \varepsilon}} \\
=\,& \int_{z_-}^{\frac{1}{2}}{\rm d}z \biggl\{\hat m^2\biggl[\frac{1}{\varepsilon} + 2
- \log\biggl(\frac{s}{\mu^2}\biggr) - \log\bigr[(1 - z) z \, - \hat{m}^2\bigl]\biggr]
- \frac{9 + 14 \hat{m}^2}{10}\biggr\} M_V^1 (\hat{m},z) \nonumber \\
& -2\hat m^2 \int_{z_-}^{\frac{1}{2}}{\rm d}z \log[y_{\rm top}(\hat m, z)]M_V^1 (\hat{m},z)
+ \mathcal{O}(\varepsilon)\,,\nonumber
\end{align}
where in the first line we have already discarded a term that vanishes as $\varepsilon\to 0$, see discussion
after Eq.~\eqref{eq:feAngC}. Expanding in $\varepsilon$ one gets
$\int_0^{y_{\rm top}}\! {\rm d}y\,y^{- 1 - 2 \varepsilon} = -1/(2\varepsilon) + \log(y_{\rm top})+\mathcal{O}(\varepsilon)$ for the
$y$ integral. The divergent term
does not depend on the upper integration limit, simplifying the subsequent computations. The $z$ integrals in
the second line can be carried out analytically, and for that we shall only need the following two integrals:
\begin{align}
\int_{z_-}^{\frac{1}{2}} {\rm d} z M_V^1 (\hat{m},z) =\,& \beta - 2 (1 - 2 \hat{m}^2) L_{\beta}\,,\\
\int_{z_-}^{\frac{1}{2}} {\rm d} z M_V^1 (\hat{m},z) \log\bigr[(1 - z) z \, - \hat{m}^2\bigl]= \,&
2L_\beta - 2 \beta \log(\beta) + \frac{1+\beta^2}{2}\biggl[{\rm Li}_2\biggl(\frac{2\beta}{\beta-1}\biggr)\nonumber\\
&-2{\rm Li}_2\biggl(\frac{2\beta}{\beta+1}\biggr)\! + 4 \log(\beta) L_\beta \biggr]. \nonumber
\end{align}
The result in the first line of the previous equation shows that the $1/\varepsilon$ pole cancels against its
virtual counterpart, along with the $\mu$ dependence and the term which does not vanish in the massless limit.

\begin{figure}[t]\centering
\includegraphics[width=0.4\textwidth]{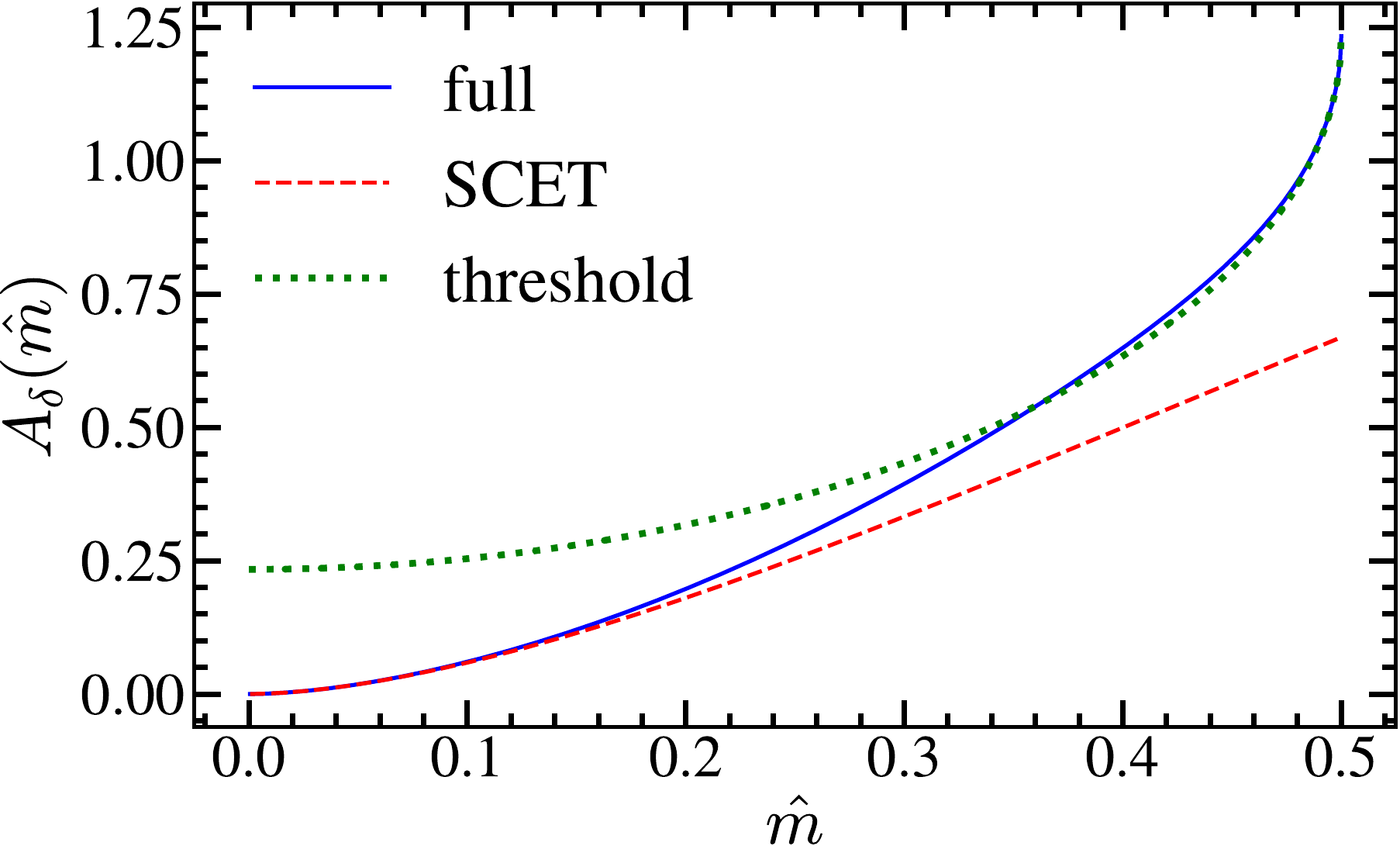}
\caption{Coefficient $A_{\delta}$ (solid blue) and its SCET (dashed red) and threshold (dotted green)
approximations.
\label{fig:Adelta}}
\end{figure}

We are now in position to show the final expression for the total angular cross-section at $\mathcal{O}(\alpha_s)$
for the vector current. We split the result in a term dubbed $A_{\delta} (\hat{m})$, which contains the virtual-radiation
contribution from the second line of Eq.~\eqref{eq:virtual}, and the analytical $z$ integrals on the second line of
Eq.~\eqref{eq:ARdelta} combined in an IR-free coefficient, plus terms in which the $z$ integrals do not admit a
simple analytical form and hence, in practice, are computed numerically:
\begin{align}\label{eq:RhadV}
R_1^{{\rm ang},V} (\hat{m}) = \,& \frac{3C_F}{4}\biggl\{\!A_{\delta} (\hat{m})
+ \!\int_{z_-}^{1/2}\! {\rm d} z\! \tilde{A}_q^V[\hat{m}, y_{\rm top} (\hat m, z), z]
+\! \int_{\hat{m}}^{\frac{1}{2}}\! {\rm d} z\!
\tilde{A}_g^V [\hat{m}, y_{\max} (\hat{m},z), y_{\tau} (\hat{m},z),z]\!\biggr\},\nonumber\\
A_{\delta} (\hat{m}) = \, &\hat{m}^2 \biggl\{ 2 \beta [\log
(\hat{m}) - 1] + \frac{1 + \beta^2}{2} \biggl[ \pi^2 - 2 L^2_{\beta} +
{\rm Li}_2 \biggl( \frac{2 \beta}{\beta - 1} \biggr)
- 3\, {\rm Li}_2 \biggl( \frac{2 \beta}{\beta + 1} \biggr) \nonumber\\
&\qquad -\! 4 L_{\beta} [\log
(\hat{m}) - 1] \biggr] \biggr\} ,\nonumber\\
\tilde{A}_q^V(\hat m, y, z)= \,& \tilde{V}^{\rm fin} (\hat m, y, z) + \log (y) V^{\rm div} (\hat{m}, z, 0)\,,\nonumber\\
\tilde{A}_g^V(\hat{m}, y_{\max}, y_{\tau},z)= \,& 2 \biggl[ y_{\tau} - y_{\max} - M_V^1 (z, \hat{m}) \log \biggl(
\frac{\text{$y_{\max}$}}{y_{\tau}} \biggr) \biggr]\,,\nonumber\\
\tilde{V}^{\rm fin} (\hat m, y, z) = \,&\frac{1}{4 \hat{m} (1 - z)^2 z^3} \biggl\{ (1 - z - 2 \hat{m}^2 z) \biggl[ (1 -
2 \hat{m}) (1 - z - m)^2 \log \biggl( 1 - \frac{y z}{1 - 2 \hat{m}}
\biggr)\nonumber\\
& - (1 + 2 \hat{m}) (1 - z + \hat{m})^2 \log \biggl( 1 - \frac{y z}{1 + 2 \hat{m}} \biggr) \biggr] - 4 \hat{m} \, y (1 - z)^2 z \biggr\}\,,
\end{align}
where the definition of $\tilde{A}_g^V$ can be found in Eq.~\eqref{eq:AtildeDef} and $\tilde{V}^{\rm fin}$ is defined as
\begin{equation}
\tilde{V}^{\rm fin}(\hat m, y, z) = \int_0^y {\rm d} h h V^{\rm fin} (\hat{m}, h, 1 - z)\,.
\end{equation}
The function $A_{\delta}$ can be expanded for small $\hat m$, and the leading term will be referred to as its SCET limit,
and also around $\beta = 0$, whose leading approximation is the threshold limit:
\begin{align}
A_{\delta} (\hat{m})=\,& \frac{1}{3} \hat m^2 [\pi ^2 -6 \log(\hat m) - 6] + \mathcal{O}(\hat m^4)\,,\\
A_{\delta} (\hat{m})=\,& \frac{\pi^2}{8}-\beta+ \mathcal{O}(\beta^2)\,.\nonumber
\end{align}
The coefficient $A_{\delta}$ and its SCET approximation are shown in Fig.~\ref{fig:Adelta}, where
an enhancement towards $\hat m = 1/2$ can be observed.

The dependence of $R_1^{{\rm ang},V}$ on $\hat m$ is shown in Fig.~\ref{fig:R-loop} where one can observe that
the $\mathcal{O}(\alpha_s)$ total angular cross-section does not vanish neither in the massless limit nor at
threshold. Vector and axial-vector currents agree for $m=0$ and reproduce the analytic result quoted in Ref.~\cite{Mateu:2013gya}, that is
$R_1^{{\rm ang},C}(\hat m = 0)=3C_F/8[\log(3/2)-3]\simeq0.12186$. We can
use the same arguments as for the axial-vector current to show that all terms in $R_1^{{\rm ang},V}$ except for
$A_{\delta}$ vanish in the limit $\hat m \to 1/2$. Hence we can provide an analytic result for the total angular cross
section at threshold: \mbox{$R_1^{{\rm ang},V}(1/2)=3\pi^2C_F/36\approx1.09662$}. In fact, $A_{\delta}$ is responsible
for the vector current cross section being larger than the axial-vector one over most of the spectrum. In
particular, this non-vanishing result (which is also found for the total unoriented cross section) seems to
point into a Sommerfeld enhancement at higher orders, which would imply the need for NRQCD resummation. This,
a priory, indicates that $R_{\rm ang}$
might be an interesting viable observable to measure the top quark mass at a future linear collider through threshold scans.

\section{Event-shape differential Distributions}\label{sec:differential}
In this section we combine the real- and virtual-radiation results and project out the differential distributions.
For three particles in the final state, one has that the measurement for any event-shape (and some other
less inclusive observables involving a jet algorithm and even trimming or grooming) is a function of the reduced
mass and the kinematic variables $y$ and $z$. As discussed
in Ref.~\cite{Lepenik:2019jjk}, expanding the measurement function in the soft limit is useful to analytically
obtain the plus and Dirac delta function coefficients:
\begin{equation}
\hat{e} (\hat m, y, z) = e_{\min}(\hat m) + y f_e (\hat m, z) + \mathcal{O} (y^2) \equiv
\overline{e} (\hat m,y, z) + \mathcal{O} (y^2) \, .
\end{equation}
This function has the property that
$\hat{e} (\hat m, y, z) = e_{\min}(\hat m)$ if and only if $y=0$.
We can use the measurement function to write down a formal integral expression for the angular-differential
distribution which is valid for both currents:
\begin{align}\label{eq:feAngC}
&f_{e,1}^{{\rm ang},C}\!(\hat m, e) = \frac{3C_F}{4}\biggl\{ R_{21}^{{\rm ang},C}\delta[e - e_{\min}(\hat m)]
+ \!\! \int_{\hat m}^{\frac{1}{2}}\!\!{\rm d}z\!\int_{y_\tau(\hat m, z)}^{y_{\rm max}(\hat m, z)}\!{\rm d}y\,y A_g^C(\hat m,y, z)
\delta[e - \hat{e} (\hat m, z, y)] \!\\
&\qquad\quad\, +\! \frac{1}{(1 - \varepsilon)\Gamma(2 - \varepsilon)}
\biggl( \frac{\mu^2e^{\gamma_E}}{s}\biggr)^{\!\!\varepsilon}\!\!
\int_{z_-}^{\frac{1}{2}}\!\!{\rm d}z\!\int_0^{y_{\rm top}(\hat m, z)}\!\!\frac{{\rm d}y\, y^{1-2\varepsilon}}
{[(1 - z) z \, - \hat{m}^2 ]^{\varepsilon}}
A_{\bar q}^C(\hat m,y,z) \delta[e - \hat{e} (\hat m, y, z)] \biggr\},\nonumber
\end{align}
where we have set $\varepsilon=0$ already in the $A_g^C$ integration as that term has no support in the
soft part of the Dalitz region, and also $y=0$ in $\xi(\hat m, x_1,x_2)/y^2$ since keeping a non-zero $y$
yields the same result plus a term that vanishes in $d=4$ dimensions.

\subsection{Axial-vector current}

For the axial-vector current, given that there are no soft singularities, one can use $\varepsilon=0$ right
away as long as $R_{21}^{{\rm ang},A}$ is also set to zero. This implies that the $\varepsilon$-dependent
factor out front the integral becomes $1$ and also that $f_{e,1}^{{\rm ang},A}(\hat m, e)$ is purely non-singular:
it is an integrable function as
\mbox{$e\to e_{\rm min}(\hat m)$}.\footnote{This does not imply that $f_{e,1}^{{\rm ang},A}[\hat m, e_{\rm min}(\hat m)]$
is finite. As seen in Ref.~\cite{Lepenik:2019jjk}, $f_{e,1}^{{\rm ang},A}(\hat m, e)$ can have a logarithmic
divergence in the dijet limit. This happens for those event shapes for which $e_{\rm min}(\hat m)=0$, which
includes any observables in the E- or P-schemes.} The $y$ and $z$ integrals can be carried out analytically for some
simple event shapes such as $2$-jettiness or heavy jet mass~\cite{Clavelli:1979md, Chandramohan:1980ry, Clavelli:1981yh} (see next section for explicit expressions), and can
be integrated numerically yielding unbinned distributions with machine precision in fractions of a second using
the algorithm introduced in Ref.~\cite{Lepenik:2019jjk}. Finally, for completeness, we connect with the notation
of Eq.~\eqref{eq:general-diff}: $C_F F^{\rm ang}_{A,e}(\hat m, e)=f_{e,1}^{{\rm ang},A}(\hat m, e)$.

\subsection{Vector current}

For the vector current one has to proceed with care, as there are soft singularities that need special treatment.
To that end, following the same strategy as for the computation of the total angular cross-section, we single out the $A_{\bar q}^V$
terms inversely proportional to $y^2$ in the integral on the second line, as those are the only
ones, together with the virtual radiation, that can yield singular structures:
\begin{align}
F_{\rm sing}^{\rm real} =\,& \frac{1}{(1 - \varepsilon)\Gamma (2 - \varepsilon)}\biggl( \frac{\mu^2e^{\gamma_E}}{s}
\biggr)^{\!\!\varepsilon}\!\!
\int_{z_-}^{\frac{1}{2}}\!{\rm d}z\frac{V^{\rm div} (\hat{m}, z, \varepsilon)}{[(1 - z) z \, - \hat{m}^2 ]^{\varepsilon}}
\!\int_0^{y_{\rm top}(\hat m, z)}\!\!\frac{{\rm d}y}{y^{1+2\varepsilon}}
\delta[e - \hat{e} (\hat m, y, z)] \\
=\,&\delta[e - e_{\min}(\hat m)] \!\!\int_{z_-}^{\frac{1}{2}}\!{\rm d}z\biggl\{\hat m^2\biggl[\frac{1}{\varepsilon} +2
- \log[(1 - z) z \, - \hat{m}^2 ] - \log\biggl(\frac{s}{\mu^2}\biggr)\!\biggr] - \frac{9 + 14 \hat{m}^2}{10}\biggr\} M_V^1 (\hat{m},z)\nonumber\\
& -\hat m^2 \!\!\int_{z_-}^{\frac{1}{2}}\!{\rm d}z M_V^1 (\hat{m},z) \!\int_0^{y_{\rm top}(\hat m, z)}\!{\rm d}y
\biggl[\frac{1}{y}\biggr]_+ \delta[e - \bar{e} (\hat m, y, z)]\nonumber\\
& -\hat m^2 \!\!\int_{z_-}^{\frac{1}{2}}\!{\rm d}z M_V^1 (\hat{m},z) \!\int_0^{y_{\rm top}(\hat m, z)}\!
\frac{{\rm d}y}{y} \{\delta[e - \hat{e} (\hat m, y, z)] - \delta[e - \bar{e} (\hat m, y, z)]\}\,,\nonumber
\end{align}
where to get to the second line we have used the identity
$y^{- 1 - 2 \varepsilon} = - 1/(2 \varepsilon) \delta (y) + [1/y]_+ +\mathcal{O} (\varepsilon)$
and expanded in $\varepsilon$. When one adds the virtual-radiation contribution to the first line, the IR singularity and $\mu$ dependence disappear
and the coefficient $A_{\delta}$ defined in Eq.~\eqref{eq:RhadV} is found. The term in the last line is regular when $y\to 0$ and does not yield any
distribution. It is important to add and subtract $\delta[e - \bar{e} (\hat m, y, z)]$ and not simply $\delta[e - e_{\rm min} (\hat m)]$ since otherwise the
subtracted term would still contain singular structures.

To fully disentangle the coefficient of the plus and Dirac delta functions we proceed as follows with the term in the third line:
\begin{align}
& \int_{z_-}^{\frac{1}{2}}\!{\rm d}z \frac{M_V^1 (\hat{m},z)}{f_e(\hat m, z)} \int_0^1 {\rm d}y\,\theta[y_{\rm top}(\hat m, z) - y]
\biggl[\frac{1}{y}\biggr]_+ \delta\biggl[y - \frac{e-e_{\rm min}(\hat m)}{f_e(\hat m, z)}\biggr]\\
& = \int_{z_-}^{\frac{1}{2}}\!{\rm d}z \frac{M_V^1 (\hat{m},z)}{f_e(\hat m, z)} \biggl[\frac{f_e(\hat m, z)}{e-e_{\rm min}(\hat m)}\biggr]_+
\,\theta[f_e(\hat m, z)y_{\rm top}(\hat m, z) - e + e_{\rm min}(\hat m)] \nonumber\\
& = \int_{z_-}^{\frac{1}{2}}\!{\rm d}z M_V^1 (\hat{m},z) \biggl\{\biggl[\frac{1}{e - e_{\rm min}(\hat m)}\biggr]_+
-\delta[e - e_{\rm min}(\hat m)] \log[f_e(\hat m, z)]\nonumber\\
& \qquad\qquad\qquad\qquad\quad -\!\frac{\theta[e - e_{\rm min}(\hat m) - f_e(\hat m, z) y_{\rm top}(\hat m, z])}{e - e_{\rm min}(\hat m)} \biggr\},\nonumber
\end{align}
where we have followed the same steps as in Eqs.~(3.23) to (3.27) of Ref.~\cite{Lepenik:2019jjk}. We can now write down the analytic form for the
Dirac delta and plus distribution coefficients defined in Eq.~\eqref{eq:general-diff}:
\begin{align}
A^{\rm ang}_{e}({\hat m}) = \,& \frac{3}{4}[A_\delta(\hat m) - 2 \hat{m}^2 I_e (\hat{m})]\,,\qquad\quad
B^{\rm ang}_{\rm plus}({\hat m}) = \frac{3\hat m^2}{2} [2 (1 - 2 \hat{m}^2) L_{\beta} - \beta] \,,\\
I_e (\hat{m}) =\, &\! -\!\! \int_{z_-}^{\frac{1}{2}} {\rm d} z M_V^1 (\hat{m},z) \log [f_e (z)]\,,\nonumber
\end{align}
where both coefficients vanish in the massless limit and the event-shape-dependent function $I_e$ was already defined in Ref.~\cite{Lepenik:2019jjk},
and analytically computed for a large number of event shapes in various schemes. We close this section writing down an expression
for the non-singular distribution
\begin{align}
F^{\rm ang}_{V,e} (e, \hat{m}) = \, & \frac{3}{4} \biggl[ \int_{\hat{m}}^{\frac{1}{2}}\!\!{\rm d}z\!\!
\int_{y_\tau(\hat m, z)}^{y_{\rm max}(\hat m, z)}\!\!{\rm d}y\,y A_g^{C\!}(\hat m,y, z)
+\!\! \int_{z_-}^{\frac{1}{2}}\!\!{\rm d}z\!\!\int_0^{y_{\rm top}(\hat m, z)} \!\!{\rm d}y\, y V^{\rm fin} (\hat{m}, y, 1 - z) \!\biggr]
\delta[e - \hat{e} (\hat m, z, y)] \nonumber\\
&-\!\frac{3\hat m^2}{4} \!\!\int_{z_-}^{\frac{1}{2}}\!{\rm d}z M_V^1 (\hat{m},z) \!\int_0^{y_{\rm top}(\hat m, z)}\!
\frac{{\rm d}y}{y} \{\delta[e - \hat{e} (\hat m, y, z)] - \delta[e - \bar{e} (\hat m, y, z)]\} \nonumber\\
& +\!\frac{3\hat m^2}{4} \!\!\int_{z_-}^{\frac{1}{2}}\!{\rm d}z M_V^1 (\hat{m},z)
\frac{\theta[e - e_{\rm min}(\hat m) - f_e(\hat m, z) y_{\rm top}(\hat m, z])}{e - e_{\rm min}(\hat m)}\,.
\end{align}
All results in this section have been accurately reproduced by an independent computation performed by some of us
in which the angular term is only projected from the $\theta_T$-differential cross section after adding real- and
virtual-radiation contributions, that is, after having cancelled the IR singularities. This computation is fundamentally
different since, at intermediate steps and upon expanding in $\varepsilon$, an angular structure different from those in
Eq.~\eqref{eq:decomposition} appear (this evanescent structure is even divergent for $\theta_T=\pm \pi$).
This is yet another artifact of dimensional regularization that disappears when adding all terms (the proof in
Ref.~\cite{Mateu:2013gya} implicitly assumes $4$ space-time dimensions).

\begin{figure*}[t!]
\subfigure[]
{\includegraphics[width=0.46\textwidth]{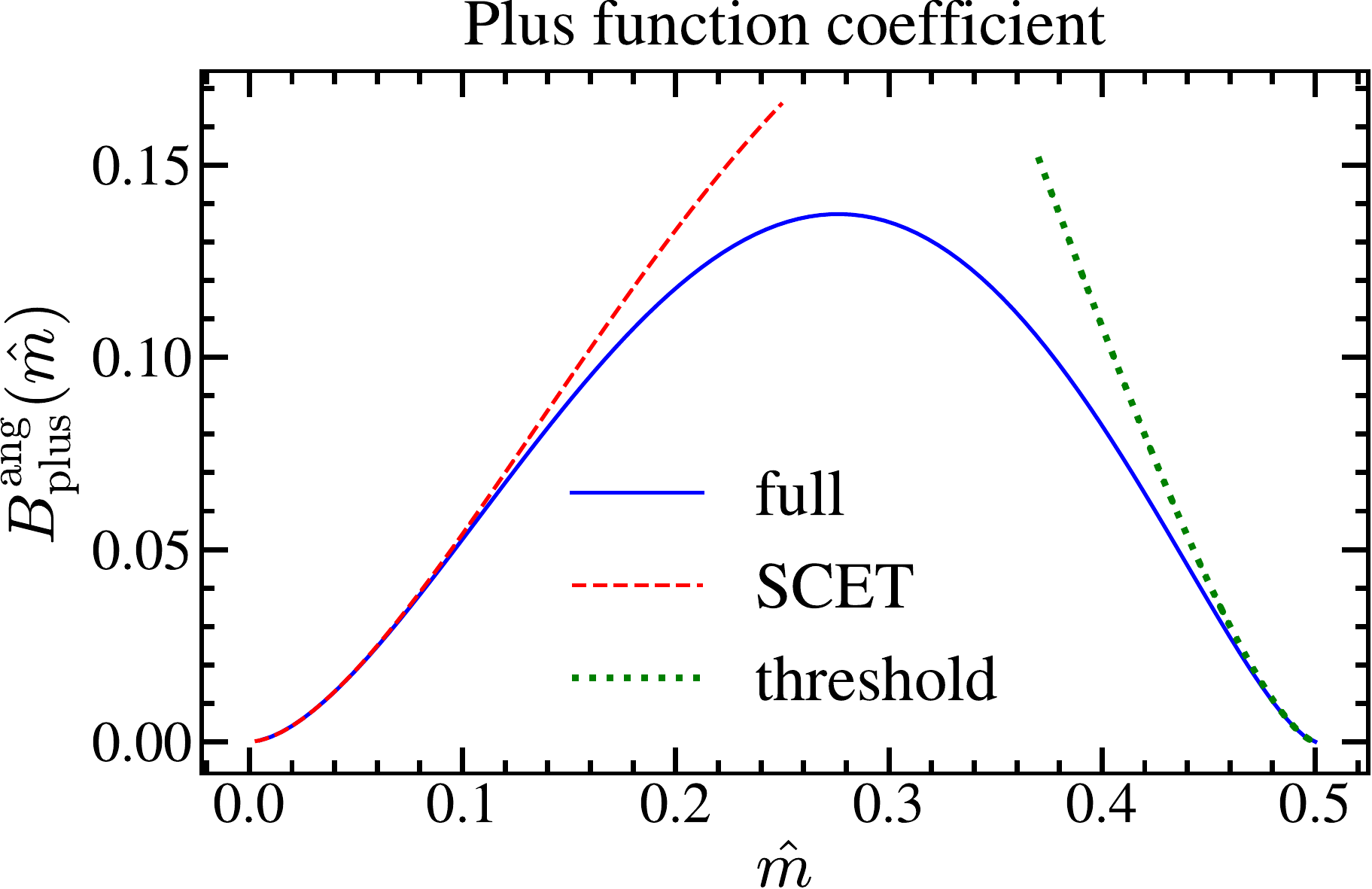}
\label{fig:Bplus}}~~~~~~
\subfigure[]{\includegraphics[width=0.45\textwidth]{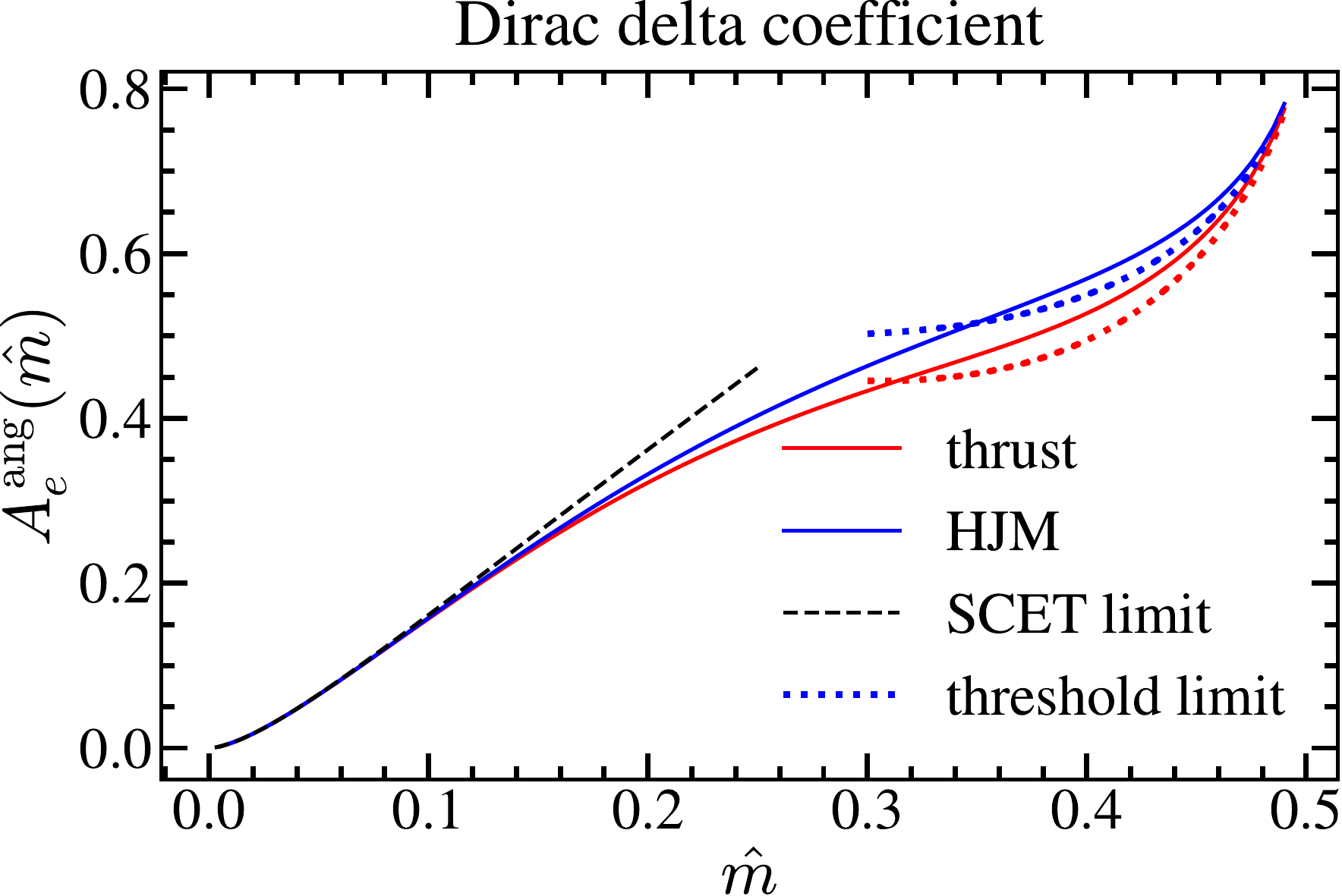}
\label{fig:DeltaCoef}}
\caption{Panel (a): Plus function coefficient $B_{\rm plus}^{\rm ang}(\hat m)$ (solid blue) and its leading SCET (dashed red) and threshold (dotted green) approximations.
Panel (b): Dirac delta function coefficient $A_e^{\rm ang}(\hat m)$ for $2$-jettiness $e=\tau_J$ (blue) and heavy
jet mass $e=\rho$ (red) as solid lines for the full result, dotted lines for the respective threshold expansions, and as
a black dashed line for the common SCET expansion.}
\label{fig:DeltaPlus}
\end{figure*}

Even though we have a formal expression for the non-singular terms, in practice it is however simpler to compute
(numerically or analytically, depending on the event shape) the complete distribution (singular plus non-singular)
for $e>e_{\rm min}$ such that one can drop the plus prescription from $1/[e - e_{\rm min}(\hat m)]_+$ and the delta function
is simply zero. Furthermore, one can set $\varepsilon=0$ and work only with the real-radiation contribution in $d=4$
dimensions. Since the coefficient of the plus distribution has been computed analytically, the non-singular distribution is then obtained
by simply subtracting the radiative tail:
\begin{equation}
C_F F^{\rm ang}_{V,e} (e, \hat{m}) = f_{e,1}^{{\rm ang},V}[\hat m, e>e_{\rm min}(\hat m)]|_{\varepsilon=0}
- C_F B^{\rm ang}_{\rm plus}({\hat m}) \biggl[\frac{1}{e - e_{\rm min}(\hat m)}\biggr].
\end{equation}
Finally, we provide the leading term of the plus function coefficient when expanded around the massless
limit, $B_{\rm plus}^{\rm ang}(\hat m)\approx -3\hat m^2[1+2\log(\hat m)]/2$, which we again call the
SCET approximation, and around $\beta=0$ (threshold approximation),
$B_{\rm plus}^{\rm ang}(\hat m)\approx\beta^3/2$. Therefore, $B_{\rm plus}^{\rm ang}$ vanishes both in the
massless limit and at threshold. We have already argued that such behavior is expected for $\hat m=0$:
for massless quarks the distribution is purely non-singular. In Ref.~\cite{Lepenik:2019jjk} it was stated that
at threshold there is not enough energy to emit an extra particle and therefore there is no radiative tail,
causing a null value for $B_{\rm plus}^{\rm ang}$ (one can however have a non-zero coefficient for the
delta function). In Fig.~\ref{fig:Bplus} $B_{\rm plus}^{\rm ang}$ and its approximations are shown.

Although all our results have been expressed in terms of the quark's pole mass, with a single simple modification we
can obtain $\overline{\rm MS}$ results. At the order that we are working we only need the relation between these two
mass schemes at leading order:
\begin{equation}
m_{\rm pole} = \overline{m}(\mu)\biggl\{1 + \biggl[\frac{\alpha_s(\mu)}{\pi}\biggr]C_F
\biggl[1 - \frac{3}{2}\log\biggl(\frac{\overline{m}(\mu)}{\mu}\biggr)\biggr]\biggr\} + \mathcal{O}(\alpha_s^2)\,.
\end{equation}
None of the results for the axial-vector current need any modification: one simply replaces the pole mass by $\overline m(\mu)$.
For the vector current, one proceeds in the same way and corrects the $\mathcal{O}(\alpha_s)$ angular cross section and delta
function coefficient. Defining $\tilde{m}_\mu\equiv \overline{m}(\mu)/Q$ and $\tilde \beta_\mu \equiv \sqrt{1 - 4\tilde{m}_\mu^2}$,
and using the notation that quantities with a bar on top are expressed in the $\overline{\rm MS}$ scheme, one has:
\begin{align}
\overline{R}_1^{{\rm ang},V} (\tilde{m}_\mu) = \, & R_1^{{\rm ang},V} (\tilde{m}_\mu)
+ \frac{3C_F}{4} \delta_{\rm ang}^{\overline{\rm MS}}\,,\\
\bar A^{\rm ang}_{e} (\tilde{m}_\mu) = \, & A^{\rm ang}_{e}(\tilde{m}_\mu)
+ \delta_{\rm ang}^{\overline{\rm MS}}\,,\nonumber\\
\delta_{\rm ang}^{\overline{\rm MS}} \equiv\,& \frac{\tilde m_\mu^2}{\tilde\beta_\mu}
\biggl\{2 - 3\log\biggl[\frac{\overline{m}(\mu)}{\mu}\biggr]\!\biggr\}\,.\nonumber
\end{align}
These two modifications can be encompassed in the following single substitution: $A_\delta \to A_\delta + (4/3) \delta_{\rm ang}^{\overline{\rm MS}}$.
\subsection{Computation of Moments}
The algorithm to numerically compute moments introduced in Ref.~\cite{Lepenik:2019jjk} can be easily adapted
for those of the angular distribution. We define the displaced angular moments as
\begin{equation}
\langle (e-e_{\rm min})^n\rangle _{\rm ang}^C = \frac{1}{\sigma_0}\int_0^{e_{\rm max}}
{\rm d}e \frac{{\rm d}\sigma^C_{\rm ang}}{{\rm d}e}(e-e_{\rm min})^n\,.
\end{equation}
To any order one has $\langle (e-e_{\rm min})^0\rangle^C_{\rm ang}=R^C_{\rm ang}$, while for higher moments,
at $\mathcal{O}(\alpha_s)$ one uses the following expression in which all terms have $\varepsilon$ set to $0$:
\begin{align}
\langle (e-e_{\rm min})^n\rangle _{\rm ang}^C =\,& \frac{3C_F}{4}\frac{\alpha_s}{\pi}\biggl\{\int_{\hat m}^{\frac{1}{2}}\!\!{\rm d}z\!\int_{y_\tau(\hat m, z)}^{y_{\rm max}(\hat m, z)}\!{\rm d}y\,y A_g^C(\hat m,y, z)
[\hat{e} (\hat m, z, y) - e_{\rm min}(\hat m)]^n\\
&+\int_{z_-}^{\frac{1}{2}}\!\!{\rm d}z
\!\int_0^{y_{\rm top}(\hat m, z)}\!\!{\rm d}y\, y
A_{\bar q}^C(\hat m,y,z) [\hat{e} (\hat m, z, y) - e_{\rm min}(\hat m)]^n \biggr\}+\mathcal{O}(\alpha_s^2).\nonumber
\end{align}
To compute regular moments one can similarly modify Eq.~(4.6) of Ref.~\cite{Lepenik:2019jjk}. One can adapt the
methods described in Ref.~\cite{Lepenik:2019jjk} to compute differential and cumulative cross sections either
with a (slow and unprecise) Monte Carlo strategy or following a (fast and accurate) ``deterministic'' algorithm.
Since that would be too repetitive, and it is relatively straightforward, we shall not describe how this is done
and show instead some numerical results in the following sections.

\section{Thrust and Heavy Jet Mass Distributions}\label{sec:analytic}
In Ref.~\cite{Lepenik:2019jjk} a lot of emphasis was put into describing a general method for numerically obtaining
unbinned event-shape distributions in a fast and precise way. In this section we take a different route and discuss
some analytic (or partially analytic) results. To that end,
we compute the differential and cumulative cross sections for two mass-sensitive event shapes:
$2$-jettiness (a generalization of thrust useful for massive particles) and heavy jet mass (HJM). While we
are capable of obtaining fully analytical results for the differential distribution, for the cumulative versions
we are left with a one-dimensional numerical integral.

The definition of $2$-jettiness depends on the thrust axis defined around Eq.~\eqref{eq:axisDef} and reads
\begin{equation}\label{eq:tauDef}
\tau_J = \frac{1}{Q} \min_{\hat{n}} \sum_i (E_i - | \hat{n} \cdot \vec{p}_i|) \,.
\end{equation}
For three partons, one of them massless, it can be shown that the measurement function can be expressed as a
minimum condition:
\begin{equation}
\tau_J = \min \left\{ 1 - y, 1 - \, {\rm mod} (\hat{m}, y, 1 - z), 1 - \,{\rm mod} (\hat{m},y, z) \right\},
\end{equation}
with ${\rm mod} (\hat{m},y, z) \equiv \sqrt{(1 - y z)^2 - 4 \hat{m}^2}$. The three values in the list correspond
to the thrust axis parallel to the $3$-momentum of the gluon, quark and anti-quark, respectively. On the other hand,
heavy jet mass is the largest invariant mass of the two hemispheres defined
by the plane orthogonal to the thrust axis. For the configuration just described the measurement
is best written as a piecewise function:
\begin{align}\label{eq:HJMdef}
& z_- \leqslant z \leqslant \frac{1}{2}\,, & 0 \leqslant & \,y \leqslant y_{\rm top} (\hat{m},z)\,,
& \rho & = \hat{m}^2 + y z\,,
\\
& \frac{1}{2} \leqslant z \leqslant z_+\,, & 0 \leqslant &\, y \leqslant y_{\rm top}
(\hat{m}, 1 - z)\,, & \rho& = \hat{m}^2 + y (1 - z)\,, \nonumber\\
& \hat{m} \leqslant z \leqslant 1 - \hat{m}\,, & y_{\rm low} (\hat{m},z)\leqslant & \,y \leqslant y_{\max} (\hat{m}, z)\,, & \rho& = 1 - y\,, \nonumber
\end{align}
where we have defined $y_{\rm low} (\hat{m},z)=\max [y_{\tau} (\hat{m}, z), y_{\tau} (\hat{m},1 - z)]$. Again, the
three regions correspond to the thrust axis pointing into the anti-quark, quark and gluon $3$-momentum directions,
respectively. It is therefore trivial to see that in the massless limit heavy jet mass and $2$-jettiness, with either two or three
partons, are identical.

\subsection{Thrust}
\begin{figure*}[t!]
\subfigure[]
{\includegraphics[width=0.46\textwidth]{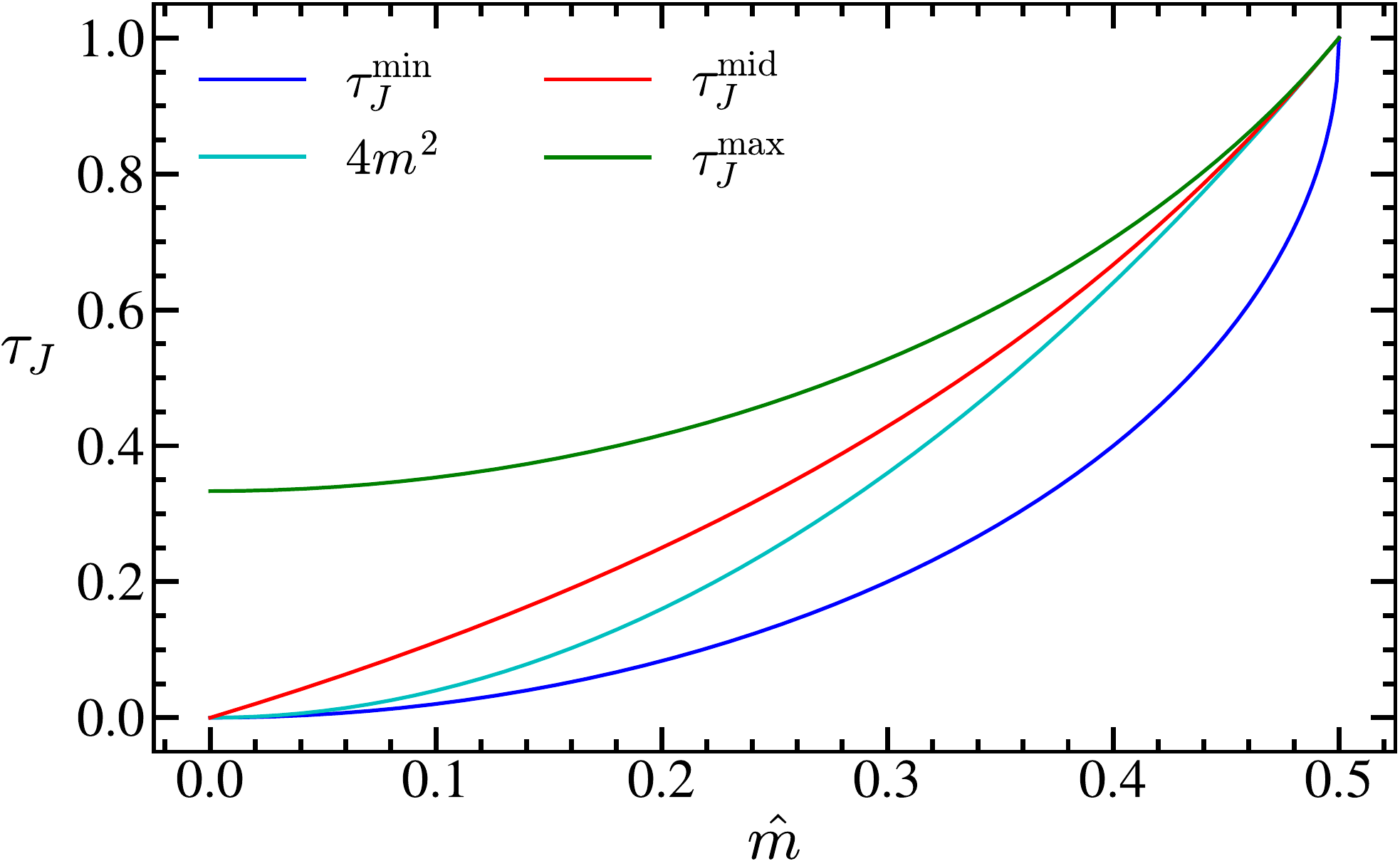}
\label{fig:tauVal}}~~~~
\subfigure[]{\includegraphics[width=0.46\textwidth]{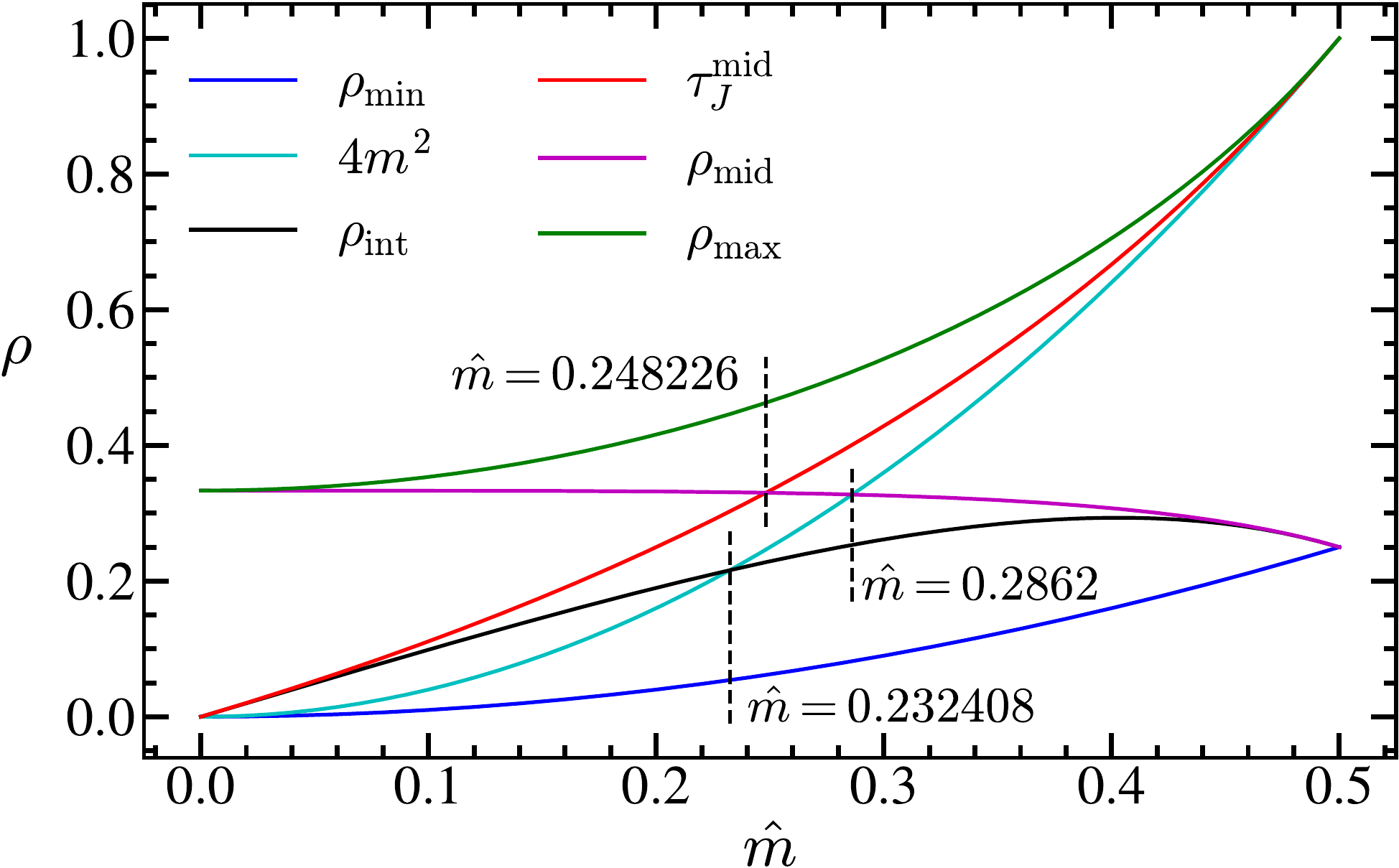}
\label{fig:rhoVal}}
\caption{Values of $2$-jettiness [\,panel (a)\,] and heavy jet mass [\,panel (b)\,] for which the contour lines of constant event shape
reach a limiting value: either they take their maximum or minimal values or transition from hitting the phase space boundary
to enter the Dalitz region.}
\label{fig:tauRhoVal}
\end{figure*}
\begin{figure*}[t!]
\subfigure[]
{\includegraphics[width=0.46\textwidth]{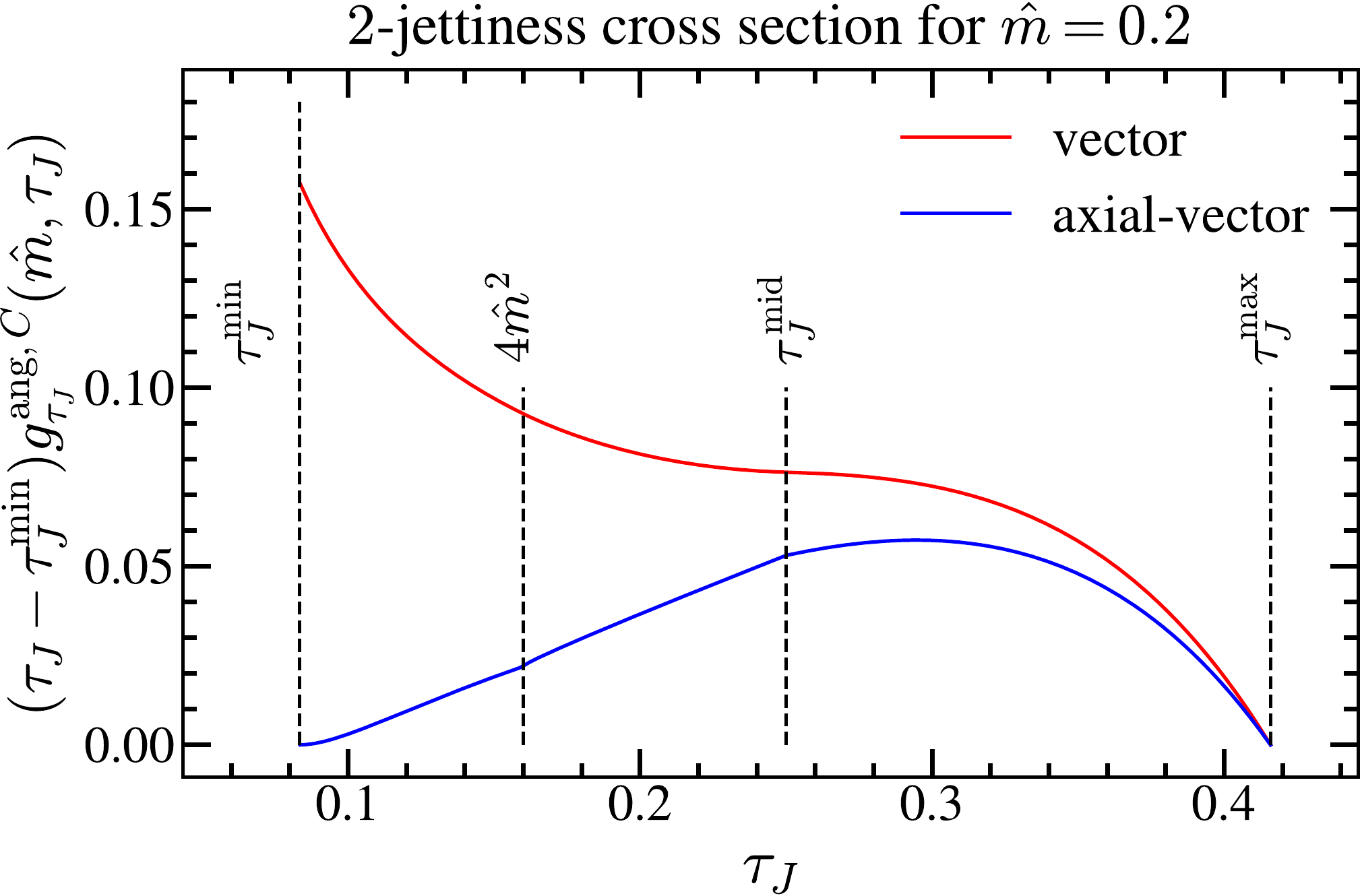}
\label{fig:tau-cont}}
\subfigure[]{\includegraphics[width=0.46\textwidth]{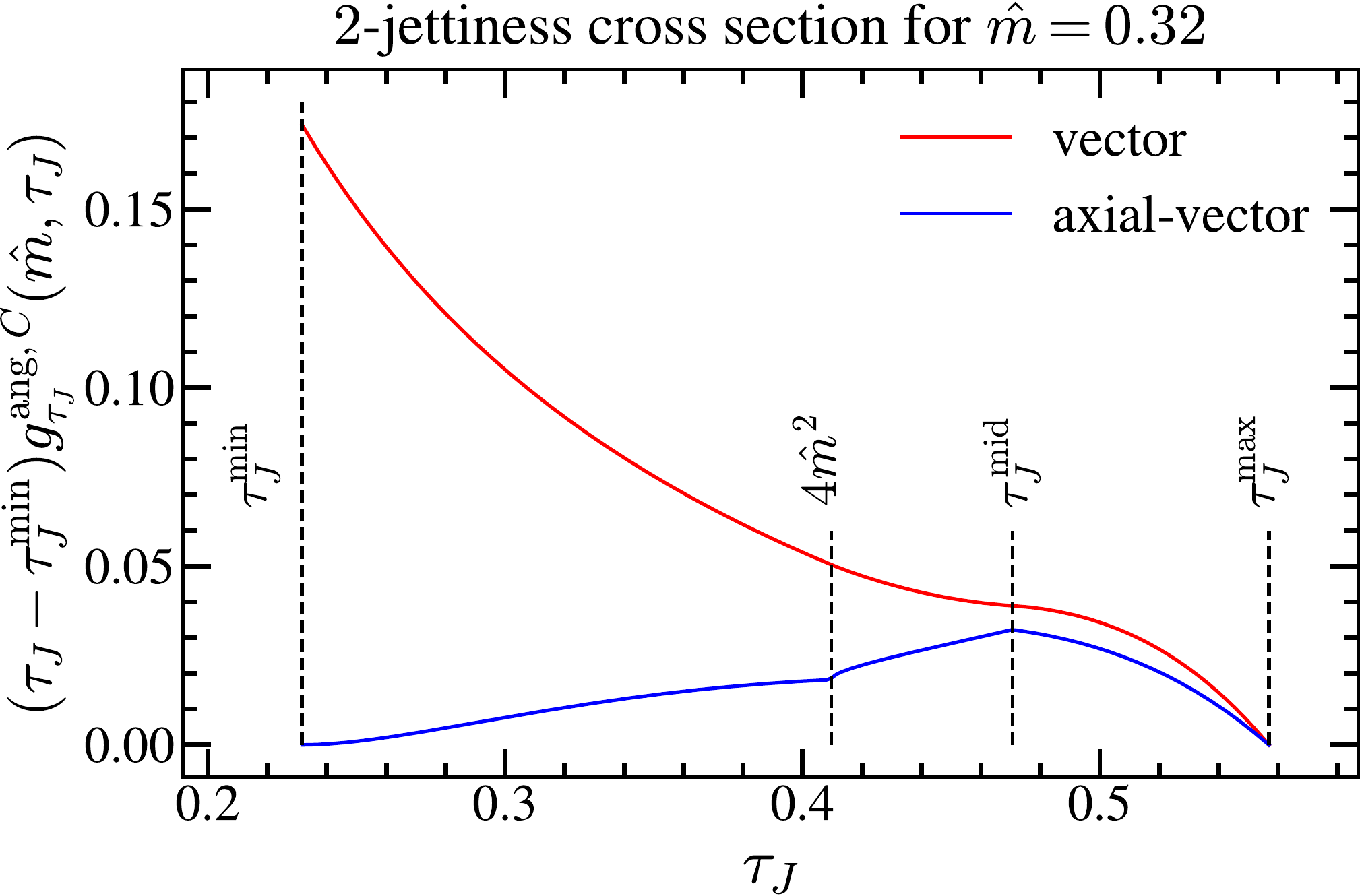}
\label{fig:tau-inc}}
\subfigure[]
{\includegraphics[width=0.465\textwidth]{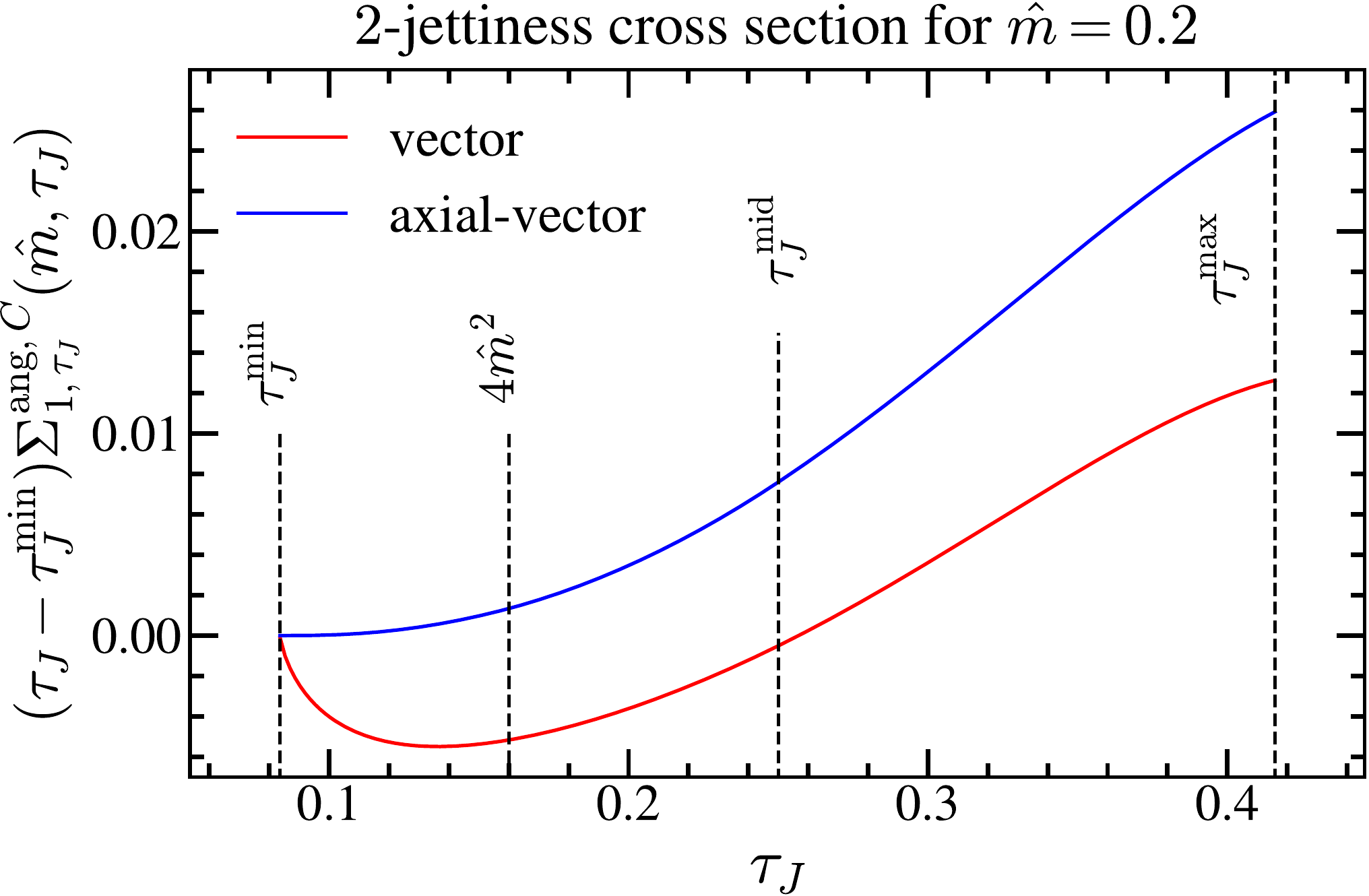}
\label{fig:cumul-tau-cont}}~~~~
\subfigure[]{\includegraphics[width=0.465\textwidth]{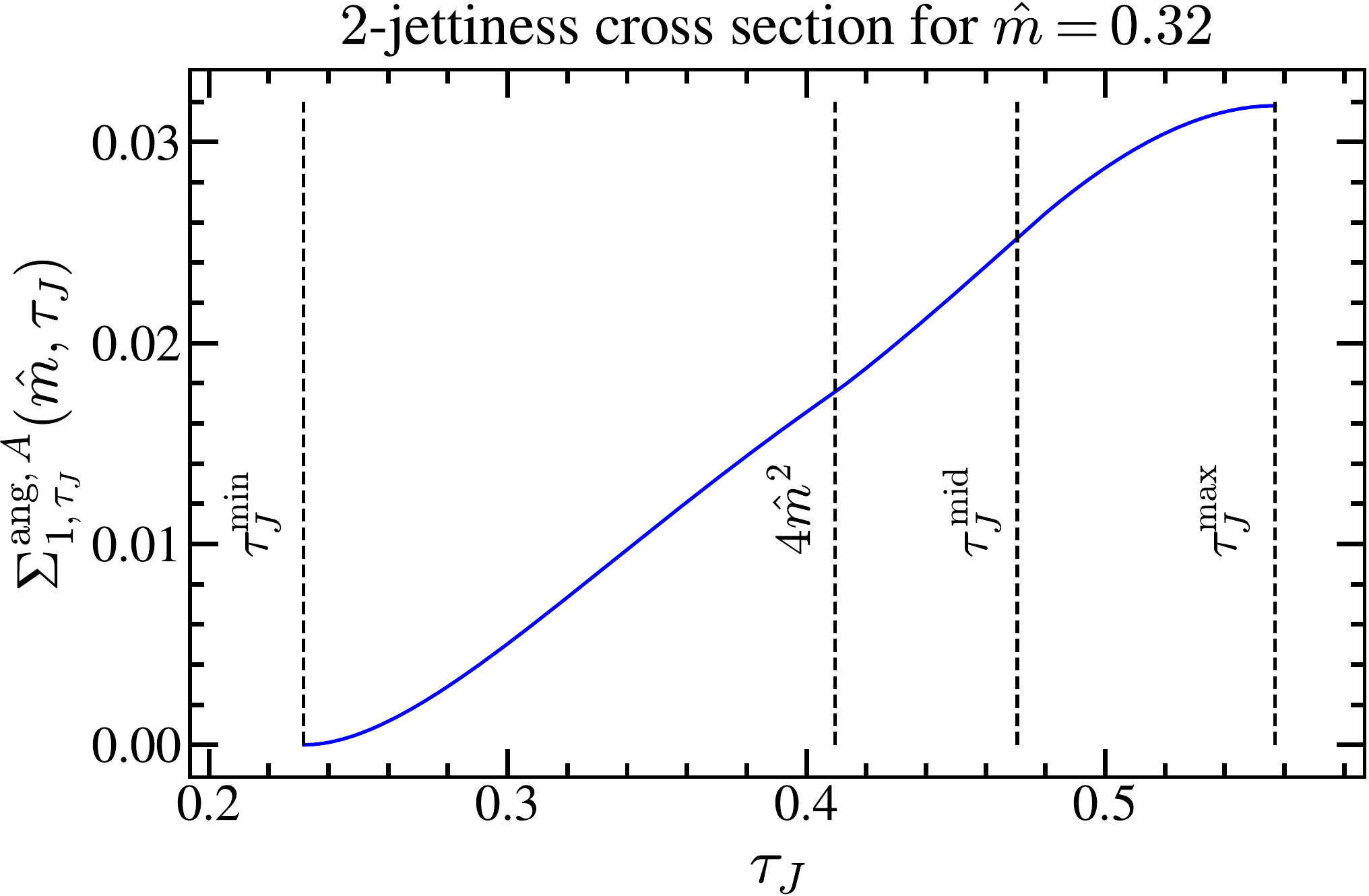}
\label{fig:cumul-tau-inc}}

\caption{Differential (upper panels) and cumulative (lower panels) $2$-jettiness
distribution for $\hat m = 0.2$ (left panels) and $\hat m = 0.32$ (right panels) for the
vector (red) and axial-vector (blue) currents. In panels (a) to (c) we multiply the cross
section by $\tau_J - \tau_J^{\rm min}$ to achieve a finite result across the
whole spectrum. In panel (d) we show only the axial-vector current and do not multiply by
$\tau_J - \tau_J^{\rm min}$. Vertical dashed black lines signal the limiting values of
$\tau$ shown in Fig.~\ref{fig:tauVal}, where one can observe either cusps or discontinuities
in the cross sections.} \label{fig:tau-dif-cum}
\end{figure*}

Even though we have provided a very compact expression for the $\tau_J$ measurement in Eq.~\eqref{eq:tauDef},
for an analytic computation it is more practical to use the regions displayed in Eq.~\eqref{eq:HJMdef}. The
Dalitz region mirror symmetry simplifies the discussion, since it restricts the integration to $z<1/2$ such that it
is enough to consider the anti-quark and gluon regions only, for which the corresponding measurement delta
functions read
\begin{equation}
\delta^{\tau_J}_{\bar q} = \frac{t(\tau_J)}{z \xi(\tau_J)} \delta \biggl[y - \frac{1 - \xi(\tau_J)}{z} \biggr]\,,\qquad
\delta^{\tau_J}_g = \delta[y - t(\tau_J)]\,,
\end{equation}
where we have defined $t (\tau_J) = 1 - \tau_J$ and $\xi(\hat m, \tau_J) = \sqrt{t (\tau_J)^2 + 4 \hat{m}^2}$.
Since the highest possible value of $y$ in the Dalitz region is $y_{\rm max}(1/2)=\beta^2$ (in the gluon region),
the contour line of constant $\tau_J$ lives only in the anti-quark region for $\tau_J < 4 \hat{m}^2$,
where it meets the phase space boundary at \mbox{$z_1(\hat m, \tau_J) = [1 + \tau_J - \xi(\hat m, \tau_J)] / 2$}.
From the limiting
condition $z_1(\hat m, \tau_J^{\rm min}) = z_-(\hat m)$ one obtains the minimal value of $2$-jettiness:
$\tau_J^{\rm min}(\hat m) = 1- \sqrt{1 - 4 \hat m^2}$. From the condition $z_1(\hat m, \tau_J^{\rm mid})=\hat m$
(that is, the contour line hits the point at which the phase-space boundary meets the
line that separates the anti-quark and gluon regions) one can see that for
$4 \hat{m}^2 < \tau_J < \hat{m} / (1 - \hat{m})\equiv \tau_J^{\rm mid}(\hat m)$ the contour line
also has a patch in the gluon region, which cuts the phase space boundary at
\mbox{$z_2(\hat m, \tau_J) = [ 1 - \sqrt{1 - 4 \hat{m}^2 / \tau_J} ] / 2$} [\,one can imagine that the contour line
``leaves'' the Dalitz region (anti-quark patch) through $z_1$ and re-enters it in $z_2$ (gluon patch)\,].
One can easily check that $z_2(\hat m, \tau_J^{\rm mid})=\hat m$.
Finally, if $\tau_J > \tau_J^{\rm mid}(\hat m)$ the contour line becomes continuous (although not smooth),
lives both in the anti-quark and gluon regions, never exits the Dalitz region but meets the thrust
axis boundary at $z_3(\hat m, \tau_J) = [1 - \xi(\hat m, \tau_J)] / t(\tau_J)$. From the limiting condition
$z_3(\hat m, \tau_J^{\rm max})=1/2$ one obtains the
maximum value of $2$-jettiness: $\tau_J^{\rm max}(\hat m) = (\,5- 4\sqrt{1 - 3 \hat m^2}\,)/3$.
One can easily see that for physical values of $\hat m$ the hierarchy
$\tau_J^{\rm min} (\hat m) \leqslant 4\hat m^2\leqslant \tau_J^{\rm mid} (\hat{m}) \leqslant \tau_J^{\rm max} (\hat m)$ holds,
as can be checked graphically in Fig.~\ref{fig:tauVal}.
Defining $z_{ij}(\hat m, \tau_J) = \max[z_i(\hat m, \tau_J), z_j(\hat m, \tau_J)]$ we obtain
\begin{align}
\frac{1}{\sigma_0} \frac{{\rm d} \sigma^C_{\rm ang}}{{\rm d} \tau_J} =\,
& \frac{3 \alpha_s C_F}{4 \pi} g^{{\rm ang}, C}_{\tau_J} (\hat{m},\tau_J)\,, \\
g^{{\rm ang}, C}_{\tau_J} \!(\hat{m},\tau) = \,& \frac{[1 - \xi(\hat m, \tau)] t(\tau)}{\xi(\hat m, \tau)}\!\!\!
\int^{\frac{1}{2}}_{z_{13}(\hat m, \tau)} \!\!\frac{{\rm d} z}{z^2}
A_{\bar q}^C \!\biggl[\hat{m}, \!\frac{1 - \xi(\hat m, \tau)}{z}, z \biggr]\!\!
+\! t(\tau) \theta (\tau \!- 4 \hat{m}^2)\!\!\!
\int^{\frac{1}{2}}_{z_{23}(\hat m, \tau)}\!\!\!\!\!\!\!\!\! {\rm d} z A_g^{C}[\hat{m}, t(\tau), z]
\nonumber\\
= \,& \frac{t(\tau)}{\xi(\hat m, \tau)} g^C_{\bar q} [\hat m, z_{13}(\hat m, \tau), 1 - \xi(\hat m,\tau)]
+ \theta(\tau - 4 \hat{m}^2) g^C_g[\hat m, z_{23}(\hat m, \tau),\tau] \,, \nonumber
\end{align}
where the functions $g^C_{g,\bar q}$ can be computed analytically:
\begin{align}\label{eq:gsAna}
g_{\bar q}^V (\hat m, z, \xi) = \,& \frac{1}{2 \xi [(1 - \xi)^2 - 4 \hat{m}^2]} \biggl[ 2
\hat{m}^2 \bigl[2 + 3 (\xi - 2) \xi + 4 \hat{m}^2 (4 \xi - 3) + 16 \hat{m}^4\bigr]\!
\log \biggl( \frac{1}{z} - 1 \biggr) \\
& - \!\frac{1 - 2 z}{(1 - z)^2 z} \bigl\{ \xi^2 (1 - z) [\xi
- 2 (1 - \xi) z] + \hat{m}^2 \xi (1 - z) [4 (1 - \xi) z + \xi]
\nonumber\\
& +\! 4 \hat{m}^4 z (1 - 6 \xi + 4 \xi
z) - 16 \hat{m}^6 z\bigr\} \biggr], \nonumber\\
g_{\bar q}^A (\hat m, z, \xi) =\, & \frac{\xi}{2 z^2_e (1 - z) [(1 - \xi)^2 - 4
\hat{m}^2]} \biggl[ - 2 \hat{m}^2 z^2 (1 - z) (2 - 4 \hat{m}^2 - \xi^2)
\log \biggl( \frac{1}{z} - 1 \biggr) \nonumber\\
& - \!(1 - 2 z) \{ \xi (1 - 2 \hat{m}^2) (1 - z) (1 + 2 z) - 2 z
[1 - z + 2 \hat{m}^2 (\hat{m}^2 + 2 z - 2)] \} \nonumber\\
& -\! \hat{m}^2 \xi^2 (1 - z) (1 + 4 z)
\vphantom{\frac{1}{z}} \biggr], \nonumber\\
g_g^V (\hat m, z,\tau_J) =\, & \frac{1}{1 - \tau_J}\biggl[\tau_J (1 - 2 z) - 2 \hat{m}^2 \log \biggl(
\frac{1}{z} - 1 \biggr)\!\biggr], \nonumber\\
g_g^A (\hat m, z,\tau_J) =\, & \frac{1}{1 - \tau_J}\biggl\{(1 - 2 z) \biggl[ \tau_J + \frac{2
\hat{m}^4}{(1 - z) z} \biggr] + \hat{m}^2 (\tau_J^2 -
4 \tau_J - \beta^2) \log \biggl( \frac{1}{z} - 1 \biggr)\!\biggr\} .
\nonumber
\end{align}
A graphical representation of the $2$-jettiness angular differential distribution at $\mathcal{O}(\alpha_s)$ is to be found in Fig.~\ref{fig:tau-dif-cum}.
For both currents, implementing the SCET counting $\tau \propto \mathcal{O}
(\lambda^2)$ and $m \propto \mathcal{O} (\lambda)$ one finds:
\begin{align}
g^V_{\tau_J} (\hat{m}, \tau_J) \simeq \,& \frac{8 \hat{m}^4 - 8 \hat{m}^2
\tau + \tau^2 - 4 \hat{m}^2 (\tau - \hat{m}^2) \log (\tau -
\hat{m}^2)}{2(\tau - \hat{m}^2) (\tau - 2 \hat{m}^2)} + \mathcal{O}
(\lambda)\,, \\
g^A_{\tau_J} (\hat{m}, \tau_J) \simeq \,& \frac{\tau(\tau - 2 \hat{m}^2)}{2(\tau - \hat{m}^2)^2} + \mathcal{O} (\lambda)\,, \nonumber
\end{align}
which is significantly different to the case of the unoriented cross section, for which both currents coincide at leading order in the SCET power
counting.

One can compute the oriented cumulative distribution, which is defined as
\begin{equation}
\Sigma_e^{{\rm ang},C}(\hat m, e_c) = \frac{1}{\sigma_0^C}\int_{e_{\rm min}}^{e_c} {\rm d}e
\frac{{\rm d}\sigma_{\rm ang}^C}{{\rm d}e}
\equiv R_0^{{\rm ang},C}(\hat m) \theta[e_c-e_{\rm min}(\hat m)]+ \sum_{n=1}
\biggl[\frac{\alpha_s(\mu)}{\pi}\biggr]^n\Sigma_{e,n}^{{\rm ang},C}(\hat m, \mu, e_c)\,,
\end{equation}
following the same logic as in Ref.~\cite{Lepenik:2019jjk}. Since $\Sigma_e^{{\rm ang},C}$ obeys an homogeneous
renormalization group equation, there is no $\mu$ dependence in $\Sigma_{e,1}^{{\rm ang},C}$.
At $\mathcal{O}(\alpha_s)$ one ends up in the following compact result:
\begin{align}
\Sigma^{{\rm ang},C}_{\tau_J,1}(\hat m, \tau_c) =\, & R_1^{{\rm ang},C}(\hat m) -\frac{3C_F}{4}\!
\int^{\frac{1}{2}}_{z_{13}(\hat m,\tau_c)} {\rm d} z \biggl\{ \tilde{A}_{\bar q}^C [\hat{m}, y_{\rm top} (\hat{m}, z), z] - \tilde{A}_{\bar q}^C\!
\biggl[\hat{m}, \frac{1 - \xi(\hat m,\tau_c)}{z}, z \biggr]\! \biggr\}\nonumber\\
& - \frac{3C_F}{4}\! \int_{\max[\hat{m}, z^3(\hat m,\tau_c)]}^{\frac{1}{2}} {\rm d} z\, \tilde{A}_g^C
[\hat m, \min [y_{\max} (\hat{m},z), 1 - \tau_c], y_{\tau} (\hat{m},z),z]\,,
\end{align}
where the analytic expressions for ${\tilde A}^C_{{\bar q}, g}$ have been already given in Eqs.~\eqref{eq:RangAx}
and \eqref{eq:RhadV} for the axial-vector and vector currents, respectively. The $z$ integrals are in practice
computed numerically with high accuracy even for values very close to $\tau_J^{\rm min}(\hat m)$.
In Fig.~\ref{fig:tau-dif-cum} we show the NLO pieces for the \mbox{$2$-jettiness} differential and cumulative cross sections for
two values of $\hat m$. For $\hat m = 0.32$ one can observe small kinks in $\tau = 4 \hat m^2$ and
$\tau = \tau_J^{\rm mid}$. Finally, we observe a negative cumulative cross section for the vector current and
$\hat m=0.2$, indicating the necessity of Sudakov log resummation.

\subsection{Heavy Jet Mass}
\begin{figure*}[t!]
\subfigure[]
{\includegraphics[width=0.46\textwidth]{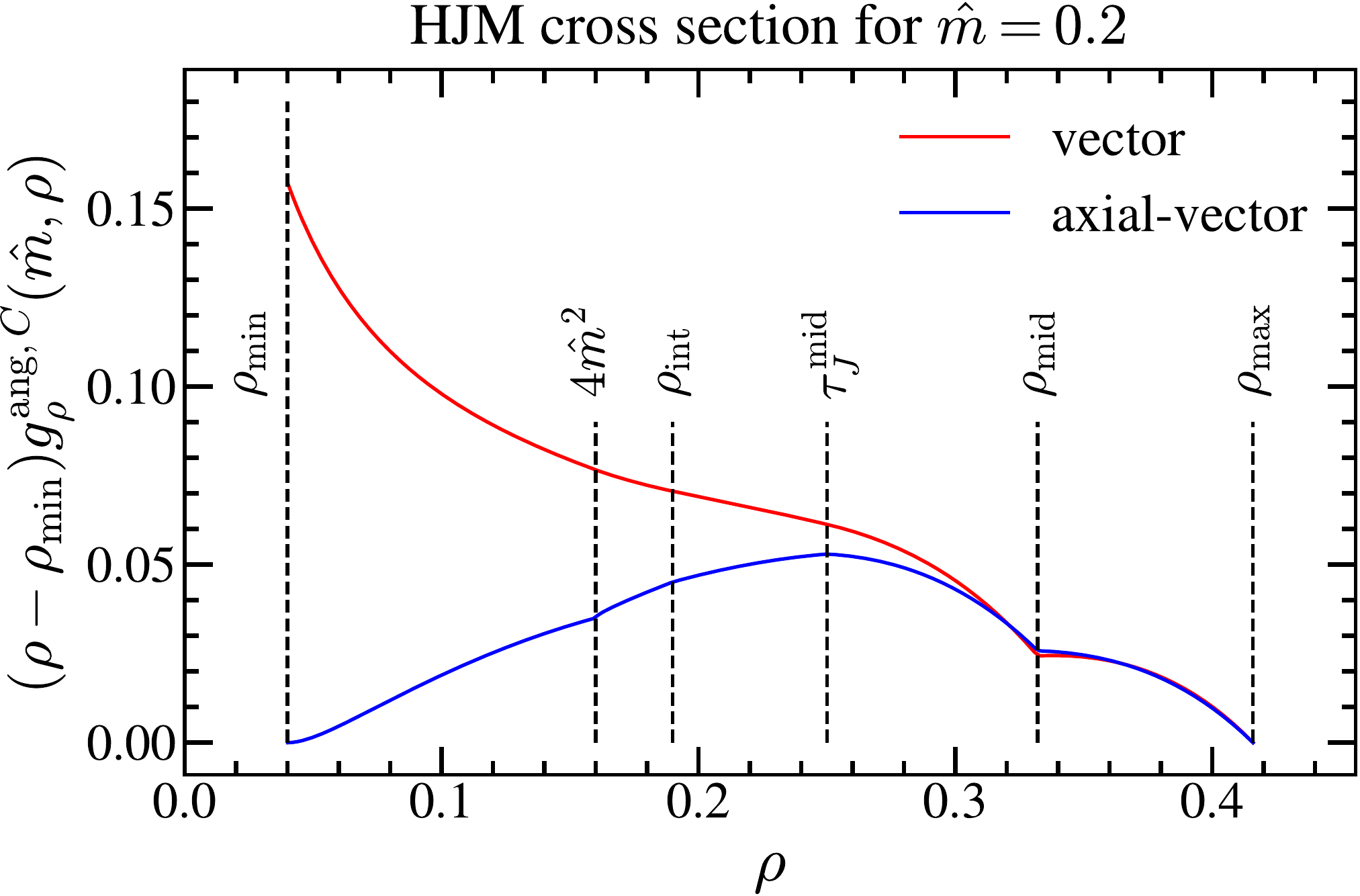}
\label{fig:rho-cont}}
\subfigure[]{\includegraphics[width=0.46\textwidth]{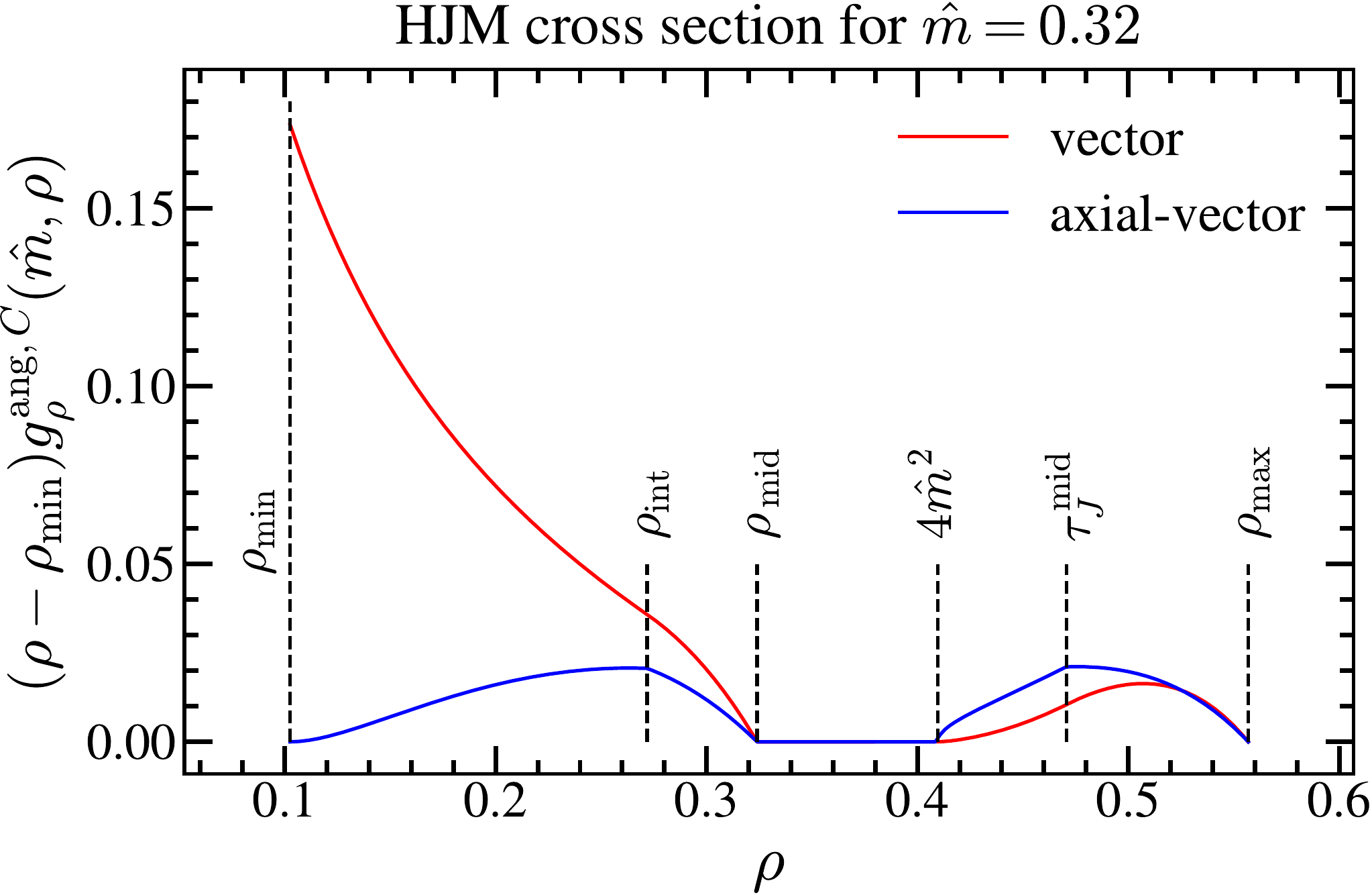}
\label{fig:rho-inc}}
\subfigure[]
{\includegraphics[width=0.465\textwidth]{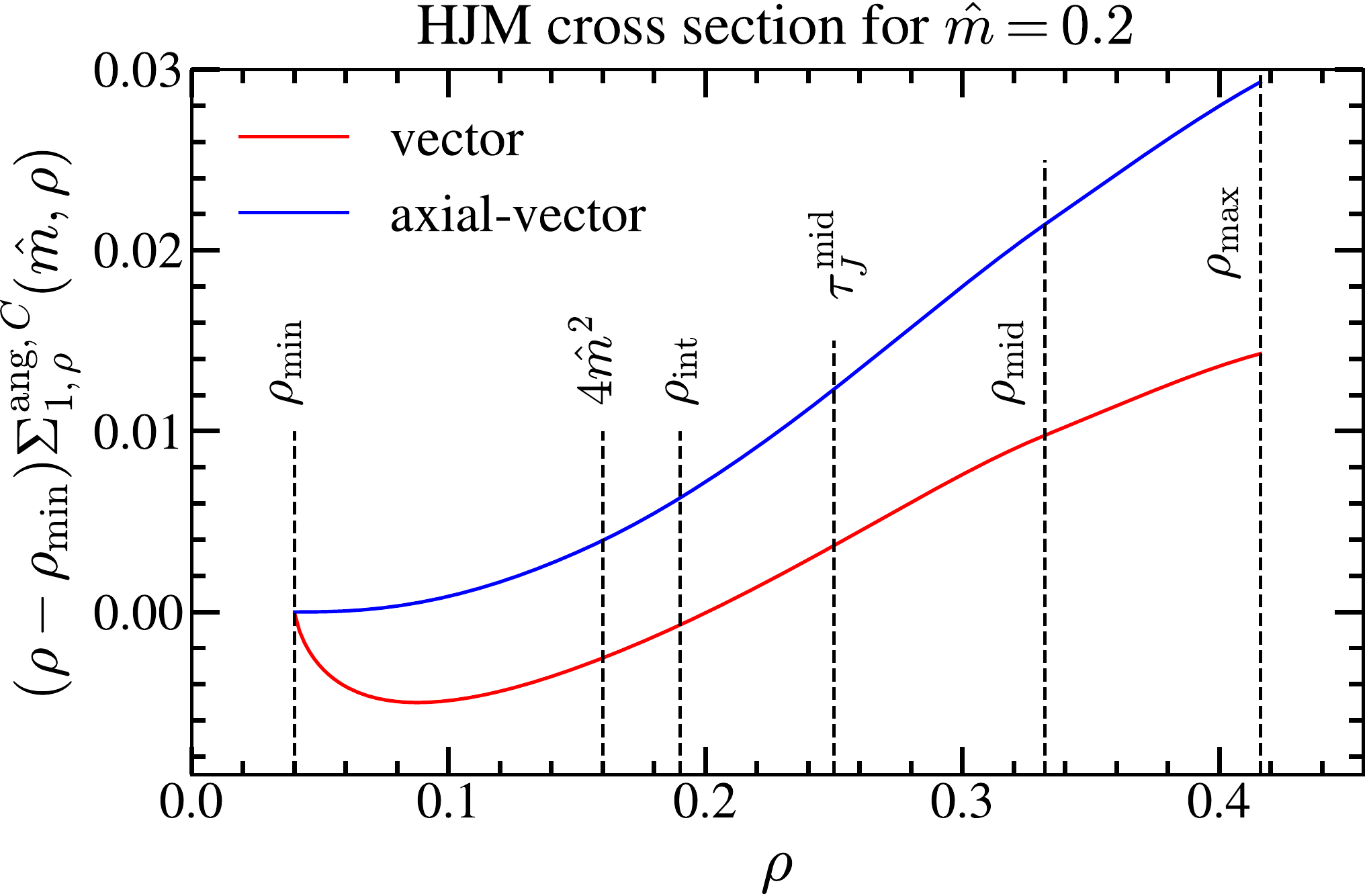}
\label{fig:cumul-rho-cont}}~~~~
\subfigure[]{\includegraphics[width=0.465\textwidth]{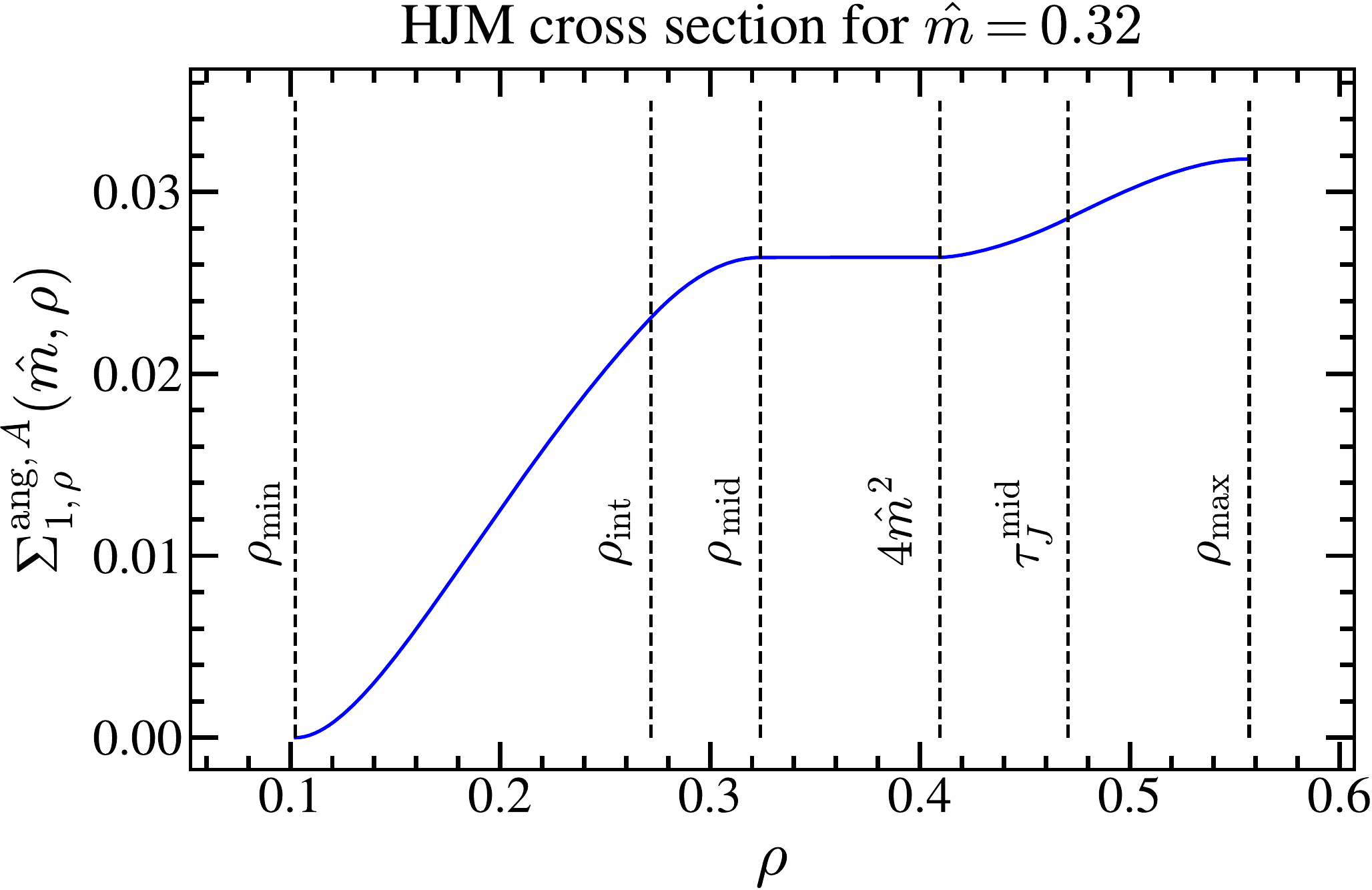}
\label{fig:cumul-rho-inc}}
\caption{Same as Fig.~\ref{fig:tau-dif-cum} for heavy jet mass. Panel (d) shows only the results for the
axial-vector current to highlight the region of constant cumulative cross section.}
\label{fig:rho-dif-cumul}
\end{figure*}

The differential and cumulative cross sections for heavy jet mass can be expressed in terms of functions already computed. In fact, for the region
in which the thrust axis is collinear to the gluon momentum heavy jet mass and $2$-jettiness are identical. For the region of $\hat n$ pointing into
the same direction as $\vec{p}_{\bar q}$ the measurement delta function reads:
\begin{equation}
\delta_{\bar q}^\rho = \frac{1}{z} \delta \biggl( y - \frac{\rho - \hat{m}^2}{z} \biggr).
\end{equation}
It is useful to define $r(\hat m,\rho) = 1 + \rho - \hat{m}^2$ and
$\chi(\hat m, \rho) = \sqrt{r(\hat m, \rho)^2 - 4\rho}$. Since the patch of the contour line in the gluon region has been
discussed at length in the previous subsection, we now focus on the $\bar q$ region exclusively.
We anticipate that there is no value of $\rho$ for which the full contour line becomes continuous.
For $\hat m^2 < \rho < \hat{m} (1 - \hat{m} - \hat{m}^2) / (1 - \hat{m})\equiv \rho_{\rm int}(\hat m)$ it
hits the phase space boundary at $z_4(\hat m, \rho)=[r(\hat m, \rho) - \chi(\hat m, \rho)] / 2$ in the $\bar q$ region.
From the limiting condition $z_4(\hat m,\rho_{\rm min})=z_-(\hat m)$ we obtain the minimal value of heavy jet mass
$\rho_{\rm min}(\hat m)=\hat m^2$.
The value $\rho_{\rm int}(\hat m)$ is obtained from the condition $z_4(\hat m,\rho_{\rm int}) = \hat m$.
For \mbox{$ \rho_{\rm int}(\hat m)< \rho < \hat{m}^2 + 2 \sqrt{1 - 3 \hat{m}^2}/3 - 1/3\equiv \rho_{\rm mid}(\hat m)$}
the contour line hits
the boundary of the $g$ and $\bar q$ regions at
$z_5(\hat m,\rho) = [r(\hat m,\rho) - 1] / \sqrt{(1 - \rho)^2 - 2 \hat{m}^2 (1 + \rho) + \hat{m}^4}\neq z_3(\hat m, \rho)$.
The expression for $\rho_{\rm mid}(\hat m)$ is obtained from
the condition $z_5(\hat m,\rho_{\rm mid})=1/2$. Finally, for $\rho_{\rm mid}(\hat m) <\rho<\tau_J^{\rm max}(\hat m)$
the contour line exists only in the gluon region. Therefore the maximum value of heavy jet mass is
$\rho_{\rm max}(\hat m)=\tau_J^{\rm max}(\hat m)$. One can easily see that for
$0 \leqslant \hat{m} \leqslant 1/2$ one has
\mbox{$\rho_{\rm min}(\hat{m}) \leqslant \rho_{\rm int} (\hat{m}) \leqslant \{\tau_J^{\rm mid}(\hat m), \rho_{\rm mid} (\hat m)\} \leqslant \rho_{\rm max} (\hat m)$}
as can be checked graphically in Fig.~\ref{fig:rhoVal}.
For $\hat m \lessgtr 0.248226$ one has \mbox{$\rho_{\rm mid}(\hat m) \gtrless \tau_J^{\rm mid}(\hat m)$}, while $\hat m \lessgtr (5-\sqrt{13})/6\approx 0.232408$
implies $4\hat m^2 \lessgtr \rho_{\rm int}(\hat m)$, although these have no implications. On the other hand,
if $\hat m \lessgtr \sqrt{(2 \sqrt{13}-5)/27}\approx 0.286169$ one has $4\hat m^2 \lessgtr \rho_{\rm mid}(\hat m)$, and this entails the cross section
is zero for $\rho_{\rm mid}(\hat m) < \rho < 4\hat m^2$ as can be seen in Figs.~\ref{fig:rho-cont} and \ref{fig:rho-inc}.
\begin{figure*}[t!]
\subfigure[]
{\includegraphics[width=0.46\textwidth]{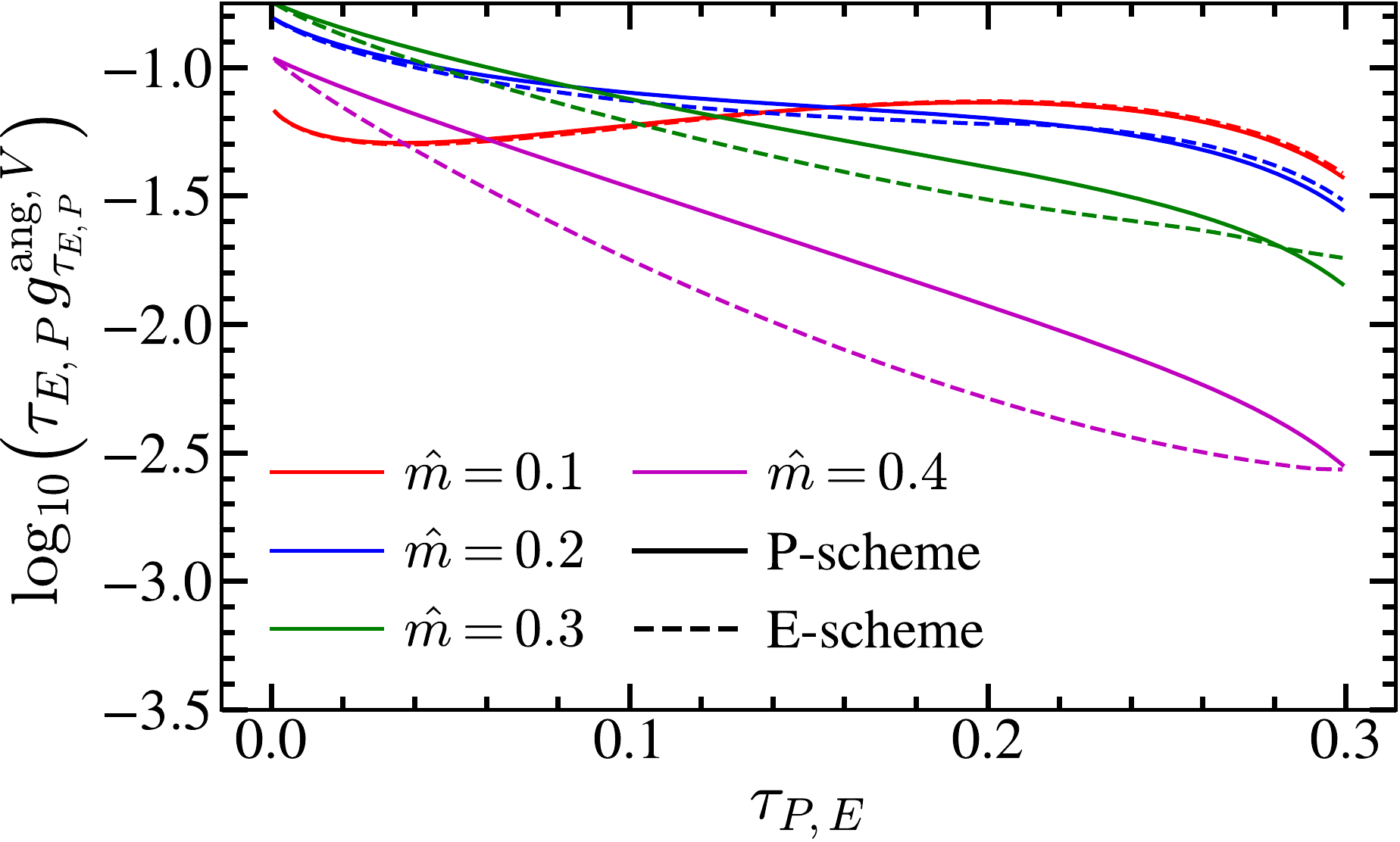}
\label{fig:tau-EP}}
\subfigure[]{\includegraphics[width=0.46\textwidth]{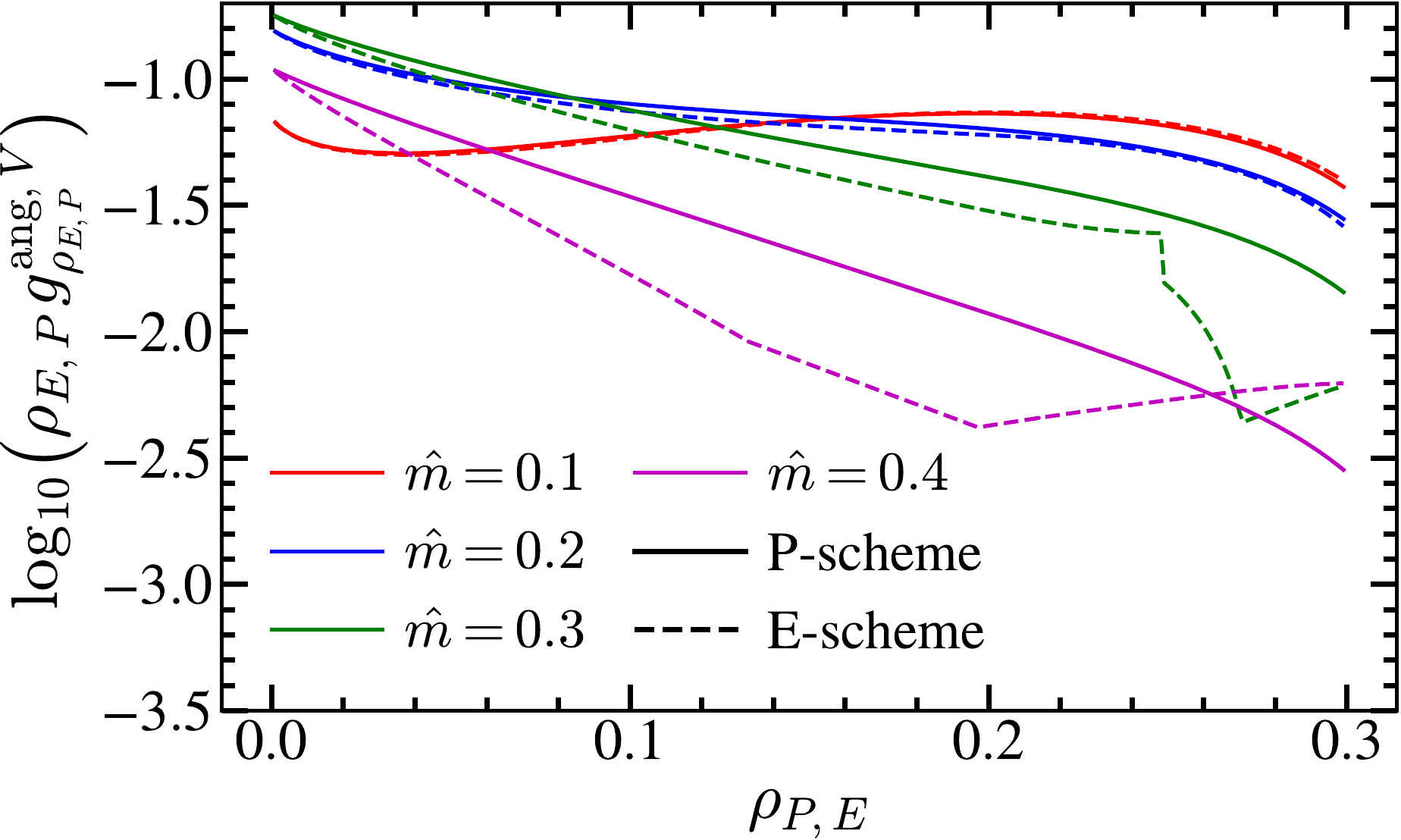}
\label{fig:rho-EP}}
\subfigure[]
{\includegraphics[width=0.465\textwidth]{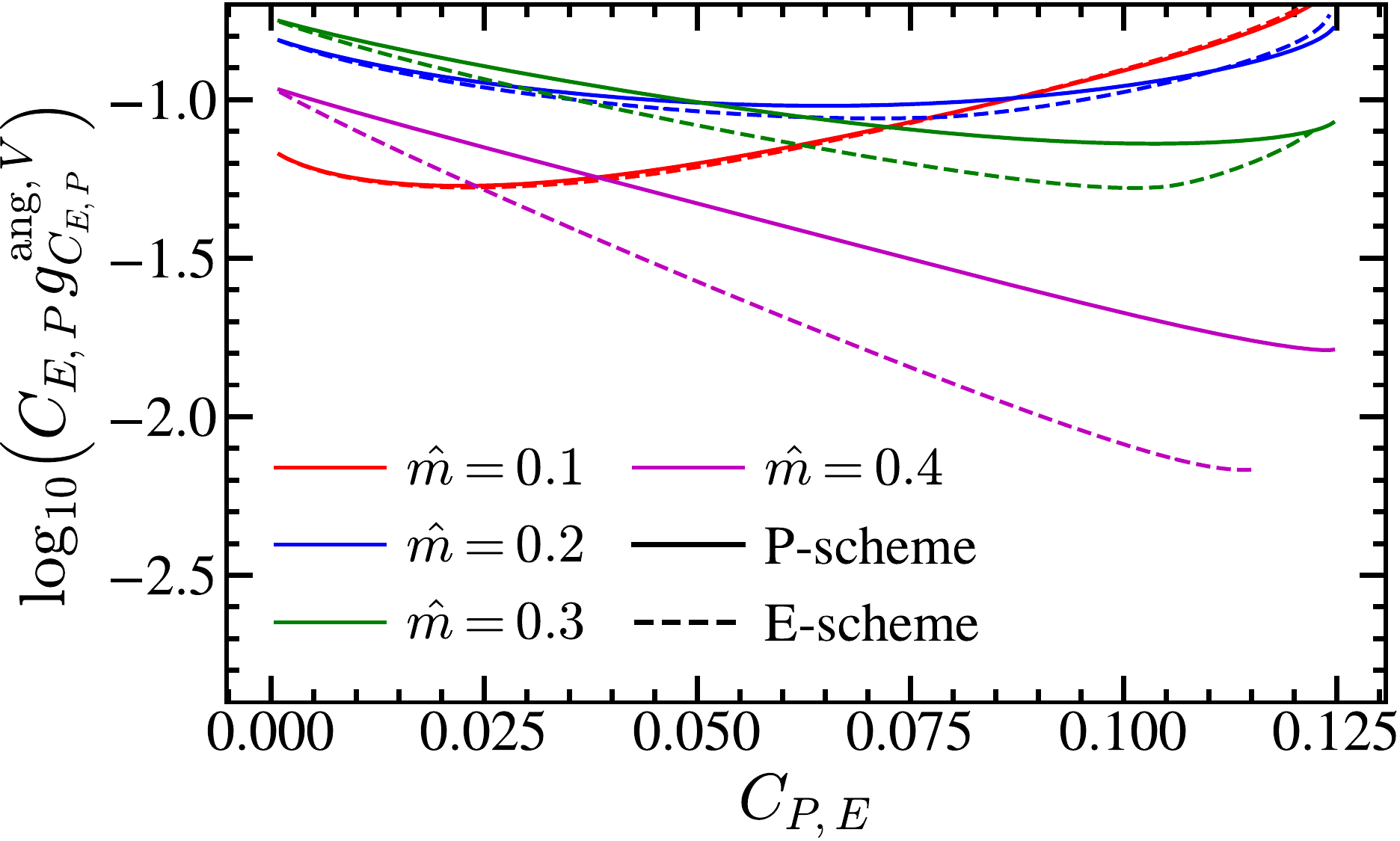}
\label{fig:C-EP}}~~~~
\subfigure[]{\includegraphics[width=0.465\textwidth]{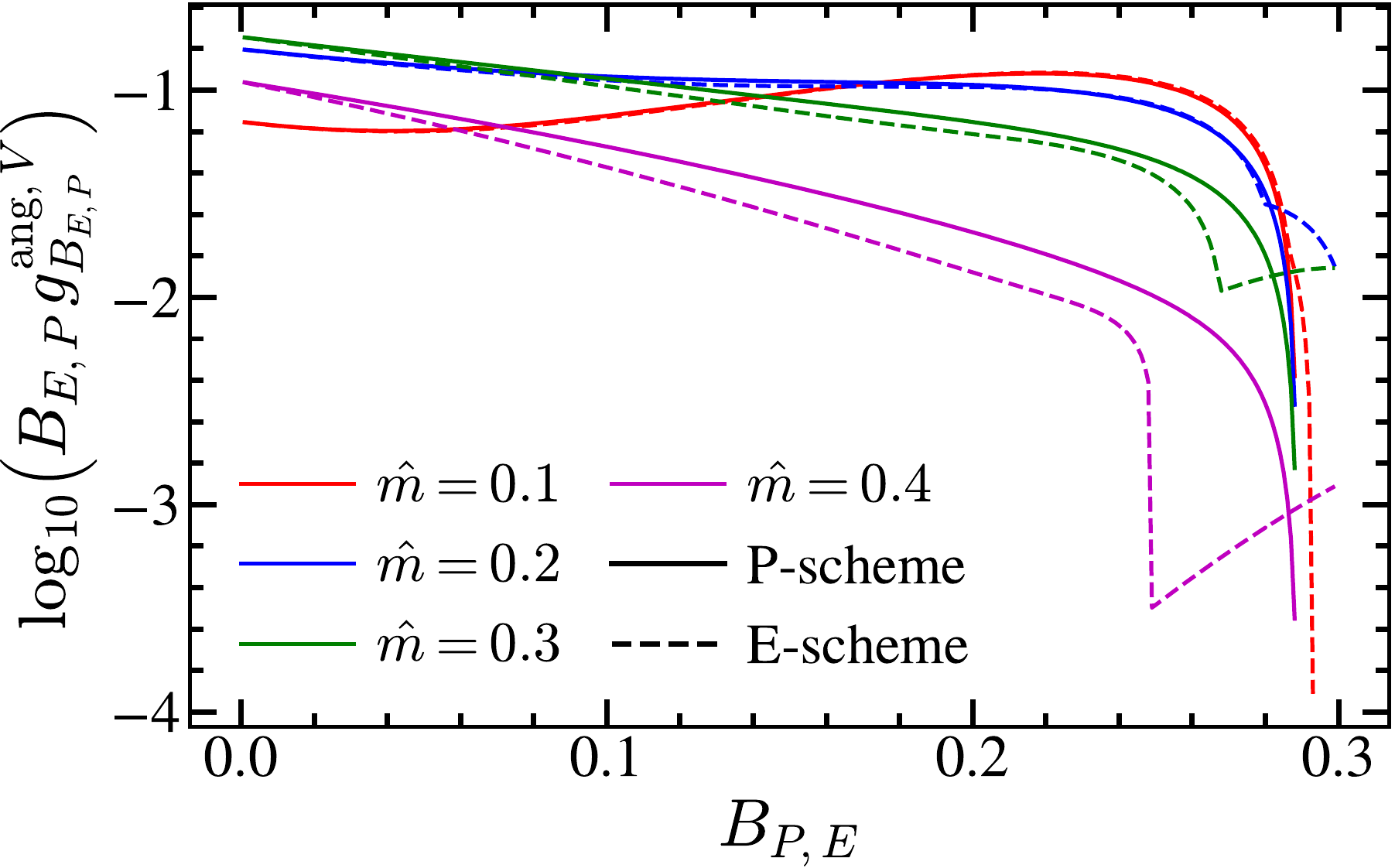}
\label{fig:B-EP}}
\caption{Differential cross sections for the vector current in the P- (solid lines) and E-scheme (dashed lines) for $\hat m = 0.1, 0.2, 0.3$ and $0.4$
in red, blue, green and magenta, respectively. Panels (a), (b), (c) and (d) correspond to thrust, heavy-jet-mass,
C-parameter and jet broadening, respectively.}
\label{fig:EP}
\end{figure*}
A simple computation yields the following result for the differential cross section
\begin{align}
\frac{1}{\sigma_0} \frac{{\rm d} \sigma^C_{\rm ang}}{{\rm d} \rho} =\,
& \frac{3 \alpha_s C_F}{4 \pi} g^{{\rm ang}, C}_{\tau_J} (\hat{m},\rho)\,, \\
g^{{\rm ang}, C}_{\rho} \!(\hat{m},\rho) = \,& \theta[\rho_{\rm mid}(\hat m) - \rho]\!
\int^{\frac{1}{2}}_{z_{45}(\hat m,\rho)} \!\!\frac{{\rm d} z}{z^2}
A_{\bar q}^C \!\biggl[\hat{m}, \frac{\rho - \hat m^2}{z}, z \biggr]\!
+ t(\rho) \theta (\rho \!- 4 \hat{m}^2)\!\!
\int^{\frac{1}{2}}_{z_{23}(\hat m,\rho)}\!\!\! {\rm d} z A_g^{C}[\hat{m}, t(\rho), z]
\nonumber\\
= \,& \theta[\rho_{\rm mid}(\hat m) - \rho] g^C_{\bar q} [\hat m, z_{45}(\hat m,\rho), \rho - \hat m^2]
+ \theta(\rho - 4 \hat{m}^2) g^C_g[\hat m, z_{23}(\hat m,\rho),\rho] \,, \nonumber
\end{align}
where the analytic form of $g^C_{g,\bar q}$ for both currents has been given already in Eq.~\eqref{eq:gsAna}.
Implementing the SCET counting $\rho \propto \mathcal{O}(\lambda^2)$ and $m \propto \mathcal{O} (\lambda)$ one finds:
\begin{align}
f^V_{\rho} (\hat{m}, \rho) =\, & \frac{\hat{m}^4 - 6 \hat{m}^2 \rho - 4
\hat{m}^2 \rho \log (\rho) + \rho^2}{2\rho (\rho - \hat{m}^2)} +\mathcal{O} (\lambda)\,, \\
f^A_{\rho} (\hat{m}, \rho) =\, & \frac{\rho^2 - \hat{m}^4}{2\rho^2} +\mathcal{O} (\lambda)\,, \nonumber
\end{align}
where again we find different limits for vector and axial-vector currents. This implies that the subleading
jet function appearing in the factorization theorem derived in Ref.~\cite{Hagiwara:2010cd}, for
massive quarks, depends on the current. Despite appearances, the ``jet function'' is the same for $2$-jettiness
and heavy-jet-mass once the endpoints $e_{\rm min}$, expanded in the SCET limit, have been shifted away:
\begin{equation}
f^C_{\rho} (\hat{m}, e+\hat m^2)+\mathcal{O} (\lambda) = f^V_{\tau_J} (\hat{m}, e + 2\hat m^2)+\mathcal{O} (\lambda)\,.
\end{equation}
Finally, for $\hat m=0$ the limits of both currents coincide.
For the cumulative distribution we have
\begin{align}
\Sigma^{{\rm ang},C}_{\rho,1}(\hat m, \rho_c) =\, & R_1^{{\rm ang},C}(\hat m) - \frac{3C_F}{4}
\biggl\{ \int_{\max
[\hat{m}, z_3(\hat m, \rho_c)]}^{\frac{1}{2}}\!\! {\rm d} z \tilde{A}_g^C [\hat{m}, \min [y_{\max} (\hat{m},z), 1 - \rho_c], y_{\tau} (\hat{m},z)]\nonumber \\
& - \theta [\rho_{\rm mid}(\hat m) - \rho_c] \!
\int^{\frac{1}{2}}_{z_{45}(\hat m, \rho_c)}\!\!{\rm d} z \biggl[ \! \tilde{A}_{\bar q}^C [\hat{m}, y_{\rm top} (\hat{m},z), z] - \tilde{A}_{\bar q}^C
\biggl(\hat{m}, \frac{\rho_c - \hat{m}^2}{z}, z \biggr) \!\biggr]\!\biggr\},
\end{align}
where again all pieces are known and we compute the $z$ integration numerically. In Figs.~\ref{fig:cumul-rho-cont} and
\ref{fig:cumul-rho-inc} we show $\Sigma^{{\rm ang},C}_{\rho,1}$ for two values of $\hat m$. In general we find that
cusps for cumulative cross sections are less pronounced than for their differential counterparts. In particular,
for $\hat m=0.32$ one can
see that for $\rho_{\rm mid} < \rho < 4\hat m^2$ the cumulative cross section is constant, as a result of the
differential cross section being zero in that patch.

\section{Monte Carlo Strategy}\label{sec:MC}
\begin{figure}[t]\centering
\includegraphics[width=0.3\textwidth]{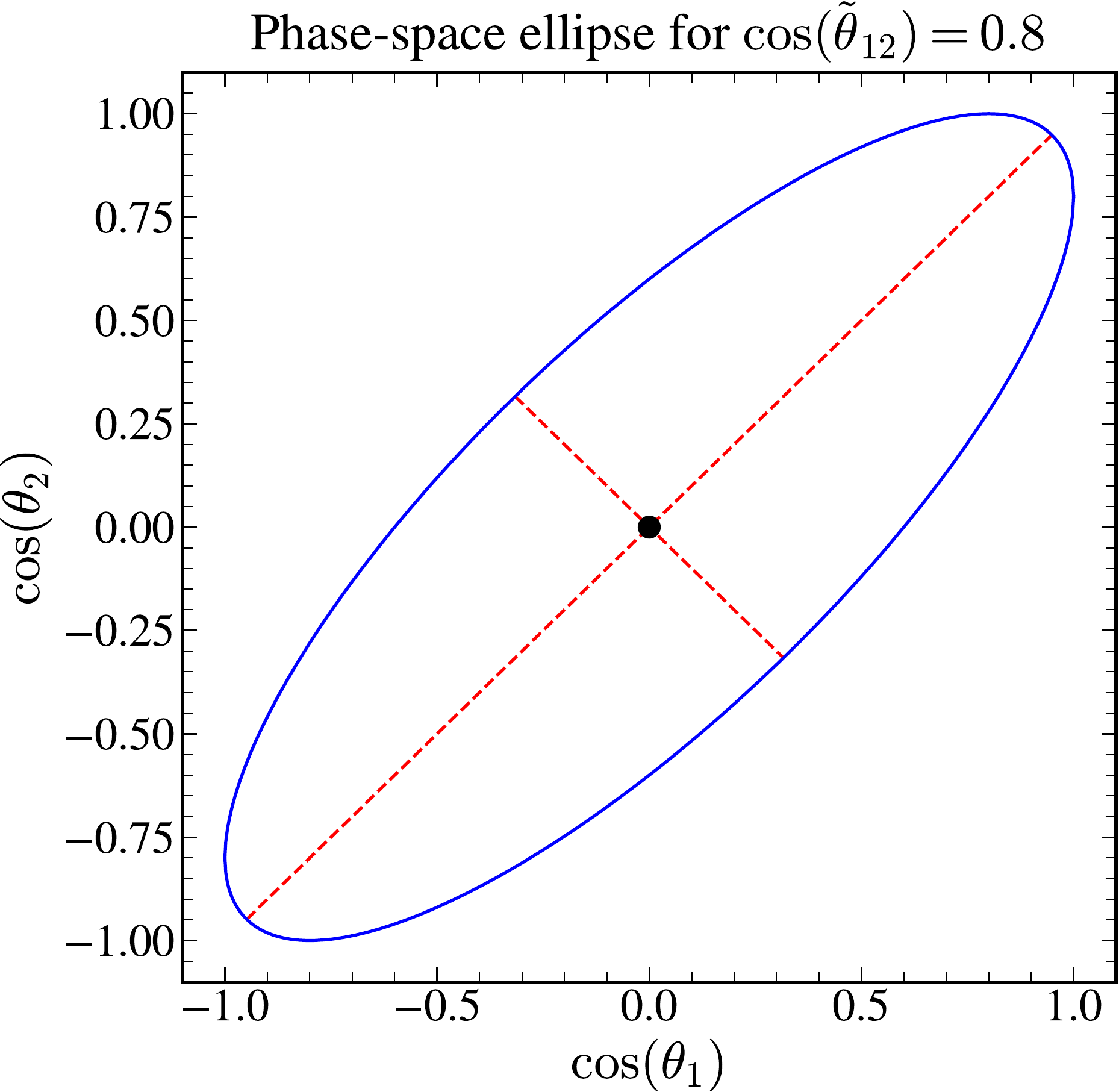}
\caption{Dalitz region in the polar variables $\cos(\theta_1)$ and $\cos(\theta_2)$ for fixed values of parton energies. The blue solid line corresponds to
the phase-space boundary, the black dot is the ellipse's center, and the dashed red lines correspond to the major and minor
axes. The figure has been generated with the value $\cos(\tilde\theta_{12})=0.8$.\label{fig:ellipse}}
\end{figure}
In Sec.~\ref{sec:real} it has been discussed how to project out the angular differential distribution by analytically integrating the two
polar angles $\theta_{1,2}$ with appropriately chosen weights. Using that result ---\,specialized to four dimensions\,--- in Sec.~\ref{sec:analytic}
the differential distribution's radiative tail was analytically computed for $2$-jettiness and heavy jet mass. We have also discussed how
the results of Ref.~\cite{Lepenik:2019jjk} can be adapted to numerically compute unbinned differential distributions for any
event shape. Using existing automated tools such as \texttt{MadGraph5}~\cite{Alwall:2014hca} one should, in principle, be able to obtain numerically
the tail of the angular distribution, since it can be computed in $d=4$.
This possibility shall be explored in this section, but as an additional check we have designed and
coded our own Monte Carlo integrator, which in turn has been used to show that such approach is very inefficient. To that end, we devise a
strategy to numerically carry out the angular projection and, at the same time,
obtain binned distributions for any event shape. The idea can be trivially generalized to any other angular measurements one can come
up with.

Since the mapping of the $(y,z)$ phase space to the unit square was already discussed in Ref.~\cite{Lepenik:2019jjk}, the reader is referred to
that article for details and we focus instead in the integration of both polar angles. In the following we consider $(y,z)$ fixed (that is, we integrate
$\theta_{1,2}$ before $y,z$) and describe the mathematical steps followed to map the angular phase space to the unit square. We start by noting that
the condition $h_{12}>0$ defines an ellipse centered at the origin with semi-major and semi-minor axes given by $\max\{r_s,r_c\}$ and
$\min\{r_s,r_c\}$, respectively, with $r_s = \sqrt{2} \sin (\tilde{\theta}_{12}/2)$ and $r_c = \sqrt{2} \cos (\tilde{\theta}_{12}/2)$, as shown in
Fig.~\ref{fig:ellipse}. It is easy to see
that $r_c \gtrless r_s$ if $\cos (\tilde{\theta}_{12})\gtrless0$. The area of such ellipse is $S=\pi \sin (\tilde{\theta}_{12})$, and it can be mapped
to the unit circle with the following change of variables:
\begin{equation}
\cos (\theta_{1, 2}) = \cos \biggl(\frac{\tilde{\theta}_{12}}{2} \biggr)c_+ \pm \sin \biggl(\frac{\tilde{\theta}_{12}}{2} \biggr) c_-\,,
\end{equation}
which furthermore implies $h_{12} = (1 - c_+^2 - c_-^2) \sin^2 (\tilde{\theta}_{12})$ and
${\rm d}\! \cos (\theta_1) {\rm d}\! \cos (\theta_2) = \sin (\tilde{\theta}_{12}) {\rm d} c_+ {\rm d} c_-$. Switching to polar coordinates
$c_+ = r \cos(\alpha)$, $c_- = r \sin(\alpha)$ we have
\begin{equation}
\int \frac{{\rm d}\! \cos (\theta_1) {\rm d} \!\cos (\theta_2)\theta(h_{12})}{h_{12}^{1 / 2 + \varepsilon}}
= \sin^{- 2 \varepsilon} (\tilde{\theta}_{12}) \int_0^{2 \pi} {\rm d} \alpha
\int_0^1 \frac{{\rm d} r r}{(1 - r^2)^{1 / 2 + \varepsilon}} .
\end{equation}
\begin{figure*}[t!]
\subfigure[]
{\includegraphics[width=0.46\textwidth]{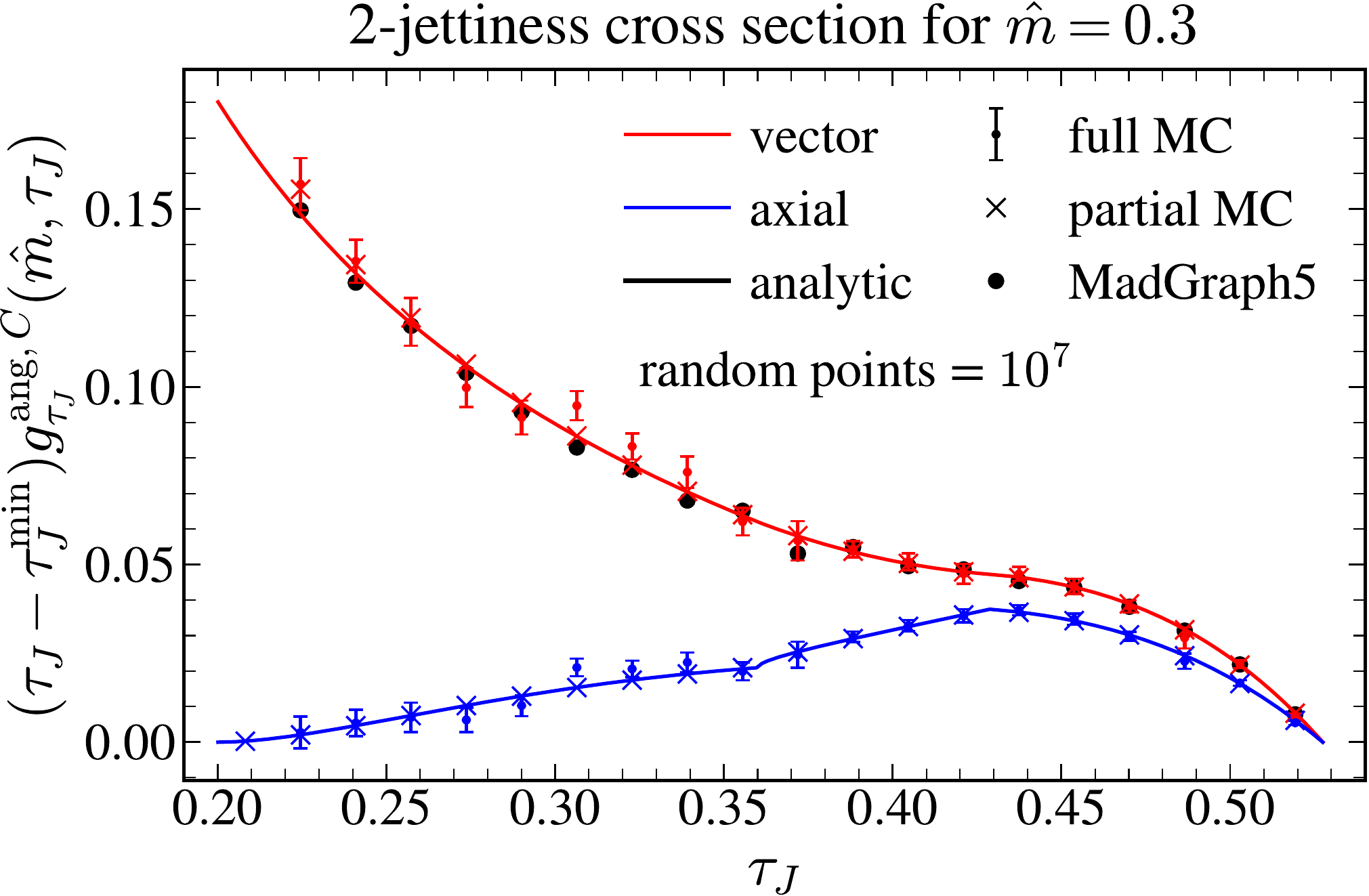}
\label{fig:tauMC}}~~~~
\subfigure[]{\includegraphics[width=0.46\textwidth]{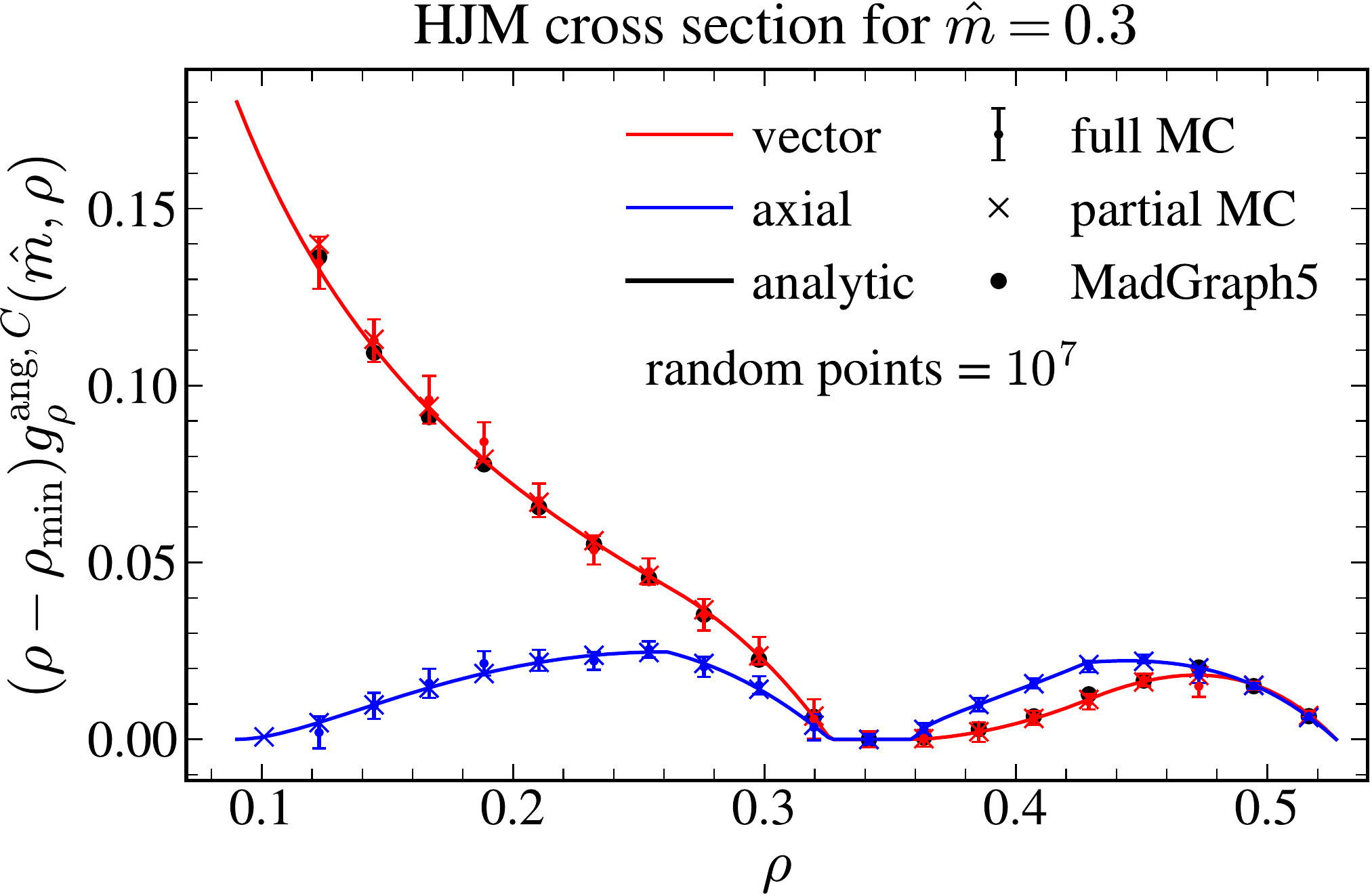}
\label{fig:rhoMC}}
\caption{Differential angular distributions for $2$-jettiness (left panel) and heavy jet mass (right panel), for $\hat m = 0.3$. We show analytic
results with solid lines while the outcome of the Monte Carlo binned results appear as dots with error bars for the 4D phase-space integration
(dubbed ``full MC'') and crosses if the angular projections has been carried out analytically (labeled as ``partial MC''). \texttt{MadGraph5} results
appear as black dots. We employ 10 million points in both Monte Carlo's and \texttt{MadGraph5}, distributed in histograms with 20 equally-spaced bins.}
\label{fig:MC}
\end{figure*}
The trivial rescaling $\alpha = 2\pi \eta$ maps the last integral to the unit square, amenable for a Monte Carlo treatment. For completeness,
we provide a master integral that can be used to reproduce the relations given in Eqs.~\eqref{eq:cosInt} using the change of variables just
presented:
\begin{equation}
\int_0^1 \frac{{\rm d} r r^{2 n + 1}}{(1 - r^2)^{1 / 2 + \varepsilon}} = \frac{n!}{(1 - 2\varepsilon) \bigl( \frac{3}{2} - \varepsilon \bigr)_{\!n}} .
\end{equation}

These ideas and some relevant results given in this article have been coded in \texttt{Python} to carry out the full $4$-dimensional phase-space
integration numerically. To make our point clear it is enough to consider binned distributions for $2$-jettiness and heavy jet mass. Furthermore,
we do not implement importance sampling, but this could be easily done for instance using VEGAS~\cite{Lepage:2020tgj} or with the Metropolis
algorithm. The way in which the program works can be summarized as follows: at each step of the integration, four random numbers
between $0$ and $1$ are generated, out of which the values of $y,z$ and $\cos(\theta_{1,2})$ are determined. From these, one computes the
numerical values for $\cos(\theta_T)$, the squared matrix element $|M|^2$, the Jacobian $J=\beta^2\sqrt{1 - 4\hat m^2/(1 - y)}$,
and the two event shapes under consideration: $\tau_J$ and $\rho$. At this point, two histograms are filled according to the event-shape
values, with weights given by $w = r\, [2 - 5\cos^2(\theta_T)]\,J\,|M|^2/\sqrt{1-r^2}$, such that the angular part is projected out. After
normalizing to the number of random points and bin-sizes, and accounting for the relevant normalization factors, the distributions are obtained.
We show the results in Fig.~\ref{fig:MC} (dots with error bars), along with the result of our previous analytic computations (solid lines).
We take $20$ evenly spaced bins between the minimal and maximal values of the
event shapes, and employ $10$ million random points. As can be seen in the plots, the error bars are much larger that one should expect for such
sampling, and the quality degrades towards threshold. The reason for this misbehavior is that there are positive and negative weights. Furthermore,
one is numerically projecting the angular cross section, which is much smaller than the total one, such that numerical inaccuracies are highly magnified.
To better appreciate this deficiency we have coded another Monte Carlo in which the analytically-projected squared matrix element is numerically
integrated in $y$ and $z$, producing binned distributions. The results, which use the same number of random points and bins, are shown in
Fig.~\ref{fig:MC} with crosses (error bars are negligibly small and therefore are not shown), and one can observe the prediction is equally robust in the whole
spectrum.\footnote{The apparent discrepancy close to threshold is caused by the fact that the cross section is rapidly varying such that the
binned distribution differs a bit from the differential one. Smaller bins can be used to avoid this issue.}
We find that statistical uncertainties for the vector (axial-vector) current are in average $60$ ($400$) times larger for the 4D Monte Carlo.
Therefore we conclude that a numerical, MC-based, projection of the angular cross section should be avoided if an analytic computation
is available.

We end this section discussing the results obtained with \texttt{MadGraph5}. We use the latest long-term stable version \texttt{MG5aMC\_LTS\_2.9.13}.
For this exploratory study we have focused on the vector current only, as it can be easily isolated in \texttt{MadGraph5} fixing the s-channel particle
---\,that is, a photon. We use a center-of-mass energy of $1\,$TeV, adjust the quark mass such that the reduced mass is $\hat m=0.3$, and
produce binned cross sections for 2-jettiness and heavy jet mass. We use fixed renormalization and factorization scales, considering the
bottom quark as massive instead of the top to make sure no decay products are being produced. The way in which \texttt{MadGraph5} numerically
integrates the phase space is intrinsically different to our in-house Monte Carlo, since its aim is producing events (mimicking a real experiment)
with the appropriate likelihood (hence no event-weighting is necessary). For each set of events \texttt{MadGraph5} generates a global weight
corresponding to the total cross section of the process. To obtain the angular binned distribution, for each event we compute the values of
$\tau_J$, $\rho$ and $\cos(\theta_T)$ and fill the bins of our histograms with the weight $2 - 5\cos^2(\theta_T)$. To prevent biases in the
generation of events, the standard settings have been modified: no lower or upper cutoff is applied to $p_\perp$, rapidity or energy,
neither for jets nor for individual particles, except for the following exception. We have observed that if the parameter \texttt{ptj} (which corresponds
to the minimum transverse momentum of the jets) is set to $0$
the run always crashes. This is possibly related to a numerical regularization of IR divergences. If the default value ($20\,$GeV) is used, the
program runs steadily but the differential cross section notoriously undershoots the theoretical result in the dijet region. For \texttt{ptj} values
of order a few GeV there are eventual crashes that prevent collecting enough statistics unless one splits the task in several runs ---\,which are
recombined at the end\,--- and implements an error handling strategy. We have also observed that a)~the execution time increases as \texttt{ptj}
decreases, and b)~the absolute normalization strongly depends on the value of \texttt{ptj}\footnote{This is not surprising as only the real radiation
contribution is accounted for. One should expect that, once the virtual radiation diagram is included, the cutoff dependence is softened and the
total cross section is correctly reproduced.}. Therefore, only the differential cross section's shape
can be trusted. For our final comparison, a value of $2\,$GeV for \texttt{ptj} is used, and the cross section is normalized
by hand to reproduce the analytic results in the far tail, where a finite cutoff should produce no artifacts. We again generate $10^7$ events, which
are distributed in $100$ individual runs and use the exact same bins as for our previous study. We find statistical uncertainties $25\%$ larger that
those of our 4D Monte Carlo. As can be seen in Fig.~\ref{fig:MC} (black circles with no error bars) the agreement with our analytic
results if fairly good everywhere in the spectrum, but from our numerical investigations we conclude that if the extreme dijet region is to be explored
(e.g.\ to numerically determine $B^{{\rm ang},V}_{\rm plus}$) the value of \texttt{ptj} should be further lowered causing additional trouble.
At the sight of these facts, we conclude that using \texttt{MadGraph5} to obtain the angular differential distribution ---\,one of the main results of this
manuscript\,--- is far from optimal.

\section{Numerical Analysis}\label{sec:numerical}
\begin{figure*}[t!]
\subfigure[]
{\includegraphics[width=0.46\textwidth]{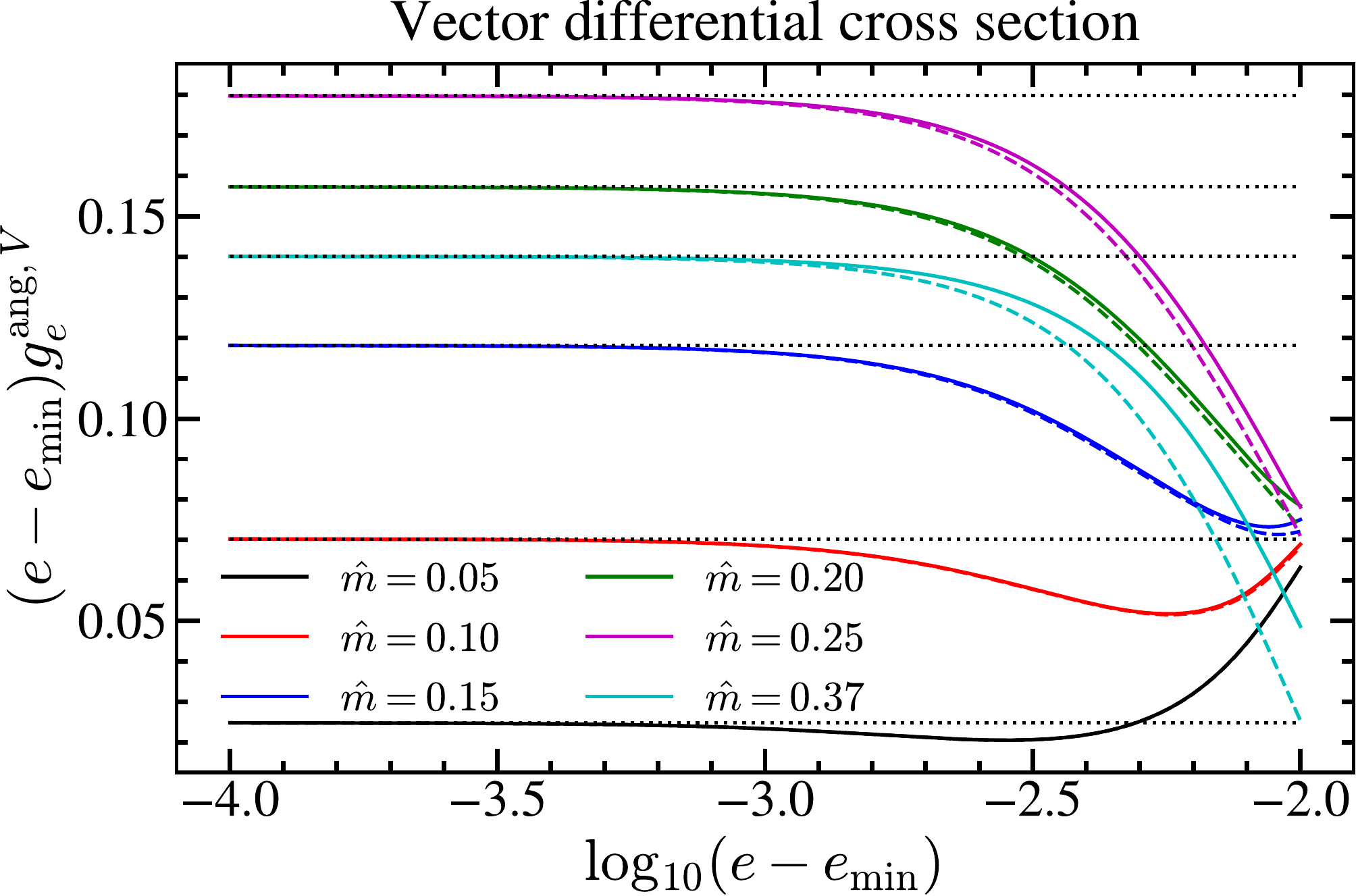}
\label{fig:VectorLim}}~
\subfigure[]{\includegraphics[width=0.46\textwidth]{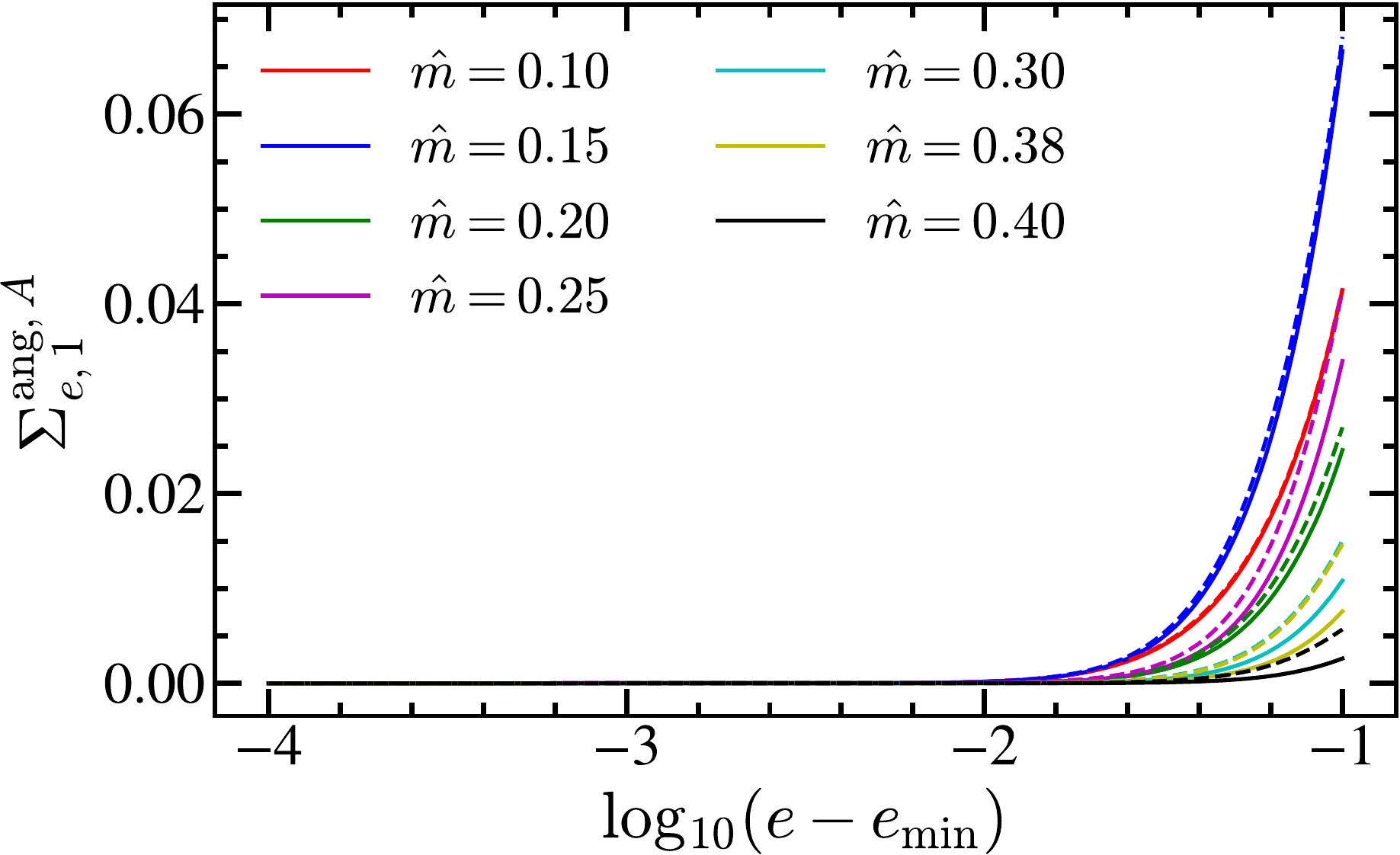}
\label{fig:AxialCum}}
\subfigure[]
{\includegraphics[width=0.46\textwidth]{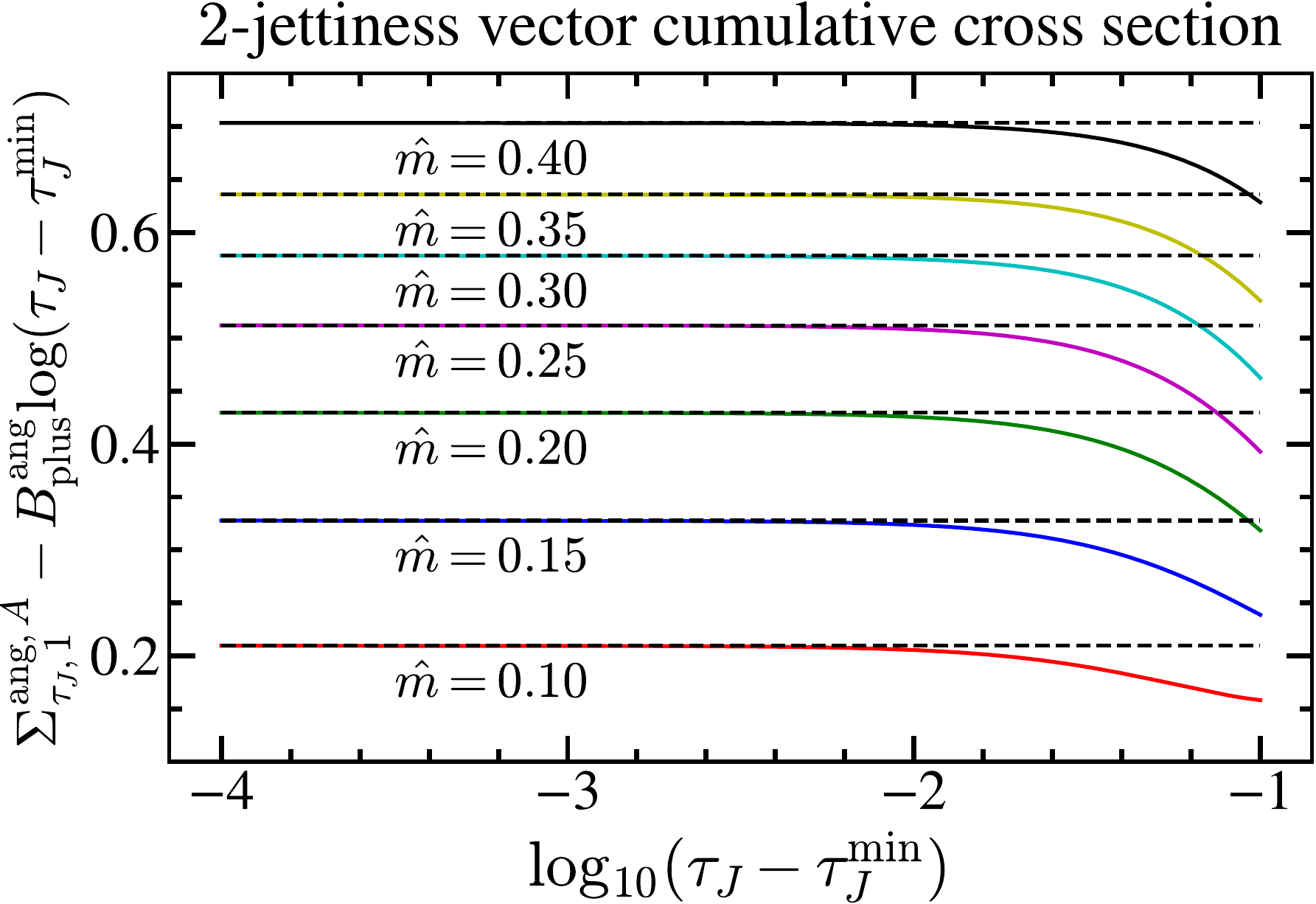}
\label{fig:VectorTauCum}}~~~~~~
\subfigure[]{\includegraphics[width=0.46\textwidth]{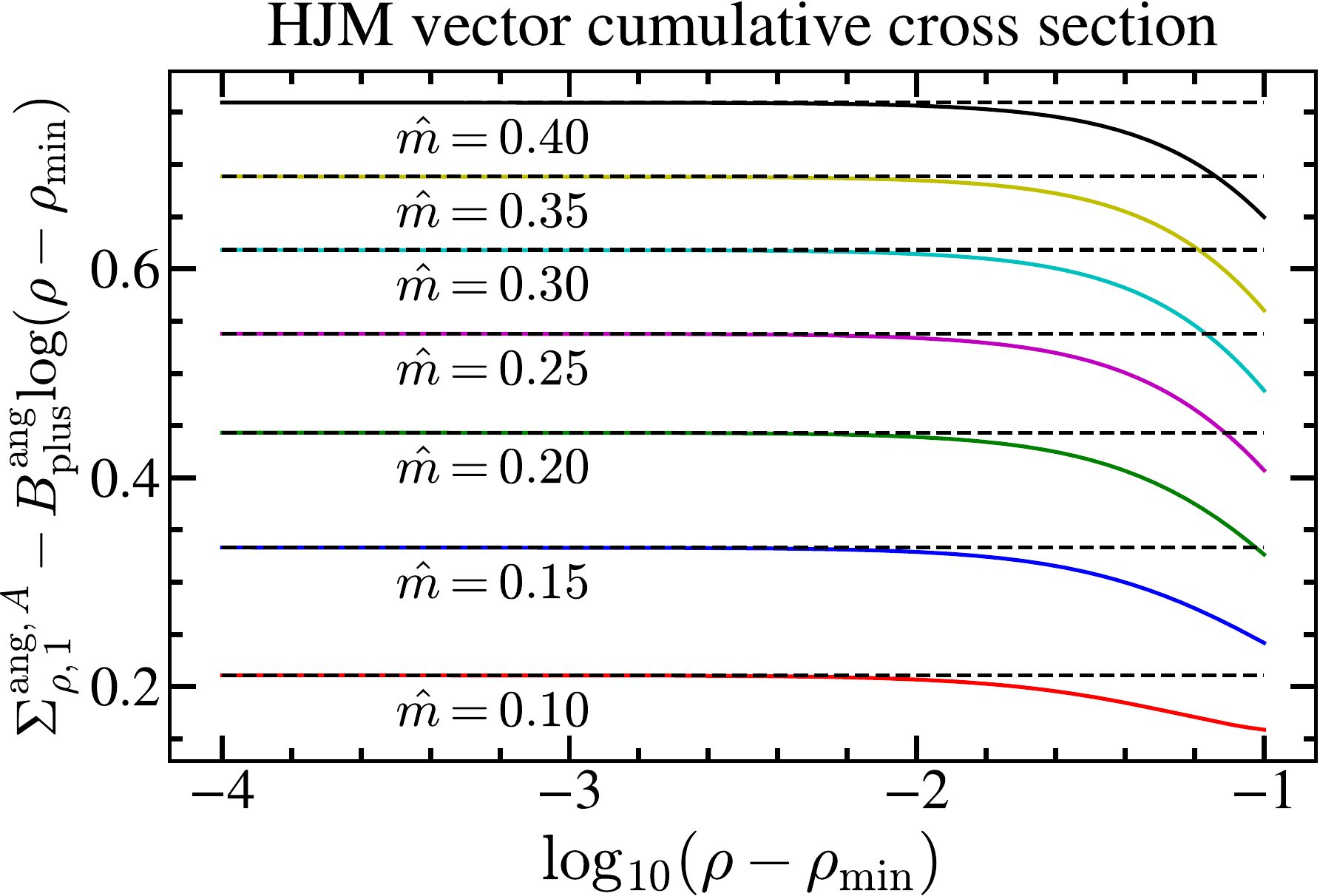}
\label{fig:VectorHJMCum}}
\caption{Panel (a): Limit of the differential cross section as taken in Eq.~\eqref{eq:lim1} for the
vector current, for $2$-jettiness (solid lines)
and heavy jet mass (dashed lines), for various values of the reduced mass $\hat m$.
The dotted lines indicate the analytic value of $(4/3)B_{\rm plus}^{\rm ang}$.
Panel (b): Limit of the axial-vector cumulative distribution as taken in the second line of
Eq.~\eqref{eq:Lim3} for $2$-jettiness (solid lines) and heavy jet mass (dashed lines), for a number of reduced
masses $\hat m$.
Lower panels: Limit of the vector current cumulative cross section as taken in the first line of Eq.~\eqref{eq:Lim3} for $2$-jettines
[\,panel(c)\,] and heavy jet mass [\,panel (d)\,] for various values of $\hat m$. Dashed
horizontal lines indicate the analytic values of $C_F A_e (\hat{m})$.}
\label{fig:Lim}
\end{figure*}
One can perform a number of tests on the numerical and analytic computations carried out in this article.
First, we have checked that both for $2$-jettiness and heavy jet mass, taking a numerical derivative of the
cumulative distribution accurately reproduces the differential one, including the kinks. Second,
the differential cross section for the vector current must verify the following condition (the
equivalent statement for the axial-vector current is that the limit is simply zero)
\begin{equation}\label{eq:lim1}
\lim_{e\to e_{\rm min}(\hat m)}[e - e_{\rm min}(\hat m)]g^{{\rm ang},V}_e(\hat m, e) =
\frac{4}{3} B_{\rm plus}^{\rm ang}(\hat m)\,.
\end{equation}
A graphical verification of this requirement can be found in Fig.~\ref{fig:VectorLim} for the vector current.

Third, integrating the differential cross section must yield the total cross section. For the vector current
this implies a constraint between the radiative tail of the differential cross section and the coefficients
of the singular distributions, while for the axial-vector current it is simply an integral condition:
\begin{align}
R_1^{{\rm ang},V} (\hat{m}) =\, &C_F\Biggl\{\int_{e_{\min}}^{e_{\max}} {\rm d} e \biggl[ \frac{3}{4}
g^{{\rm ang},V}_e(\hat m, e)
- \frac{B^{\rm ang}_{\rm plus} (\hat{m})}{e - e_{\min}} \biggr]\! +
B^{\rm ang}_{\rm plus} (\hat{m}) \log (e_{\max} - e_{\min}) + A_e^{\rm ang}
(\hat{m})\Biggr\}\,,\nonumber\\
R_1^{{\rm ang},A} (\hat{m}) =\, &\frac{3C_F}{4}\!\int_{e_{\min}}^{e_{\max}} {\rm d} e\,
g^{{\rm ang},A}_e(\hat m, e)\,.
\end{align}
We have verified that these constraints are satisfied for all event shapes discussed in this article.

An equivalent test on the cumulative cross section can be derived. To that end we note that
$\Sigma_{e,1}^{{\rm ang},V}$ can also be decomposed into singular and non-singular terms:
\begin{equation}
\Sigma_{e,1}^{{\rm ang},V} (\hat{m}, e) = C_F A^{\rm ang}_e(\hat m)\theta(e-e_{\rm min})
+ C_F B_{\rm plus}^{\rm ang} (\hat{m}) \log (e - e_{\min})
+ \Sigma_{e,{\rm NS}, 1}^{{\rm ang},V} (\hat{m}, e)\,,
\end{equation}
while for the axial-vector current one has only the non-singular term. For the cumulative non-singular cross section
(which is nothing more than the cumulative of the non-singular differential cross section)
implies $\Sigma_{e,{\rm NS}, 1}^{{\rm ang},C}(\hat m,e_{\rm min})=0$ (again, this is trivial to see
because the non-singular differential distribution is by definition integrable and therefore vanishes
if the lower and upper integration limits coincide). This can be translated into the
following constraints:
\begin{align}\label{eq:Lim3}
C_F A_{e}^{\rm ang} (\hat{m}) =\,& \lim_{e \rightarrow e_{\min}} [\Sigma_{e,1}^{{\rm ang},V} (\hat{m}, e) -
C_F B_{\rm plus}^{\rm ang} (\hat{m}) \log (e - e_{\min})]\,,\\
0 =\,&\lim_{e \rightarrow e_{\min}} \Sigma_{e,1}^{{\rm ang},A}(\hat{m}, e)\,.\nonumber
\end{align}
These conditions have been checked graphically, as can be seen for the axial-vector current in Fig.~\ref{fig:AxialCum}
(both for \mbox{$2$-jettiness} and heavy jet mass), and for the vector current in Figs.~\ref{fig:VectorTauCum}
and \ref{fig:VectorHJMCum} for \mbox{$2$-jettiness} and heavy jet mass, respectively.
\begin{figure*}[t!]
\subfigure[]
{\includegraphics[width=0.46\textwidth]{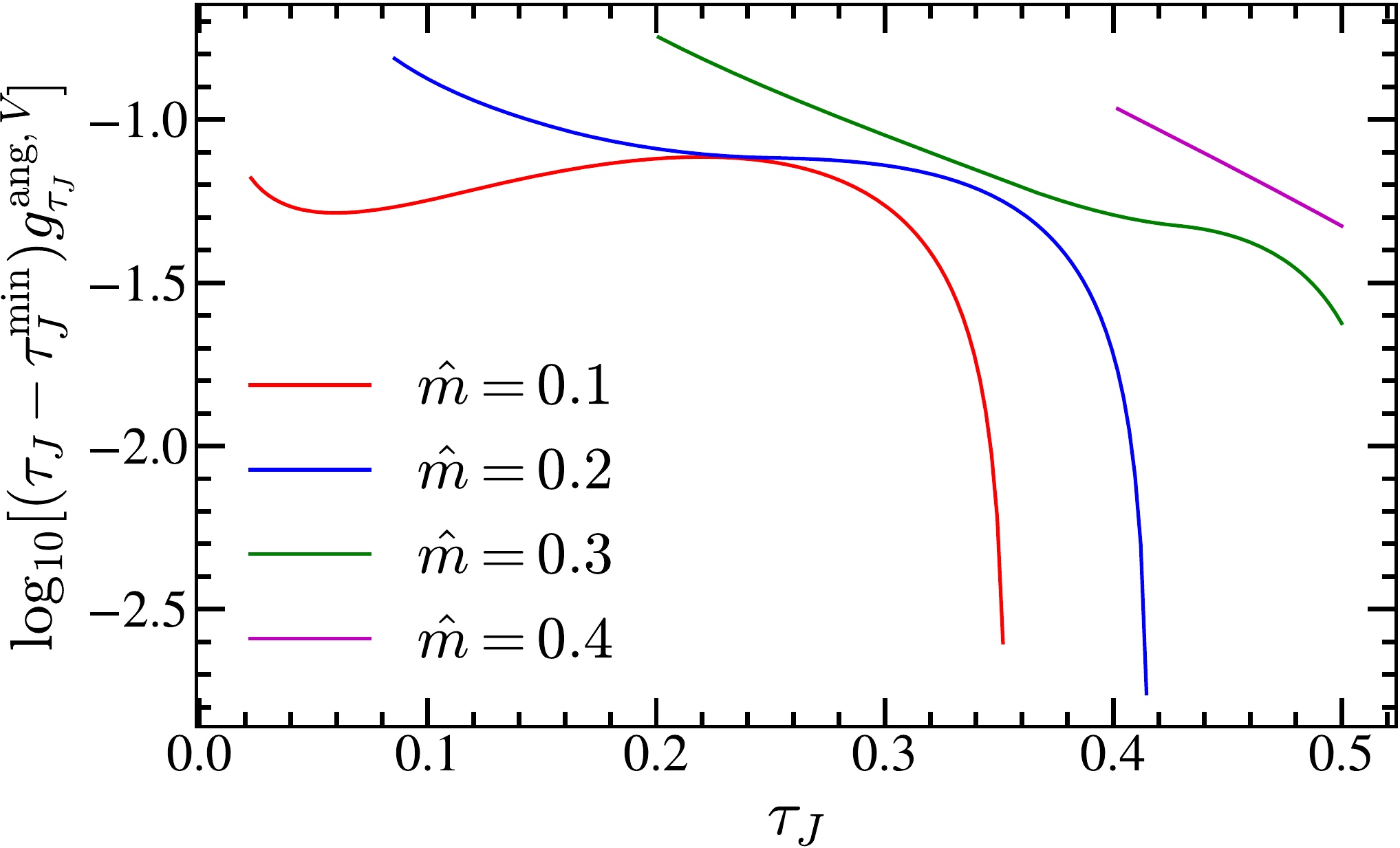}
\label{fig:tau-Diff}}
\subfigure[]{\includegraphics[width=0.46\textwidth]{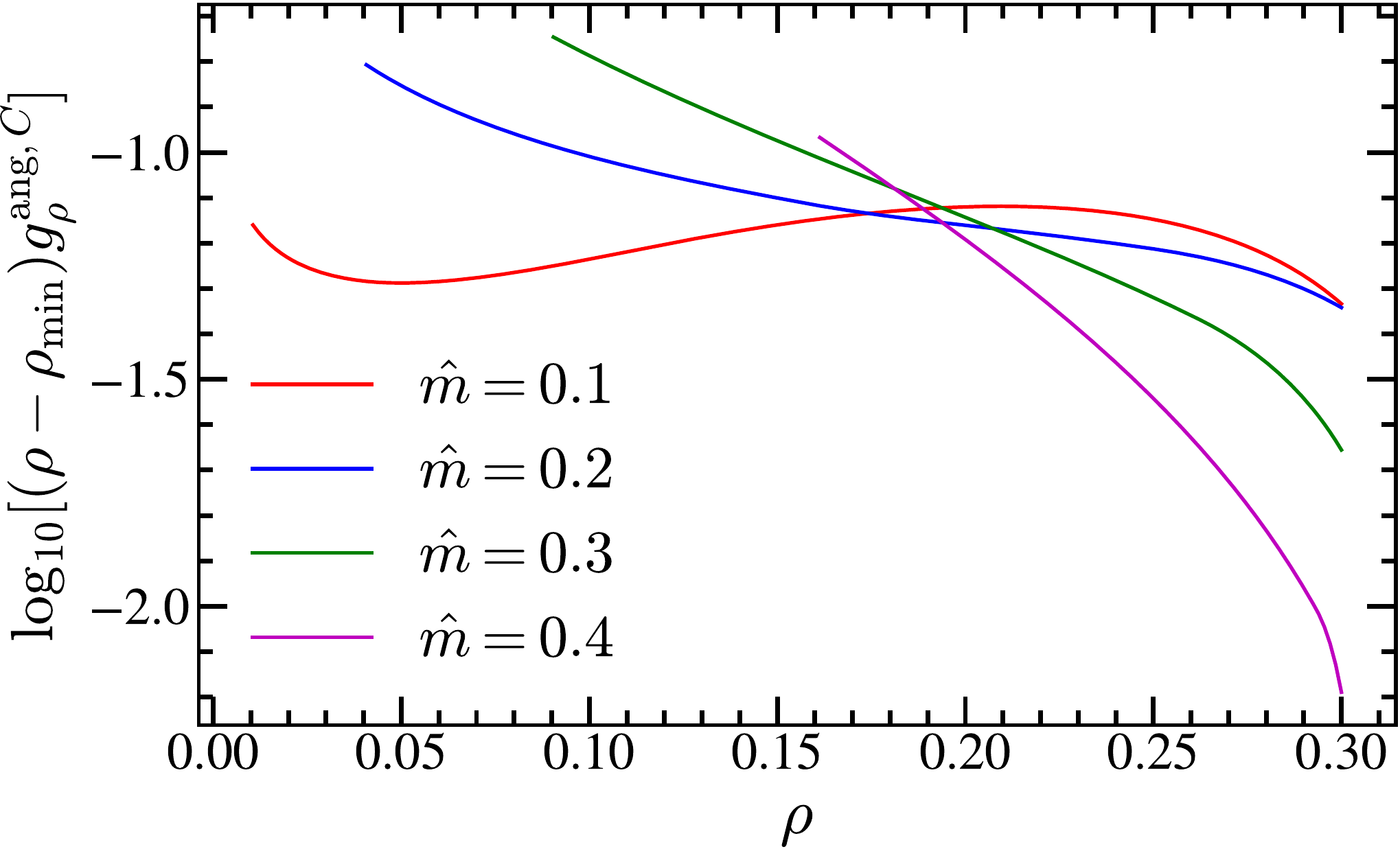}
\label{fig:rho-Diff}}
\caption{Same as Fig.~\ref{fig:EP} in the original massive scheme for $2$-jettiness and heavy jet mass.}
\label{fig:allDiff}
\end{figure*}
In Fig.~\ref{fig:EP} we show vector-current differential cross sections for a selection of event shapes in the
P- and E-schemes, for various values of the reduced mass $\hat m$. We multiply the results by the event shape value
$e$ in order to have a finite result at $e=0$. As discussed in Ref.~\cite{Lepenik:2019jjk}, both schemes have
a small sensitivity to the quark mass since $e^{E,P}_{\rm min}=0$. The E-scheme heavy-jet-mass distribution for
large values of the reduced mass presents kinks and discontinuities, while the rest of schemes, masses and
event shapes are quite smooth. In Fig.~\ref{fig:allDiff} we show the same distributions for $2$-jettiness and
heavy jet mass in their original (mass-sensitive) schemes.

\begin{figure*}[t!]
\subfigure[]
{\includegraphics[width=0.46\textwidth]{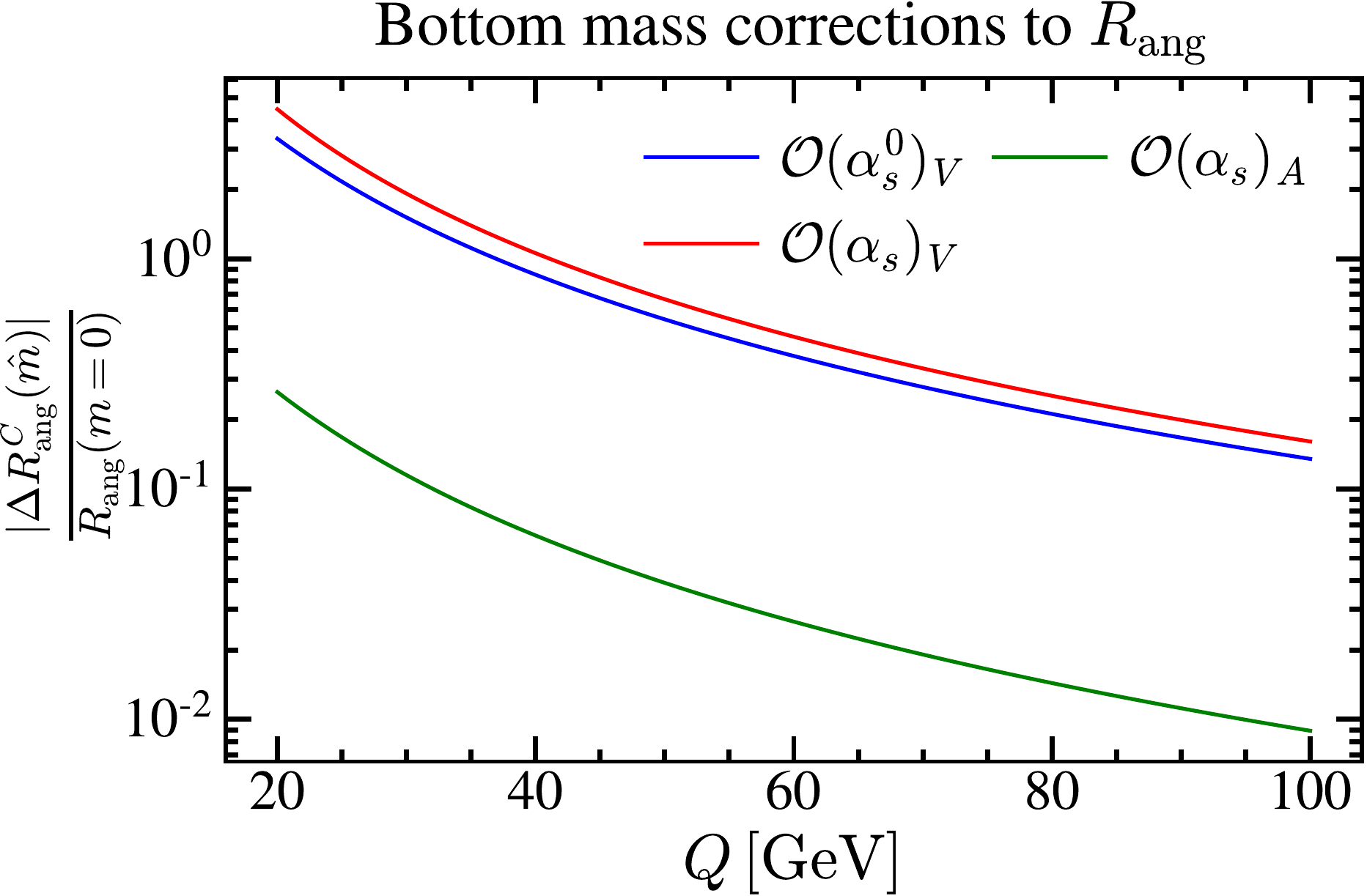}
\label{fig:tag}}~
\subfigure[]{\includegraphics[width=0.46\textwidth]{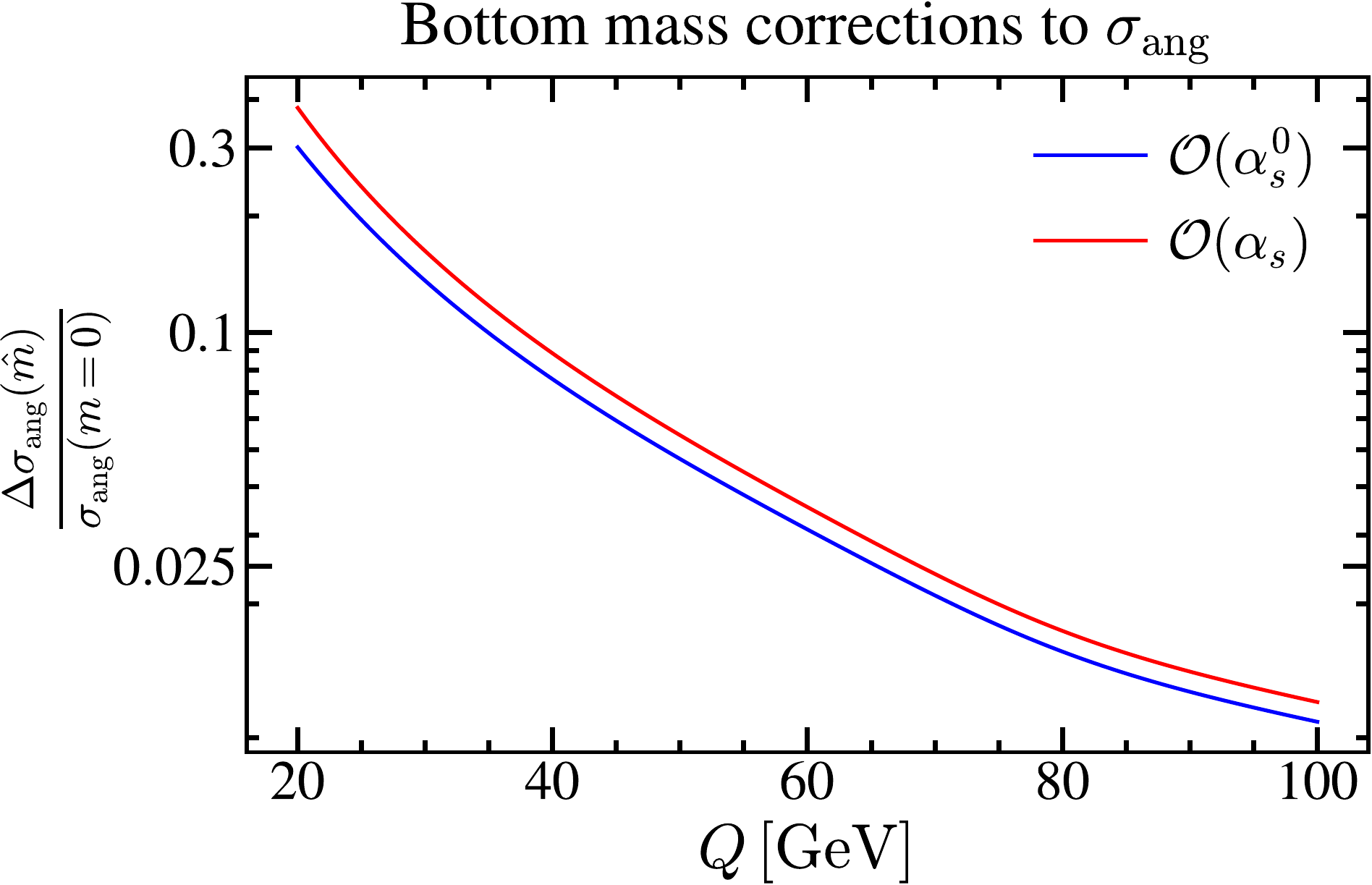}
\label{fig:notag}}
\caption{Bottom mass corrections to the total angular cross-section as a function of the center-of-mass energy $Q$
for bottom- and current-tagged measurements [\,panel (a)\,] or totally inclusive [\,panel (b)\,]. In blue and red we show
the LO and NLO results for the vector [\,panel (a)\,] or total [\,panel (b)\,] cross sections. The green line in panel (a)
corresponds to the axial-vector current, which starts only at $\mathcal{O}(\alpha_s)$.}
\label{fig:bottom}
\end{figure*}

We close this section quantifying the size of the bottom quark mass corrections to the total angular cross
section as a function of the center-of-mass energy $Q$, arguably the most relevant result in this article. For
this analysis we use the quark mass in the $\overline{\rm MS}$ scheme and set the renormalization
scale to its canonical value $\mu=Q$. Finally, we use the canonical reference values $\overline{m}_b(\overline{m}_b)=4.2\,$GeV
and $\alpha_s^{(n_f=5)}=0.1181$, which are evolved to $\mu=Q$ using \texttt{REvolver}~\cite{Hoang:2021fhn}.
First we assume that one can experimentally `tag' on bottom quarks and a given current and compute the ratio
of the bottom correction over the massless result for both currents at leading and next-to-leading order for
the vector current, and at the only available order for the axial-vector current ---\,that is, $\mathcal{O}(\alpha_s)$.
The vector current represents always a much bigger correction since it starts at $\mathcal{O}(\alpha_s^0)$ while
the massless results and the axial-vector current have no tree-level contribution. As expected, the correction is
significantly larger at smaller energies: for the NLO prediction, while at $Q=30\,$GeV the correction is
$1.85$ times larger than the massless approximation, at $Q=\{50,100\}\,$GeV it has already gone down to $\{64,16\}\%$,
as can be seen in Fig.~\ref{fig:tag}. For the axial-vector current the correction is always negative and at the three
energies just quoted amounts to $11\%$, $3.8\%$ and $9$ \textperthousand, respectively. A more realistic
comparison, presented in Fig.~\ref{fig:notag}, considers the inclusive measurement of the cross section, that is,
the incoherent sum of cross sections for all quarks lighter than the
top and including the two currents. We again consider the ratio (mass correction)/(massless approximation),
which is computed taking into account the electroweak factors of Eq.~\eqref{eq:EW}. We use the numerical values
$m_Z=91.1876\,$GeV, $\Gamma_{\!Z}=2.4952\,$GeV and $\sin(\theta_W)=0.23119$. Here the correction is milder,
but still sizable such that it has to be included in any precision analysis, in particular if it includes data at small
or intermediate energies. For the NLO prediction the massive correction is $38\%$, $4.27\%$ and $1.11\%$ for
$Q=20, 50$ and $100\,$GeV, respectively.

\section{Conclusions}\label{sec:conclusions}
Building on the earlier computations of Refs.~\cite{Mateu:2013gya,Lepenik:2019jjk} we have determined oriented
event-shape distributions for massive quarks up to $\mathcal{O}(\alpha_s)$, as well as the corresponding total angular
cross section. The tree-level computation reveals that, as opposed to the massless situation, there is a non-vanishing
result at lowest order for the vector current. Therefore we have worked out the angular-differential $2$- and $3$-body phase
space in $d=4-2\varepsilon$ dimensions, necessary to regulate the infrared singularities present in the virtual- and
real-radiation contributions, and checked that upon the integration of all angles the usual angular-inclusive phase space results
are reproduced. To construct the $d$-dimensional $3$-body phase space in a coherent way we use the Gram-Schmidt
procedure.

Our strategy is to project out the angular structure (in $d$ dimensions) at very early stages of the computation. This
causes that at intermediate steps some spurious terms, artifact of dimensional regularization, show up,
they cancel out when all terms are added. Other regularization methods such as giving the gluon a small mass would
not have this problem. However, having an additional energy scale would significantly complicate the computations,
on top of spoiling gauge invariance, so we have discarded this possibility. After adding real- and virtual-radiation contributions we end
up with a result free from infrared divergences. Moreover, we have checked that upon sending the quark mass to zero
the massless results of Ref.~\cite{Mateu:2013gya} are recovered. For the vector current we find a universal coefficient
multiplying the plus function, and derive a closed form for the coefficient of the Dirac delta function that has dependence
on the observable only through an integral, which was already encountered (and solved analytically for many event
shapes) in Ref.~\cite{Lepenik:2019jjk}. We also provide results for the total angular cross-section in terms of
integrals that can be easily computed numerically, and find that this observable is enhanced when the massive quarks
are slow for the vector current. This indicates that the total angular cross-section might be an interesting observable
to determine the top quark mass at a future linear collider through threshold scans.

Our results have been used to compute the differential and cumulative distributions for any observable with a simple adaption
of the algorithm described in Ref.~\cite{Lepenik:2019jjk}. We have however focused in deriving and discussing in
detail theoretical expressions for the most prominent event-shapes: $2$-jettiness and thrust. We compute analytically the differential
cross section for both, which can be expressed in terms of two functions, common for the two observables discussed.
For the cumulative cross sections we obtain results written as one-dimensional integrals of functions that appeared
already in the expressions for the total angular cross-section. For heavy jet mass we discovered that, for reduced masses
larger than a certain value, the distribution is zero on a finite patch which is located between the minimal and maximal
values of $\rho$. In this ``island'', the cumulative cross section is constant.

We have described how to treat the polar-angle phase-space integrals such to make them easily implementable in a Monte Carlo
program. We have indeed coded such a Monte Carlo to numerically project out the angular cross section and compute, at the
same time, binned distributions. We observe that, due to the fact that weights are not necessarily positive, the convergence
of the integration is quite slow, resulting in large error bars and jumpy central values unless huge statistics are used. We have also
generated tail binned cross sections modifying the default parameters of \texttt{MadGraph5} and conclude that, on top of this problem, its
non-zero IR regularization cutoff causes a bias in the dijet part of the spectrum.

We have made extensive analytic (carrying out the computations independently in two different approaches) and numerical
(having independent codes that agree to machine precision) tests, as well as some sanity checks on our analytic and
numeric results, all of them successful. Finally, we have numerically explored the size of the massive corrections, finding
that for a realistic environment in which one is flavor blind, the corrections due to the non-zero mass of the bottom quark
are as important as $16\%$ percent at center-of-mass energies of about $30\,$GeV, and therefore must be included in
any analysis that aims for high precision.

An obvious immediate application of our computation is determining the strong coupling from fits to experimental data
on the total angular cross-section from LEP and other colliders. This observable is highly convenient, since it is less
affected by hadronization effects than differential event-shape distributions, but at the same time (for massless quarks)
is directly proportional to $\alpha_s$. Finally, due to its inclusive nature, it should be free from large logarithms.
On the theory side, possible extensions of our work include generalizing the factorization theorem derived in
Ref.~\cite{Hagiwara:2010cd} to massive quarks, and computing the corresponding subleading jet function to NLO.
Lastly, one can study unstable top quarks. Since their decay products
are not always produced in the same hemisphere, the thrust axis can be modified with respect to the stable-top
approximation. Therefore, such off-shell effects will affect the event-shape distribution as well as the total angular cross-section.

\subsection*{Acknowledgements}
This work has been supported by the MECD grant PID2019-105439GB-C22, the IFT Centro de Excelencia Severo Ochoa
Program under Grant SEV-2012-0249, the EU STRONG-2020 project under the program
H2020-INFRAIA-2018-1, grant agreement No.\ 824093 and the COST Action CA16201 PARTICLEFACE.
N.\,G.\,G.\ is supported by a JCyL scholarship funded by the regional government of Castilla y Le\'on and European Social Fund,
2017 call.
A.\,B.\ is supported by an FPI scholarship funded by the Spanish MICINN under grant no.\ BES-2017-081399,
and thanks the University of Salamanca for hospitality while parts of this work were completed.

\bibliography{thrust3}
\bibliographystyle{JHEP}
\end{document}